\documentclass[elsart12,epsf,eqsecnum,graphics,cite,nofootinbib]{revtex4}

\usepackage{dsfont}
\usepackage{amsmath,amssymb}
\usepackage{graphicx}
\usepackage{color}
\usepackage{slashed}

\usepackage{epsfig,epsf}
\usepackage{graphicx,color}
\usepackage{grffile}
\usepackage{amsmath,amsfonts,amssymb,amsthm,nccmath,latexsym,mathtools} \usepackage[mathscr]{euscript}

\def\beq{\begin{equation}}
\def\eeq{\end{equation}}
\newcommand{\bea}{\begin{eqnarray}}

\newcommand{\eea}{\end{eqnarray}}
\newcommand{\nn}{\nonumber}

\usepackage{amsmath}
\usepackage{epsfig,epsf}
\usepackage{graphicx}
\usepackage{verbatim}
\usepackage{float}
\begin{document}

% Yale printer values
%\voffset1.5cm
\title{Quark correlations in the Color Glass Condensate: Pauli blocking and the ridge}

\author{Tolga Altinoluk$^{a,b}$, N\'estor Armesto$^c$, Guillaume Beuf$^d$, Alex Kovner$^e$, Michael Lublinsky$^{f,e}$}

\affiliation{$^a$ CENTRA, Instituto Superior T\'{e}cnico, Universidade de Lisboa, Av. Rovisco Pais, P-1049-001 Lisboa, Portugal\\
$^b$ Laborat\'{o}rio de Instrumenta\c{c}\~{a}o e F\'{i}sica Experimental de Part\'{i}culasÊ - LIP, Lisbon, Portugal\\
$^c$ Departamento de F\'{i}sica de Part\'{i}culas and IGFAE, Universidade de Santiago de Compostela, 15782 Santiago de Compostela, Galicia-Spain \\
$^d$ European Centre for Theoretical Studies in Nuclear Physics and Related Areas (ECT*) and Fondazione Bruno Kessler, Strada delle Tabarelle 286, I-38123 Villazzano (TN), Italy \\
$^e$ Physics Department, University of Connecticut, 2152 Hillside Road, Storrs, CT 06269, USA\\
$^f$ Physics Department, Ben-Gurion University of the Negev, Beer Sheva 84105, Israel
}

\begin{abstract}
{
We consider, for the first time, correlations between produced quarks in p-A collisions in the framework of the Color Glass Condensate.  We find a quark-quark ridge that shows a dip at $\Delta\eta\sim 2$ relative to the gluon-gluon ridge. The origin of this dip is the short range (in rapidity) Pauli blocking experienced by quarks in the wave function of the incoming projectile. We observe that these correlations, present in the initial state, survive the scattering process. We suggest that  this effect may be observable in open charm-open charm correlations at the Large Hadron Collider.
}
\end{abstract}

\bibliographystyle{h-physrev4}
\maketitle

%%%%%%%%%%%%%%%%%%%%%%%%%%%%%%%%%%%%%%%%%%%%%%%%%%%%%%%%%%%%%%%%%%%%%%
%%%%%%%%%%%%%%%%%%%%%%%%%%%%%%%%%%%%%%%%%%%%%%%%%%%%%%%%%%%%%%%%%%%%%%
\section{Introduction}
The ridge correlation observed in p-p collisions at the Large Hadron Collider (LHC) has been in the center of interest of the heavy-ion community for several years.
First seen in high-multiplicity collisions by the CMS \cite{CMS}  and ATLAS \cite{Aad:2015gqa} collaborations at the LHC, similar correlations have been subsequently observed by all four large LHC experiments in p-Pb collisions \cite{CMS:2012qk}, and much more detailed studies of the properties of these correlations are available today.
Even more exciting, recently data by ATLAS \cite{Aaboud:2016yar} and CMS \cite{Khachatryan:2016txc} suggest the existence of the ridge in p-p events with multiplicities close to those in minimum bias collisions, both at $\sqrt{s}=2.76$ and 13 TeV.

Two main lines of explanations are discussed at present. One is based on a collective (hydrodynamic?) behavior of the system produced in the collision \cite{Bozek:2012gr} in an analogous manner as in heavy-ion collisions. The other one is based on the  Color Glass Condensate (CGC) \cite{jimwlk,Bal,cgc} framework to describe high-energy Quantum Chromodynamics in a weak coupling but nonperturbative regime. Within the latter, a quantitative description of the data is achieved \cite{DV} in the ``glasma graph'' approach \cite{DMV,ddgjlv} which ascribes the origin of the correlations entirely to the structure of the initial state. Other mechanisms  within the CGC framework \cite{correlations,LR} exist as well  (see also other proposals in  \cite{Hwa:2008um}).  Though it is likely that both mechanisms, corresponding to final and initial state effects, are contributing to the correlations (probably in different transverse momentum ranges), the  new p-p data mentioned above make the hydrodynamical description somewhat questionable  and the possible initial state origin of the correlations more credible.

Within the  "glasma graph" approach,  we showed  recently \cite{Altinoluk:2015uaa} that the physics underlying this contribution is the Bose enhancement of gluons in the projectile wave function. The effect is long range in rapidity since  the CGC wave function is dominated by the rapidity integrated mode of the soft gluon field.

A natural question to ask, never addressed in detail before,  is whether quarks (or antiquarks) in the CGC  are also subject to correlations. One expects quarks to experience Pauli blocking, and thus the probability to find two identical quarks with the same quantum numbers  in the CGC state should be suppressed. Such suppression, if it exists, should be observable experimentally. One anticipates this effect to be significantly smaller than for gluons, since quarks in the CGC wave function are  generated only via  gluon splitting, and thus their number is ${\cal O}(\alpha_s)$ suppressed. This makes quark pair correlation an ${\cal O}(\alpha_s^2)$ effect. Nevertheless, since the relevant coupling constant is not very small, the effect may be   observable, and is thus a worthwhile subject of study. This is the aim of the present work.

An interesting question is, in particular, whether  the Pauli blocking effect is long range in rapidity or not. The answer is not obvious a priori, since although the quarks themselves are produced via splitting off rapidity invariant gluons, the splitting probability itself depends on the rapidity of the quark and the antiquark. This is one of the questions we want to study in this paper.
As we will show, the Pauli blocking effect is indeed present, but it is short range in rapidity.
Another interesting, albeit somewhat technical point, is what is the relevant $N_c$ dependence. We will find that the suppression of Pauli blocking with respect to Bose enhancement is not ${\cal O}(\alpha_s^2)$ but  rather ${\cal O}(\alpha_s^2N_c)$, which is quite moderate for $\alpha_s\sim 0.2$ and $N_c=3$.

A natural candidate for the observation of such effects is open charm-open charm correlations that are expected to be less gluon-dominated than light hadrons.\footnote{The heavy quark mass needs to be included in the calculation for open charm-open charm correlations. This effect adds technical complexity to the calculation. Therefore, it  is neglected in this exploratory work and left for future studies.} Data from the LHCb collaboration \cite{Aaij:2012dz,Aaij:2013mga,Aaij:2015wpa}
 exist on such process. LHCb provides the cross sections but in the forward rapidity region - while our approach is suitable for the central rapidity region, and  correlations have not been analyzed until now. These data are currently discussed in the context of single versus multiple parton interactions in collinear and $k_T$-factorization, see e.g. \cite{Likhoded:2015zna,Blok:2016lmd} and \cite{vanHameren:2015wva} respectively. Another interesting possibility would be the contribution of quark-quark correlations to the difference between the azimuthal correlations of equal and opposite sign charged particles, which have been measured to be of similar magnitude in p-Pb and Pb-Pb collisions at the LHC \cite{Khachatryan:2016got}. Naturally, one would expect Pauli blocking to contribute only to the equal sign charged particle correlations, and decrease them at $\Delta\phi=0$.

The paper is organised as follows. In Section 2, we derive the expression for the number of quark pairs in the CGC wave function to lowest order in $\alpha_s$. We show that it contains a correlated part which suppresses the number of pairs
at like values of transverse momenta - the Pauli blocking contribution. This contribution is short range, in the sense that it decreases as a function of the rapidity difference between the two quarks. However, the natural exponential decrease is tempered by a rather high power of rapidity difference. As a result, this contribution can be sizeable even for significant rapidity separations.
In Section 3, we consider the double inclusive quark production in a scattering process. We concentrate on the kinematic regime where the saturation momentum of the target is relatively small, so that  the initial state correlations  have the best chance of being reflected in the spectrum of particles produced in the final state. We show that the basic features of quark pair correlations in the wave function are indeed preserved by the production process. There are, however, some important differences, which we comment on.
Finally,  Section 4 contain a short discussion of our results. Details of the calculations are presented in the Appendices.

%\section{Quarks}

\section{Pauli blocking in the projectile wave function}
\label{sec:Pauli}

Throughout this paper we will be working in the standard CGC framework, following the conventions in \cite{Lublinsky:2016meo}. We consider a left moving target that is described by the Weizs\"acker-Williams field $\alpha^a(x)$ and its saturation scale is denoted by $Q_T$. On the other hand, the right moving projectile  whose wave function describes the distribution of the soft Weizs\"acker-Williams gluons accompanying the valence color charge density $\rho^a(x)$ and we denote the saturation scale of this projectile as $Q_s$.  The production of soft  gluons from the valence charges is  treated eikonally.
The sea quarks are produced in this wave function from the soft gluons by perturbative splitting. This splitting is not eikonal, and full perturbative kinematics is retained in the calculation.

The distribution of the color charge densities will be, for simplicity, taken from the McLerran-Venugopalan \cite{mv} model. Again for simplicity, we will assume translational invariance of the projectile wave function in the transverse space. This, as always, will lead to a spurious $\delta$-function structure of some of the correlated cross section, which in a realistic case is smeared by the inverse size of the projectile.
Additionally, we will be working in the leading $N_c$ approximation.

\subsection{Quark contribution to the wave-function}
\label{sec:quark}

 Let  $d^\dagger$ and $d$ denote quark creation and annihilation operators,
while $\bar d^\dagger$ and $\bar d$ are those of the antiquark. Perturbatively the quarks and antiquarks appear in the light-cone wave function of a valence charge either via instantaneous interaction, or via splitting of a soft gluon, see details in Appendix A. The quark-antiquark component of the light cone wave function of a "dressed" color charge density is given by\footnote{In addition, the state to this order in perturbation theory contains  one-gluon and two-gluon components. We do not indicate those explicitly, as they do not contribute to correlated quark production.}
\bea
\label{vd}
|v\rangle^D_2
&=&
(1\,-\,g^4\,\kappa_4)\,|v\rangle
%\\
%\nonumber \\
%&+&
+
g^2\,
\int {dk^+d\alpha\,d^2p\,d^2q\over (2\,\pi)^3}\ 
\left[\zeta^{\gamma\delta}_{s_1 s_2}(k^+,p,q,\alpha)\ d^{\dagger \gamma}_{s_1}(q^+,q)\,
\bar d_{s_2}^{\dagger\delta}(p^+,p)\right]\,|v\rangle,
\eea
%\end{eqnarray}
%
%\begin{eqnarray}\label{vd}
%|v\rangle^D_2&=&(1\,-\,g^4\,\kappa_4)\,|v\rangle\ \nonumber \\
%&+&\ g^2\,
%\int {d^3p\,d^3q\over (2\,\pi)^3}\ \left[\zeta^{\gamma\delta}_{s_1 s_2}(k^+,p,q,\alpha)\ d_{\gamma,s_1}^\dagger(q)\,
%\bar d_{\delta,s_2}^\dagger(p)\right]\,|v\rangle,
%\end{eqnarray}
where $|v\rangle$ denotes a valence state, $g$ is the Yang-Mills coupling, $\kappa_4$ is a  constant (virtual correction) ensuring the correct normalisation of the dressed state, and $\gamma,\delta$ are color indices. The value of $\kappa_4$ is unimportant for us in this paper.
%
%\beq
%\zeta\,=\,\zeta^1\ +\ \zeta^2
%\eeq
%
%\beq
%\zeta^{1\,\gamma\delta}_{s_1 s_2}(p,q)\,=\,{
%\langle 0| d_{\alpha\,s_1}(q)\;\bar d_{\beta\,s_2}(p)\,\delta H^{\rho\,qq}\,|0\rangle\over
%g^2\,(2\,\pi)^3\,(E_q\,+\,E_p)}\,
%=\,{\tau^a_{\alpha\,\beta}\,\rho^a(-p_\perp-q_\perp)
%\over (2\,\pi)^{3}\,(q^{+}\,+\,p^+)^{2}\,(E_q\,+\,E_p) }\,\delta_{s_1 s_2}
%\eeq
%
%
%\bea
%\zeta^{2\,\gamma\delta}_{s_1 s_2}(p,q)&=&\int d^3k\,{
%\langle 0| d_{\alpha\,s_1}(q)\;\bar d_{\beta\,s_2}(p)\,\delta H^{g\,qq}\,a^{\dagger\,a}_i(k)\,|0\rangle\
%\langle 0|\,a^a_i(k)\,\delta H^{\rho\,g}\,|0\rangle \over
%g^2\,(2\,\pi)^6\,(E_q\,+\,E_p)\ E_k}\,= \nn \\
%&=&{\tau^a_{\gamma\delta}\ \Gamma^i_{s_1 s_2}(p+q,p)\ (p_i\,+\,q_i)\ \rho^a(-p_\perp\,-\,q_\perp)\over 2\,(2\,\pi)^{3}\,
%(p_\perp\,+\,q_\perp)^2\ (p^++q^+)\, (E_{q}\,+\,E_p)}
%\eea
We define the longitudinal momentum fraction $\alpha$  as
\beq
 p^+ = \alpha k^+,\ \
q^+ = \bar \alpha k^+, \ \  \bar\alpha =1-\alpha,
\eeq
with $k$ the momentum of the parent gluon that splits into a quark and an antiquark. The splitting amplitude $\zeta$ is given by
\begin{equation}
\label{zeta_def}
\zeta^{\gamma\delta}_{s_1 s_2}(k^+,p,q,\alpha)\ =
\tau^a_{\gamma\delta}\, \int \frac{d^2k}{(2\pi)^2}\,\rho^a(k)\ \phi_{s_1 s_2}(k,p,q;\alpha),
\end{equation}
where $\tau^a$ are the generators of $SU(N_c)$ in the fundamental representation. Here, 
\beq
\phi\,=\,\phi^{(1)}\ +\ \phi^{(2)},
\eeq
where
\begin{equation}
 \phi^{(1)}_{s_1 s_2}(k,p,q;\alpha) =-\delta_{s_1s_2}\frac{2\alpha\bar\alpha}{\bar\alpha p^2+\alpha q^2}(2\pi)^2 \delta^{(2)}(k-p-q)
\end{equation}
and
\bea
\phi^{(2)}_{s_1s_2}(k,p,q;\alpha)
%&=&\frac{\alpha\bar\alpha k_i\,\Gamma^i(k,p)\,k^+}{ k^2_\perp\,\left[\bar\alpha p^2\,+\,
%\alpha q^2\right]}\delta^{(2)}(k-p-q)\\
=
\frac{1 }{ k^2\,\left[\bar\alpha p^2\,+\,
\alpha q^2\right]}
%\nn\\
%&\times&
\left\{2\alpha\bar\alpha k^2-\left(\bar\alpha k\cdot p+\alpha k\cdot q\right)+2i\sigma^3k\times p\right\} (2\pi)^2\delta^{(2)}(k-p-q).
\eea
Thus,
\begin{equation}
\phi_{s_1s_2}(k,p,q;\alpha)=\phi_{s_1s_2}(k,p;\alpha) (2\pi)^2\delta^{(2)}(k-p-q)
\end{equation}
with
\bea
\phi_{s_1s_2}(k,p;\alpha)=
\frac{1 }{ k^2\,\left[\bar\alpha p^2\,+\,
\alpha (k-p)^2\right]} 
%\nn \\
%&\times&
\Big\{-\left[\bar\alpha k\cdot p+\alpha k\cdot (k-p)\right]+2i\sigma^3k\times p\Big\}.
\eea

The $\phi^{(1)}$ term comes from the instantaneous interaction, while $\phi^{(2)}$ from the soft gluon splitting.
To probe quark-quark correlations we are interested in the two quark-two antiquark component of the dressed  state.
We will adopt the same strategy as was used in the glasma graph calculation. That is, we focus on terms enhanced
by the charge density in the wave-function. Thus, at the lowest order it is given by
\begin{eqnarray}\label{v4}
|v\rangle^D_4&=&{\rm virtual} +\ \frac{g^4}{2}\,
\int {dk^+d\alpha\,d^2p'\,d^2\bar{p}'\over (2\,\pi)^3}
{d\bar k^+d\beta\, d^2q'\, d^2\bar{q}'\over (2\,\pi)^3}\\\
&\times& \left[
\zeta^{\epsilon\iota}_{s'_1 s'_2}(k^+,p',\bar{p}';\alpha)
\zeta^{\gamma\delta}_{r_1 r_2}(\bar k^+,q',\bar{q}';\beta)\ 
d_{s'_1}^{\dagger \epsilon}(\bar{\alpha}k^+, p')\,
\bar d_{s'_2}^{\dagger\iota}(\alpha k^+,\bar{p}')\ 
d_{r_1}^{\dagger\gamma}(\bar{\beta}\bar{k}^+, q')\,
\bar d_{r_2}^{\dagger\delta}(\beta\bar{k}^+,\bar{q}')\right]|v\rangle. \nonumber
\end{eqnarray}

%%%%%%%%%%%%%%%%%%%%%%%%%%%%%%%%%%%%%%%%%%%%%%%%%%%%%%%%%%%%%%%
%%%%%%%%%%%%%%%%%%%%%%%%%%%%%%%%%%%%%%%%%%%%%%%%%%%%%%%%%%%%%%%

\subsection{ Pauli blocking}
Our first order of business is to calculate correlations between the quarks in the CGC wave function. In the next Section, we will see  how these correlations translate into correlations between particles produced in a collision.

Our aim is to calculate the average of the number of quark pairs in the wave function that is formally defined, see e.g. \cite{Greiner:1998jw}, as
\bea
\label{Npair}
{dN\over dp^+d^2pdq^+d^2q}\,=\frac{1}{(2\pi)^6}\,\left\langle ^{D}_{4}\langle v|
%\Omega\,\hat S^\dagger\,
%\Omega^\dagger\,\,\,
d^{\dagger}_{\alpha,s_1}(p^+,p)d^{\dagger}_{\beta,s_2}(q^+,q)\,d_{\beta,s_2}(q^+,q) \,\,d_{\alpha,s_1}(p^+,p)\,
%\nn\\
%&&
%\hspace{2cm}
%\times\;
 %: \,\,
%\Omega\,\hat { S}\,\Omega^\dagger
|v\rangle^D_4\, \right\rangle_P \; ,
%|0\rangle.
\eea
i.e. first, we need to calculate the expectation value of the ''number of quark pairs" in our dressed state $|v\rangle_4^D$, and then, average over the color charge densities in the projectile.  

The final result, derived in Appendix B, reads
\bea
\label{Npair_square_final}
\frac{dN}{d\eta_1d^2pd\eta_2d^2q}&=&\frac{1}{(2\pi)^4}g^8\int 
%\frac{d^2k}{(2\pi)^2} \frac{d^2\bar{k}}{(2\pi)^2} \frac{d^2l}{(2\pi)^2} \frac{d^2\bar{l}}{(2\pi)^2}
d^2k \, d^2\bar{k} \, d^2l \, d^2\bar{l} \; 
\langle\rho^a(k) \rho^c(\bar{k}) \rho^b(l) \rho^d(\bar{l}) \rangle_P \nonumber\\
&&
\times
\Bigg\{ {\rm tr}(\tau^a\tau^b) {\rm tr}(\tau^c\tau^d) \Phi_2(k,l; p) \Phi_2(\bar{k},\bar{l}; q)%\nonumber\\
%&&
-{\rm tr}(\tau^a\tau^b\tau^c\tau^d) \Phi_4(k,l,\bar{k},\bar{l};p,q)\Bigg\},
\eea
where $\rho^a(k)$ and $ \rho^b(\bar{k})$ are the color charge densities in the amplitude and $\rho^c(l)$ and $\rho^d(\bar{l})$ are the color charge densities in the complex conjugate amplitude. The rapidities are defined as $\eta_1=\ln(p^+_0/p^+)$ and $\eta_2=\ln(p^+_0/q^+)$, with $p^+_0$ some reference +-momentum. The functions $\Phi_2$ and $\Phi_4$ are defined respectively as 
\beq
\label{Phi_2_def}
\Phi_2(k,l; p)\,\equiv\,\int_0^1 \,d\alpha\int \frac{d^2\bar{p}'}{(2\pi)^2}\sum_{s_1s_2}\
\phi_{s_1,s_2}(k,p,\bar{p}';\alpha)\ \phi_{s_1,s_2}^*(l,p,\bar{p}';\alpha)
\eeq
and
\bea
\label{Phi_4_def}
\Phi_4(k,l,\bar k,\bar l;p,q)&\equiv& \sum_{s_1,s_2,\bar s_1,\bar s_2} \int_0^1 \,
{d\alpha\,d\beta\over (\beta+\bar\beta e^{\eta_1-\eta_2})  (\alpha+\bar\alpha e^{\eta_2-\eta_1})}\,  \\
&&\hspace{-1.5cm}
\times\int \frac{d^2\bar{p}'}{(2\pi)^2}\frac{d^2\bar{q}'}{(2\pi)^2} 
\phi_{s_1s_2}(k,p,\bar{p}';\alpha)\ 
\phi_{\bar s_1\bar s_2}(\bar k,q,\bar{q}';\beta)\
\phi_{s_1\bar s_2}^*(l,p,\bar{q}';\beta)\,
\phi_{\bar s_1s_2}^*(\bar l,q,\bar{p}';\alpha). \nn
\eea
%where we introduce the notation $\int_q\equiv \int d^2q$.
The integrals represent "inclusiveness" over the antiquarks.
The integrals over $\bar p, \ \bar q$ reduce the number of $\delta$-functions to two, so that in general we can write
\bea
\Phi_4(k,l,\bar k,\bar l;p,q)&=& \sum_{s_1\,s_2,\bar s_1,\bar s_2} \int_0^1 \,
{d\alpha\,d\beta\over (\beta+\bar\beta e^{\eta_1-\eta_2})  (\alpha+\bar\alpha e^{\eta_2-\eta_1})}
\,
 \\
%\hspace{-2cm}
&\times&
\phi_{s_1s_2}(k,p;\alpha)\; \phi_{\bar s_1\bar s_2}(\bar k,q;\beta)\;  \phi_{s_1\bar s_2}^*(\bar k-q+p,p;\beta)\;
\phi_{\bar s_1s_2}^*(k+q-p,q;\alpha) \nn \\
&\times&
(2\pi)^2\delta^{(2)}(\bar l-k-q+p)\; (2\pi)^2\delta^{(2)}(l-\bar k+q-p).\nn
\eea

%where the normal ordering prescription has been imposed in order to avoid counting twice the same quark.
%
%As shown in Appendix B, this quantity
The quark pair density has two contributions. One contribution is proportional to  $\Phi_4$ and the other is proportional $\Phi_2\Phi_2$.\footnote{The contribution proportional to $\Phi_4$ comes with a minus sign due to the anticommutation relations between the quark and antiquark creation and annihilation operators. Therefore it is due to Pauli blocking. This fact will also be apparent in that it results from an odd number of quark loops, in contrast to the $\Phi_2\Phi_2$ contribution.}$^,$\footnote{ 
We would like to emphasize at this point that the $\Phi_4$ contribution has rapidity dependence while the $\Phi_2$ contribution is independent of  rapidity. These rapidity dependent denominators stem from the integrations over the longitudinal momenta. The fact that $\Phi_4$ is rapidity dependent is simply because $\Phi_4$ mixes the longitudinal momenta of the quarks and antiquarks between the two different $q\bar q$-pairs in the wave function, as opposed to the factorized $(\Phi_2)^2$ contribution. This is the origin of the short range rapidity nature of the quark pair density.} 
However, in the large $N_c$ limit the  interesting part of the contribution is  given by $\Phi_4$.  The diagrams that correspond to $\Phi_2\Phi_2$ yield an uncorrelated contribution which is ${\cal O}(N_c^4)$ and correlated terms ${\cal O}(N_c^2)$. On the other hand, the leading $\Phi_4$ term is ${\cal O}(N_c^3)$, and, thus, dominates the correlations. The $N_c$ counting of the diagrams originating form $\Phi_2\Phi_2$ is illustrated on Figures 1, 2 and 3.
\begin{figure}[hbt]
% [pb]\label{}
\begin{center}
\vspace*{-0.2cm}
\includegraphics[width=6.5cm]{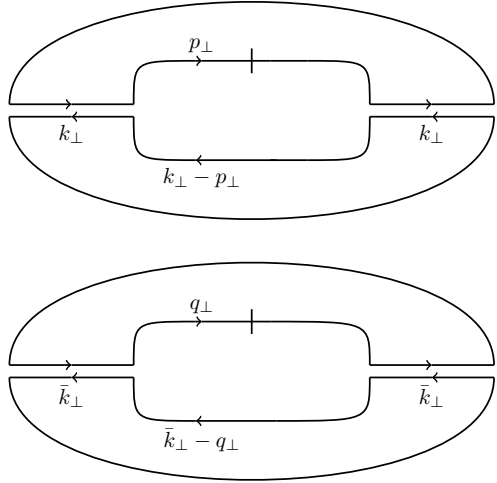}
%\includegraphics[width=8cm]{Phi2Nc3.eps}
%\vspace*{-0.12cm}
\end{center}
\caption{The uncorrelated contribution originating from $(\Phi_2)^2$. We work at large $N_c$ where gluons are represented as double lines, and the short vertical lines indicate that it corresponds to an observed particle. Arrows indicated the color flux while momenta flow from left to right.}
\end{figure}
\begin{figure}[hbt]
 %[pb]\label{}
\begin{center}
\vspace*{-0.2cm}
\includegraphics[width=6.5cm]{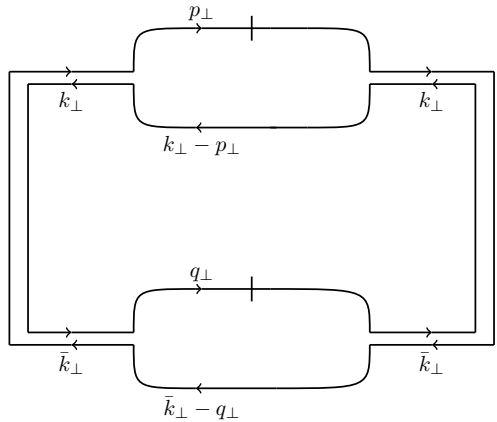}
%\includegraphics[width=8cm]{ Phi2Nc2.eps}
%\includegraphics[width=8cm]{Phi2Nc3.eps}
%\vspace*{-0.12cm}
\end{center}
\caption{The first correlated contribution of order $N_c^2$ originating from $(\Phi_2)^2$.}
\end{figure}

\begin{figure}[hbt]
 %[pb]\label{}
\begin{center}
\vspace*{-0.2cm}
\includegraphics[width=6.5cm]{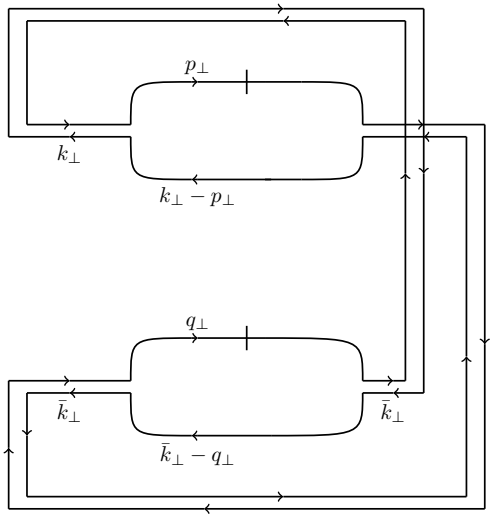}
%\vspace*{-0.12cm}
\end{center}
\caption{The second correlated contribution of order $N_c^2$ originating from $(\Phi_2)^2$.}
\end{figure}

 We will, from now on, concentrate solely on the leading $N_c$ contribution and will only consider the diagrams containing $\Phi_4$, see Figure 4.
%%%%%%%%%%%%%%%%%%%%
\begin{figure}
\label{4g}
\begin{center}
\vspace*{-0.2cm}
\includegraphics[width=12cm]{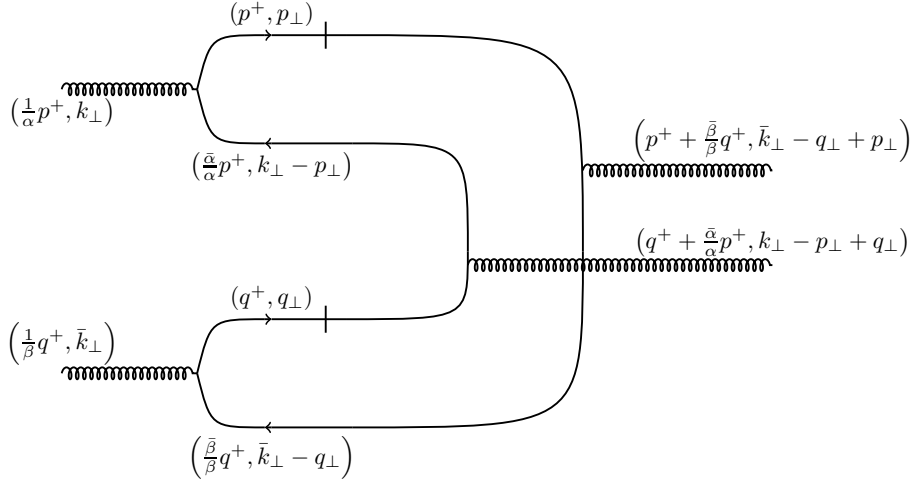}
\end{center}
\caption{The basic  graph contributing to the correlated quark production in the CGC. }
\end{figure}
%%%%%%%%%%%%%%%%%%%%%
The leading $N_c$ contribution to the correlated  quark pair density in the projectile wave function is given by
\begin{eqnarray}
\left[\frac{dN^P(p,q;\eta_1,\eta_2)}{d^2p\,d^2q \, d\eta_1\, d\eta_2}\right]_{\rm correlated}&=&-\frac{g^8}{(2 \pi)^4}\int d^2k\,d^2\bar k\,d^2l\,d^2\bar l\ 
\left\langle \rho^a(k)\rho^c(\bar k)\rho^b(l)\rho^d(\bar l)\right\rangle_P\nn\\
&\times&
\Phi_4(k,l,\bar k,\bar l; p,q)\;
{\rm tr}\{\tau^a\tau^b\tau^c\tau^d\}.
\end{eqnarray}
From this point on, we assume the McLerran-Venugopalan (MV) model \cite{mv} for averaging over color charge densities.\footnote{Note that we have also assumed the MV model for the averaging over the color charge densities in the $(\Phi_2)^2$ contribution to discuss its $N_c$ counting.} Within this model the correlators of $\rho$ factorize {\it \`a la} Wick into two point correlators. Additionally, we assume translational invariance of the CGC wave function. This is not an entirely realistic assumption, since such invariance is certainly broken on the scales of the size of the hadron. However, for relatively large transverse momenta the error introduced by this assumption should not be important. Within this framework, the basic contraction is given by
% we will not continue in the most general fashion, but will simplify our expressions using the information about the distribution of color charge density $\rho$

%
\beq
\left\langle\rho^a(k)\rho^b(p)\right\rangle_P= (2\pi)^2 \mu^2(k)\; \delta^{ab}\;\delta^{(2)}(k+p).
\eeq
We take in the following  $\mu^2(k)$ to be approximately  constant for large momenta, $\mu^2(k)=\mu^2$ for $k^2>Q_s^2$, with $Q_s$ the saturation momentum, and vanishing at small momenta,  $\mu^2(0)=0$. The latter condition is equivalent to requiring that only globally color neutral configurations contribute to the hadronic ensemble. The spatial scale of the color neutralization in our ensemble is  $Q^{-1}_s$.  We  assume that this vanishing is fast enough to regulate, at least, quadratically divergent integrals by cutting them off at $Q_s$.

There are two contractions of $\rho$ that contribute at large $N_c$ ($\propto {\cal O}(N_c^3)$), see Figures 5 and 6, and a third subleading one ($\propto {\cal O}(N_c)$) that is shown in Figure 7. The two leading contractions, to which we restrict hereafter, produce two distinct transverse momentum dependences:
\begin{equation}
\Phi_4^A\propto \delta^{(2)}(p-q)\; \delta^{(2)}(0)\, , \ \ \ \ \ \Phi_4^B\propto\delta^{(2)}(\bar k-k-q+p)\; \delta^{(2)}(0) \; .
\end{equation}
We now consider these two contributions,
\bea
\Phi_4^A (k,\bar k; p,q)&\equiv& \sum_{s_1,s_2,\bar s_1,\bar s_2} \int_0^1 \,
{d\alpha\,d\beta\over (\beta+\bar\beta e^{\eta_1-\eta_2})  (\alpha+\bar\alpha e^{\eta_2-\eta_1})}(2\pi)^4 \mu^2(k)\; \mu^2(\bar k)\,  
\delta^{(2)}(p-q) \delta^{(2)}(0)
\nn \\
&&
\hspace{2cm}
\times
 \phi_{s_1 s_2}(k,p;\alpha)\ \phi_{\bar s_1 \bar s_2}(\bar k,p;\beta)\  \phi_{s_1 \bar s_2}^*(\bar k,p;\beta)\,
\phi_{\bar s_1 s_2}^*(k,p;\alpha)
\eea
and
\bea
\Phi_4^B(k,\bar k; p,q)\equiv \sum_{s_1,s_2,\bar s_1,\bar s_2} \int_0^1 \,
{d\alpha\,d\beta\over (\beta+\bar\beta e^{\eta_1-\eta_2})  (\alpha+\bar\alpha e^{\eta_2-\eta_1})} (2\pi)^4\mu^2(k) \; \mu^2(k+q-p)
%\nn \\
%&&\times 
\delta^{(2)}(\bar k-k-q+p) \; \delta^{(2)}(0)\nn\\
% \\
%&&
\hspace{1cm}
\times  \phi_{s_1 s_2}(k,p;\alpha)\ \phi_{\bar s_1 \bar s_2}( k+q-p,q;\beta)\  \phi_{s_1 \bar s_2}^*( k,p;\beta)\,
\phi_{\bar s_1 s_2}^*(k+q-p,q;\alpha).
\eea
In both cases the spin structure becomes simple, and the trace over the spin indices can be taken explicitly. Thus,
\bea
\label{phi4A}
\Phi_4^A(k, \bar k; p,q)\!=\!\delta^{(2)}(p-q) \, \delta^{(2)}(0)  \int_0^1 \!\!\!
\frac{d\alpha\,d\beta}{ (\beta+\bar\beta e^{\eta_1-\eta_2})  (\alpha+\bar\alpha e^{\eta_2-\eta_1})}
%\nn \\
%&\times&
\frac{(2\pi)^4 \; 2\; \mu^2(k)\; \mu^2(\bar k)}{k^4\bar k^4[\bar\alpha p^2+\alpha(k-p)^2]^2[\bar\beta p^2+\beta(\bar k-p)^2]^2}\nn\\
%&&
\hspace{-1cm}
\times
\left\{\left[\bar{\alpha}k\cdot p+\alpha k\cdot (k-p)\right]^2 
+4\left[k^2p^2-(k\cdot p)^2\right]\right\}
%\nn\\
%&\times&
\left\{\left[ \bar{\beta}\bar{k}\cdot p+\beta\bar{k}\cdot(\bar{k}-p)\right]^2 
+4\left[\bar k^2p^2-(\bar k\cdot p)^2\right]\right\}
\eea
and
\bea
%&&
%\hspace{0cm}
\Phi_4^B (k, \bar k; p,q)= \delta^{(2)}(\bar k-k-q+p)\; \delta^{(2)}(0) \int_0^1
\frac{d\alpha\,d\beta}{(\beta+\bar\beta e^{\eta_1-\eta_2})  (\alpha+\bar\alpha e^{\eta_2-\eta_1})}
\frac{(2\pi)^4\;  2\; \mu^2(k)\; \mu^2(k+q-p)}{k^4(k+q-p)^4 [\bar\alpha p^2+\alpha(k-p)^2] }\nn \\
%&&
\times
\frac{1}{[\bar\beta p^2+\beta(k-p)^2]
[\bar\beta q^2+\beta(k-p)^2] \, [\bar\alpha q^2+\alpha(k-p)^2] }
%\nn \\ 
%[p^2+\alpha(k^2-2k\cdot p)]\,[p^2+\beta(k^2-2k\cdot p)]\,[\alpha(k-p)^2+\bar \alpha q^2]\,[\beta(k-p)^2+\bar \beta q^2]}\nn \\
%&&\times
%
\Big\{ \alpha\beta k^4 +\left[\alpha(\bar\beta-\beta)+\beta(\bar\alpha-\alpha)\right]k^2k\cdot p+4k^2p^2\nn\\
%&&
+\left[(\bar\alpha-\alpha)(\bar\beta-\beta)-4\right](k\cdot p)^2\Big\} 
%\nn\\
%&&\times
\Big\{ \alpha\beta (k+q-p)^4+ 
\left[\alpha(\bar\beta-\beta)+\beta(\bar\alpha-\alpha)\right](k+q-p)^2
%(\alpha+\beta-4\alpha\beta)
(k+q-p)\cdot q \nn\\
%&&
+4(k+q-p)^2q^2 
+\left[(\bar\alpha-\alpha)(\bar\beta-\beta)-4\right][(k+q-p)\cdot q]^2\Big\}.
%
%&&
%\hspace{0.5cm}
%+
%\left[(\bar\alpha-\alpha)(\bar\beta-\beta)-4\right]
%\Big((k+q-p)\cdot q\Big)^2\Big\}
%\Bigg[16 (k\times p)^2 [q\times(k-p)]^2 \\
%&&+ 4 \,[q\times(k-p)]^2 \left\{ \alpha [(4\alpha-3)k^2-2 k\cdot p]+k\cdot p\right\}\left\{ \beta [(4\beta-3)k^2-2 k\cdot p]+k\cdot p\right\}\nn\\
%&&-4\,(\alpha-\beta)\,(k\times p) [q\times(k-p)]\,\left[(4\alpha+4\beta-3)k^2-2k\cdot p\right]\nn\\
%&&\times \left\{ (\alpha-\beta)(k-q-p)\cdot (k+q-p) -4 (\alpha \bar \alpha-\beta \bar \beta) (k+q-p)^2   \right\}\nn\\
%&&+ \left[ \left\{ \alpha\, [(4\alpha-3)k^2-2 k\cdot p]+k\cdot p\right\}\left\{ \beta\, [(4\beta-3)k^2-2 k\cdot p]+k\cdot p\right\}+4 \,(k\times p)^2\right]\nn\\
%&&\times \left\{ \alpha\left[(k-p)\cdot (k+q-p)-4 \bar \alpha (k+q-p)^2\right]+\bar \alpha\, q\cdot(k+q-p)  \right\}\nn\\
%&&\times \left\{ \beta\left[(k-p)\cdot (k+q-p)-4 \bar \beta (k+q-p)^2\right]+\bar \beta\, q\cdot(k+q-p)  \right\}\Bigg].\nn
\eea

%
%%%%%%%%%%%%%%%%%%%%%%%%%%%%%%%%%%%%%%%%%
 \begin{figure}[hbt]
 %]\label{}
\begin{center}
\vspace*{-0.2cm}
\includegraphics[width=8cm]{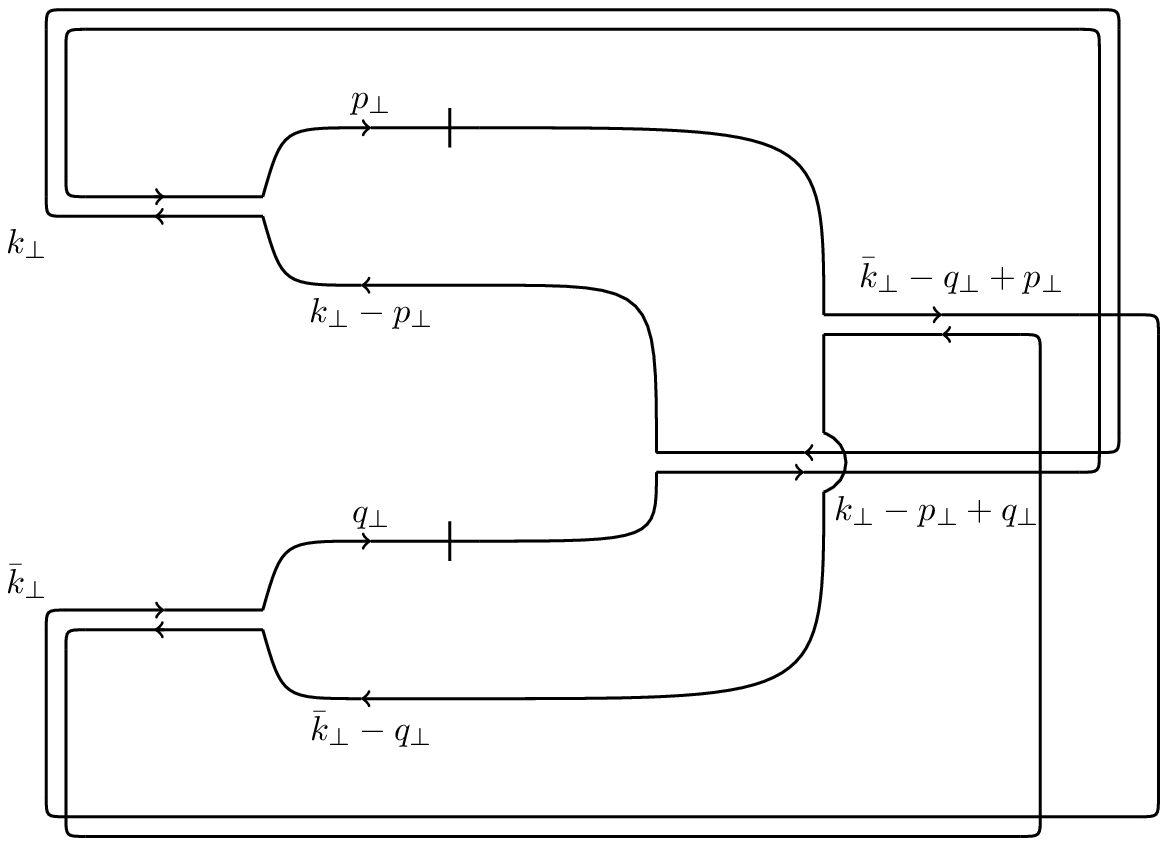}
%\vspace*{-0.12cm}
\caption{The leading order in $N_c$ source contraction that corresponds to the contribution $\Phi^A_4$ . }
\end{center}
\end{figure}
\begin{figure}[hbt]
 %[pb]\label{}
\begin{center}
\vspace*{-0.2cm}
\includegraphics[width=8cm]{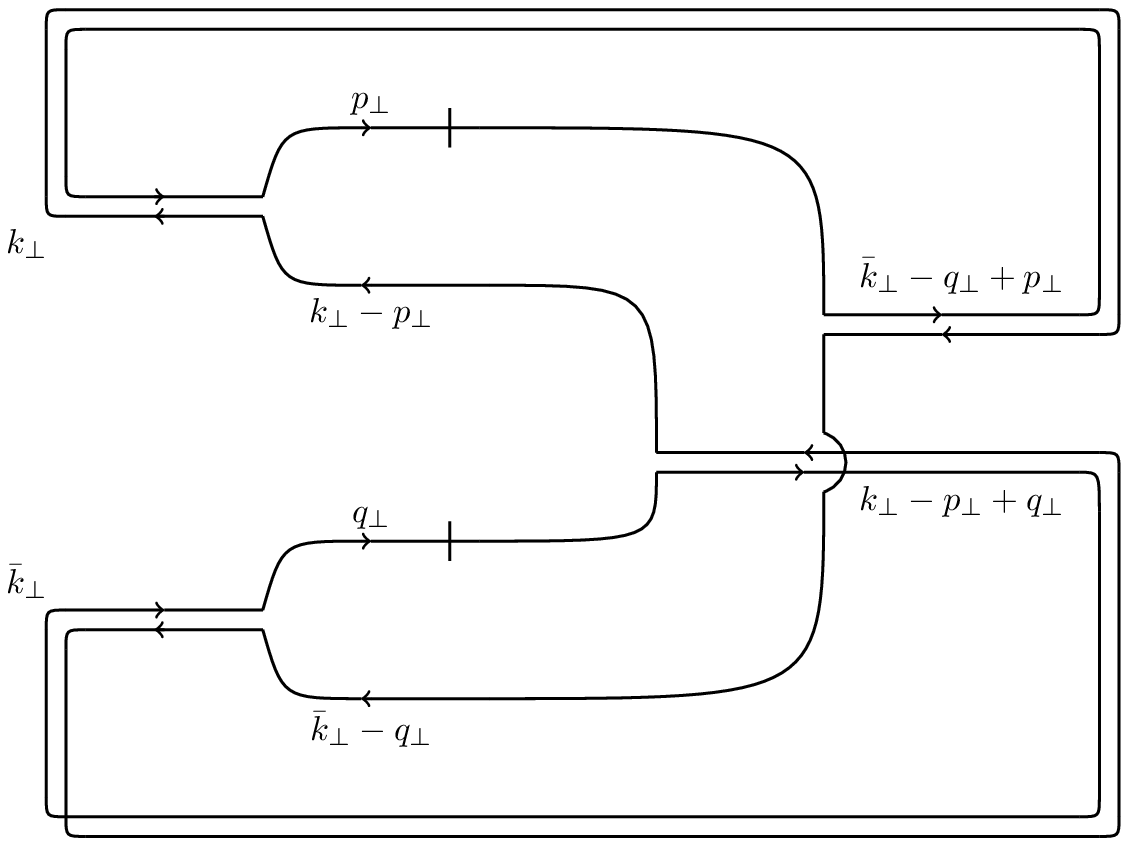}
%\vspace*{-0.12cm}
\end{center}
\caption{The leading order in $N_c$ source contraction that corresponds to the contribution $\Phi_4^B$. }
\end{figure}
%\end{figure}
\begin{figure}[hbt]
% \begin{figure}[pb]\label{}
\begin{center}
\vspace*{-0.2cm}
\includegraphics[width=8cm]{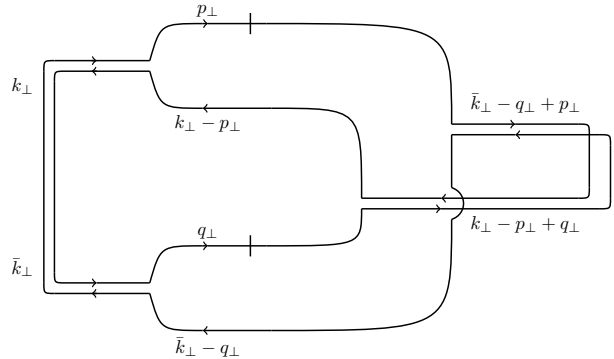}
%\vspace*{-0.12cm}
\end{center}
\caption{The subleading in $N_c$ source contraction not considered in this paper. }
\end{figure}
%%%%%%%%%%%%%%%%%%%%%%%%%%%%%%%%%%%%%%%%%

The correlated contribution clearly does not vanish. We will not calculate the integrals involved exactly. However, it is possible in a relatively simple way to estimate the result in the following kinematics. We will take the rapidity difference between the two quarks to be relatively large, $\eta_1-\eta_2\gg 1$, and the two transverse momenta to be of the same order and much larger than the saturation momentum, $|p|\sim |q|\gg Q_s$. This estimate will answer the two basic questions: what is the sign of the correlation and how far in rapidity difference does it extend?

The calculation is presented in  Appendix C. The final result is
%%
%\bea
%\label{npairsold}
%&&\left(\frac{dN^P(p,q;\eta_1,\eta_2)}{d^2pd^2q d\eta_1d\eta_2}\right)_{\rm correlated}=-Se^{\eta_2-\eta_1} (\eta_1-\eta_2)^2\frac{\mu^4}{p^4q^4}\nn\\
%&\times&\Bigg\{
%\frac{25}{2}p^4\left[\eta_1-\eta_2+\ln\frac{p^2}{Q_s^2}\right]^2\delta^2(q-p)\nn\\
%&&+ \left[3\frac{(p^2+q^2)\left[5p^2q^2-3(p\cdot q)^2-(p^2+q^2)p\cdot q\right]}{(q-p)^4}\ln\frac{(p-q)^2}{Q_s^2}+8(\eta_2-\eta_1)p\cdot q\right]\Bigg\}\nn
%\eea
%\textcolor{red}{
\bea
\label{npairs}
&&
\left[\frac{dN^P(p,q;\eta_1,\eta_2)}{d^2pd^2q d\eta_1d\eta_2}\right]_{\rm correlated}\simeq -\frac{S}{(2\pi)^2}e^{\eta_2-\eta_1} (\eta_1-\eta_2)^2\frac{\mu^4}{p^4q^4}\,g^8\; \frac{N_c^3}{4}\Bigg\{ \frac{25\pi^2}{2}q^4\left[\eta_1-\eta_2+\ln\frac{p^2}{Q_s^2}\right]^2\delta^{(2)}(q-p)\nn
\\
%&&
%\times
%
%\nn\\
&&
\hspace{3.2cm}
+ \ \pi\left[
\frac{3(p^2+q^2)}{(p-q)^4}\left\{ 5\left[p^2q^2-(p\cdot q)^2\right]-(p-q)^2p\cdot q\right\}\ln\frac{(p-q)^2}{Q_s^2}+(\eta_1-\eta_2)p\cdot q\right]\Bigg\},
%+\pi\left[
%3\frac{(p^2+q^2)\left[5p^2q^2-3(p\cdot q)^2-(p^2+q^2)p\cdot q\right]}{(q-p)^4}\ln\frac{(p-q)^2}{Q_s^2}+4(\eta_1-\eta_2)p\cdot q\right]\Bigg\},
\eea
%}
%
where $S\equiv (2\pi)^2\delta^{(2)}(0)$ is proportional to  the transverse area of the hadron.

The first thing to note is that the correlated contribution is negative, which conforms to our expectation based on the physics of the Pauli blocking. Second, the correlation is formally short range in rapidity since it decreases exponentially as a function of the rapidity difference. However, the rate of this decrease is tampered by the fourth power of $\eta_1-\eta_2$, so that in practical terms the correlation may extend fairly far in rapidity. Lastly, we note that the first term in Eqn. (\ref{npairs}) is proportional to $\delta^{(2)}(p-q)$. The technical reason for it is our assumption of translational invariance of the projectile wave function. The actual width of this $\delta$-function-like contribution should be of the order of the transverse size of the projectile. One may, in principle, expect that in the double inclusive quark production  the $\delta$-function is smeared by the saturation momentum of the target. However, as we will see and briefly discuss in the next Section, this turns out not to be the case.

\section{Pauli blocking and particle production}
In this Section, we calculate the double inclusive quark production in the CGC approach. We concentrate on the linearized approximation which is appropriate to p-p scattering and is the direct analog of the so-called "glasma graph" calculation for gluon production.

\subsection{The production cross section}
The formal expression for the inclusive quark pair production  emission reads
\bea
\label{Oq}
{d\sigma\over dp^+d^2pdq^+d^2q}\,&=& \frac{1}{(2\pi)^6}
\,\langle v| \Omega\,\hat S^\dagger\,
\Omega^\dagger\,\,\,
[\, d^{\dagger}_{\alpha,s_1}(p^+,p)\,d^{\dagger}_{\beta,s_2}(q^+,q)
%\nn\\
%&&
%\hspace{2cm}
%\times\;
d_{\beta,s_2}(q^+,q)\,d_{\alpha,s_1}(p^+,p)\, ] \,\,
\Omega\,\hat { S}\,\Omega^\dagger|v\rangle\; .
\eea
Here $\hat S$ is the eikonal $S$-matrix operator and $\Omega$ is the unitary operator which (perturbatively) diagonalizes the QCD Hamiltonian, in the CGC approximation, to the order in $\alpha_s$ in which the ground state contains two quarks as in Eqn. (\ref{v4}). The explicit form of the operator
$\Omega$ can be found in Appendix D. Note that in Eqn. (\ref{Oq}), the averaging over the projectile color charge densities and averaging over the target fields are implicit.

Let us define the coordinate space amplitudes (see Figures 8 and 9): 
\beq
\phi_{s_1,s_2}(x,z,\bar z;\alpha)\equiv \int_{k,p,\bar p}e^{ik\cdot x+ip\cdot z+i\bar p\cdot \bar z}\phi_{s_1,s_2}(k,p,\bar p;\alpha)\; ,
\eeq
\beq
\Phi_2(x,y;z_1,z_2;\bar z;k)\,\equiv\,\int_0^1 \,d\alpha\ \sum_{s_1\,s_2}\
\phi_{s_1,s_2}(x,z_1,\bar z;\alpha)\ \phi_{s_1,s_2}^*(y,z_2,\bar z;\alpha)\,e^{-i k\cdot (z_1-z_2)}
\eeq

\begin{figure}[hbt]
\label{Phi_2_coordinate}
\begin{center}
\vspace*{-0.2cm}
\includegraphics[width=9cm]{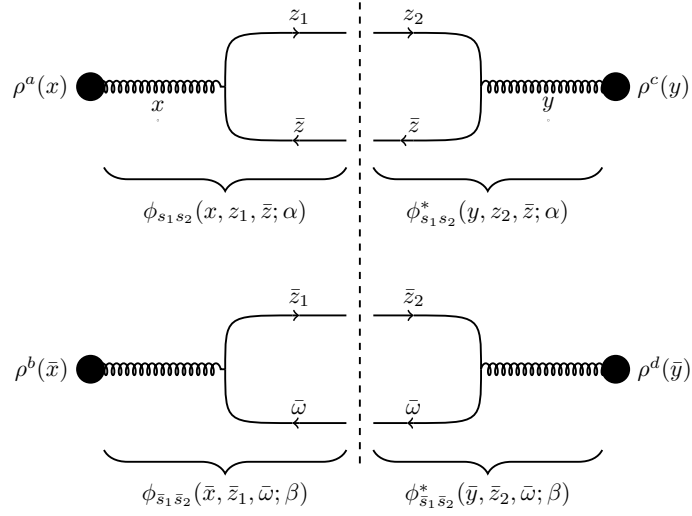}
%\includegraphics[width=8cm]{Phi2Nc3.eps}
%\vspace*{-0.12cm}
\end{center}
\caption{ The graph for the $(\Phi_2)^2$ contribution in coordinate space.
}
\end{figure}

and
\bea
\Phi_4(x,y,\bar x,\bar y;z_1,z_2,\bar z_1,\bar z_2;\bar z,\bar w; k,p) &\equiv&  \sum_{s_1,s_2,\bar s_1,\bar s_2}  e^{-i k\cdot (z_1-z_2)}\, e^{-i p\cdot (\bar z_1-\bar z_2)}
%\nn \\
%&&
%\hspace{-2cm}
%\times
\int_0^1 \,
{d\alpha\,d\beta\over (\beta+\bar\beta e^{\eta_1-\eta_2})  (\alpha+\bar\alpha e^{\eta_2-\eta_1})}\, \nn \\
&&
\hspace{-2cm}
\times \; \phi_{s_1,s_2}(x,z_1,\bar z;\alpha)\ \phi_{s_1,\bar s_2}^*(y,z_2,\bar w;\beta)\,
\phi_{\bar s_1,\bar s_2}(\bar x,\bar z_1,\bar w;\beta)\ \phi_{\bar s_1,s_2}^*(\bar y,\bar z_2,\bar z;\alpha)\; .
\eea

\begin{figure}[hbt]
\label{Phi_4_coordinate}
% [pb]\label{}
\begin{center}
\vspace*{-0.2cm}
\includegraphics[width=9cm]{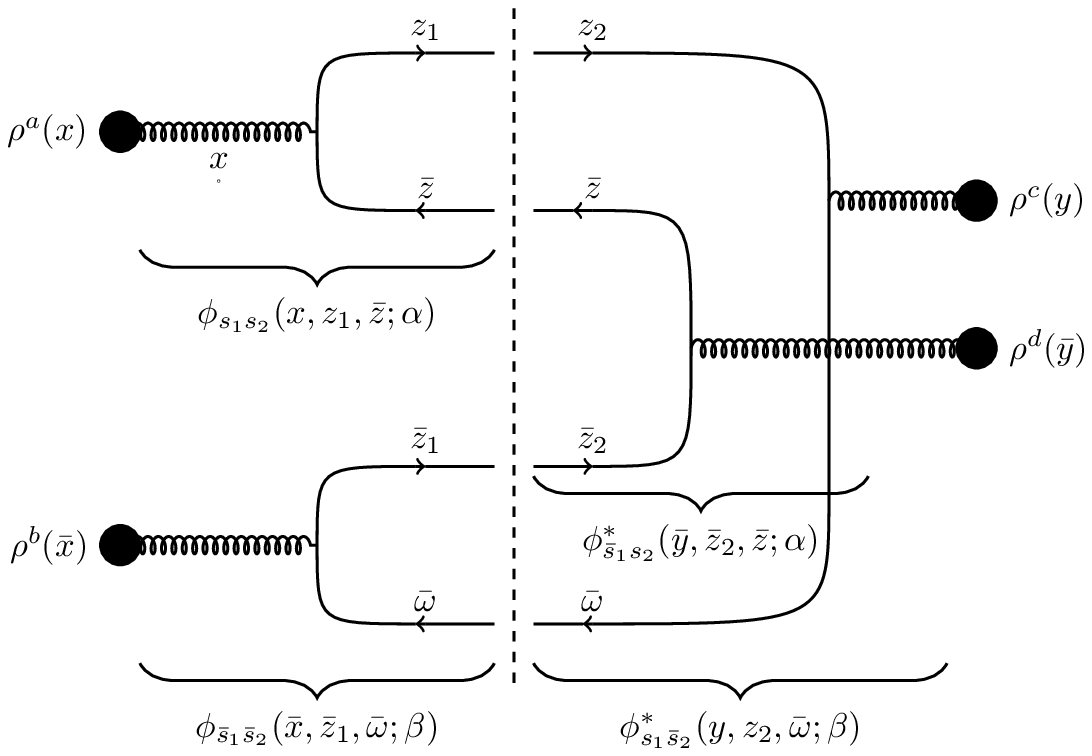}
%\includegraphics[width=8cm]{Phi2Nc3.eps}
%\vspace*{-0.12cm}
\end{center}
\caption{ The graph for the $\Phi_4$ contribution in coordinate space.
}
\end{figure}

In terms of these amplitudes the quark pair production cross section can be written as
\bea
\label{cross}
&&{d\sigma\over d\eta_1\,d^2p\, d\eta_2\,d^2q}\,=\,{g^8\over{(2\pi)^4}}\,\int_{x,y,\bar x,\bar y} \int_{z_1,z_2,\bar z_1,\bar z_2, \bar z,\bar w}
\ {1\over 2}\ \langle \rho^a(x)\rho^b(\bar x)\rho^c(y)\rho^d(\bar y)\rangle_P\nn\\
&&\times \Big\langle \Phi_2(x,y;z_1,z_2,\bar z;p)\Phi_2(\bar x,\bar y;\bar z_1,\bar z_2,\bar w;q)   \nn\\
&&     \times\, {\rm tr}\left\{ [\tau^a- S_A^{a\bar a}(x)S_F(z_1)\tau^{\bar a}S_F^\dagger(\bar z)]
                     [\tau^c- S_A^{c\bar c }(y)S_F(\bar z)\tau^{\bar c}S_F^\dagger(z_2)] \right\} \nn\\
&&\times    \,        {\rm tr}    \left\{ [\tau^b- S_A^{b\bar b}(\bar x)S_F(\bar z_1)\tau^{\bar b}S_F^\dagger(\bar w)]
                        [\tau^d- S_A^{d\bar d}(\bar y)S_F(\bar w)\tau^{\bar d}S_F^\dagger(\bar z_2)] \right\} \nn\\
%              \left\{ [S^\dagger_A( x)-Q^\dagger_A( z_1,\bar z)]
 %                      [S_A(y)-Q_A(z_2,\bar z)]\right\}^{ac}
% \left\{              [Q^\dagger(\bar z_1,\bar w)-S_A^\dagger(\bar x)][Q_A({\bar z_2,\bar w})-S_A(\bar y)\right\}^{bd}  \nn \\
&&- \,\Phi_4(x,y,\bar x,\bar y;z_1,z_2,\bar z_1,\bar z_2;\bar z,\bar w;p,q)\nn \\
&&\times\,
              {\rm tr} \left\{ [\tau^a- S_A^{a\bar a}(x)S_F(z_1)\tau^{\bar a}S_F^\dagger(\bar z)]
                     [\tau^c- S_A^{c\bar c }(y)S_F(\bar w )\tau^{\bar c}S^\dagger_F( z_2 )] \right.\nn\\
&&\times   \,          \left. [\tau^b- S_A^{b\bar b}(\bar x)S_F(\bar z_1)\tau^{\bar b}S_F^\dagger(\bar w)]
                        [\tau^d- S_A^{d\bar d}(\bar y)S_F(\bar z)\tau^{\bar d}S^\dagger _F(\bar z_2)]  \right\}
\Big\rangle_T ,
\eea
%
%{\color{red}factors $\pi$ in front of integrals}
where each of the $S$-matrices is defined in terms of the color field of the target as $S(x)=\exp\{igt^a\alpha^a(x)\}$, with $t^a$ the color matrices in the corresponding representation. Note that the color field of the target can be written in terms of its color charge density as 
\beq
\label{alpha_to_rho}
\alpha^a(x)=\frac{1}{\nabla^2}(x,y)\rho_T^a(y) \; .
\eeq

A certain disclaimer is due here.
This expression Eqn. (\ref{cross})  is not complete. It does not contain terms associated with the fragmentation of two  physical projectile gluons that scatter and split into $q\bar q$ pairs in the final state, corresponding to $\delta H^{g\, qq}$ and $\Omega_{gqq}$, see Appendices A and D. Including such terms would make the final expressions cumbersome and not very illuminating.  We do not believe that these fragmentation contributions can produce correlated pairs, and will thus work with the simplified expression Eqn. (\ref{cross}).

To get a rough idea of the actual magnitude of the correlations predicted by Eqn. (\ref{cross}), we now expand the scattering matrices to leading order in the target color charge density. This approximation is formally the same as employed in the glasma graph calculation of gluon production. Although it misses some effects, in particular due to a possible domain-like structure of the target fields, it does include  correlated production due to correlations in the projectile wave function.

The large $N_c$ counting in Eqn. (\ref{cross}) is identical to that discussed in the previous section. We thus concentrate only on the $\Phi_4$ term as before. We define $\Delta$ as
\bea
\Delta^{acbd}&=&
\Big\langle{\rm tr} \left\{ [\tau^a- S_A^{a\bar a}(x)S_F(z_1)\tau^{\bar a}S_F^\dagger(\bar z)]
                     [\tau^c- S_A^{c\bar c }(y)S_F(\bar w )\tau^{\bar c}S^\dagger_F( z_2 )] \right.\nn\\
&&\times   \,          \left. [\tau^b- S_A^{b\bar b}(\bar x)S_F(\bar z_1)\tau^{\bar b}S_F^\dagger(\bar w)]
                        [\tau^d- S_A^{d\bar d}(\bar y)S_F(\bar z)\tau^{\bar d}S^\dagger _F(\bar z_2)]  \right\} \Big\rangle_T\,.
\eea
Expanding each of the $S$-dependent factors in terms of the target color field $\alpha$ defined as $S(x)=\exp\{igt^a\alpha^a(x)\}$, with $t^a$ the color matrices in the corresponding representation, we obtain
\bea
&&
 \Delta^{acbd}=g^4\Big\langle {\rm tr}\Big[ \{\tau^a\tau^{a'}[\alpha^{a'}(x)-\alpha^{a'}(\bar z)]-\tau^{a'}\tau^{a}[\alpha^{a'}(x)-\alpha^{a'}( z_1)]\}
 %\\
%&\times&
\{\tau^c\tau^{c'}[\alpha^{c'}(y)-\alpha^{c'}( z_2)]-\tau^{c'}\tau^{c}[\alpha^{c'}(y)-\alpha^{c'}(\bar w)]\}\nn\\
&&\times
\{\tau^b\tau^{b'}[\alpha^{b'}(\bar x)-\alpha^{b'}( \bar w)]-\tau^{b'}\tau^{b}[\alpha^{b'}(\bar x)-\alpha^{b'}(\bar z_1)]\}
%\nn\\
%&\times&
\{\tau^d\tau^{d'}[\alpha^{d'}(\bar y)-\alpha^{d'}( \bar z_2)]-\tau^{d'}\tau^{d}[\alpha^{d'}(\bar y)-\alpha^{d'}(\bar z)]\}\ \Big]\Big\rangle_T.
\eea
We now consider the projectile and target color charge density contractions. The term $\Delta^{acbd}$ that enters Eqn. (\ref{cross}) is the sum of two different contractions that can be written as
\beq
\Delta^{acbd}=\delta^{ad}\delta^{cb}\left\langle\Delta_A\right\rangle_T+\delta^{ac}\delta^{bd}\left\langle\Delta_B\right\rangle_T\, .
\eeq

The type $A$ graph in the wave function calculation was obtained by contracting $\rho^a$ with $\rho^d$ and $\rho^b$ with $\rho^c$. In order to obtain the leading-$N_c$ contribution to the production cross section with this contraction on the projectile side, we have to contract the color indices with $\delta^{a'd'}\delta^{b'c'}$. This structure arises from the contractions of the target color fields and reads, at leading $N_c$,
\bea
\left\langle\Delta_A\right\rangle_T=\frac{g^4 N^5_c}{16}&&\Big\{\left\langle[\alpha(x)-\alpha(\bar z)] \cdot[\alpha(\bar y)-\alpha(\bar z)]+[\alpha(x)-\alpha(z_1)] \cdot [\alpha(\bar y)-\alpha(\bar z_2)]\right\rangle_T\Big\}\\
&\times&\Big\{\langle[\alpha(\bar x)-\alpha(\bar w)] \cdot [\alpha( y)-\alpha(\bar w)]+[\alpha(\bar x)-\alpha(\bar z_1)] \cdot [\alpha( y)-\alpha( z_2)]\rangle_T\Big\}.\nonumber
\eea
Analogously, for the type $B$ contribution we have $a=c$ and $b=d$, and therefore we need $a'=c'$ and $b'=d'$ at large $N_c$. At leading $N_c$ this gives
\bea \left\langle\Delta_B\right\rangle_T=\frac{g^4 N^5_c}{16}&&\Big\{\langle[\alpha(x)-\alpha(\bar z)] \cdot [\alpha( y)-\alpha(\bar w)]+[\alpha(x)-\alpha(z_1)] \cdot [\alpha(y)-\alpha( z_2)]\rangle_T\Big\}\\
&\times&\Big\{\langle[\alpha(\bar x)-\alpha(\bar w)] \cdot [\alpha( \bar y)-\alpha(\bar z)]+[\alpha(\bar x)-\alpha(\bar z_1)] \cdot [\alpha( \bar y)-\alpha( \bar z_2)]\rangle_T\Big\}.\nonumber
\eea
%
%Note that the term $\Delta^{acbd}$ that enters Eqn. (\ref{cross}) is the sum of two different contractions that can be written as
%\beq
%\Delta^{acbd}=\delta^{ad}\delta^{cb}\Delta_A+\delta^{ac}\delta^{bd}\Delta_B,
%\eeq
%where the contributions $\Delta_A$ and $\Delta_B$ implicitly include the target side contractions as explained above.
The expressions for $\langle\Delta_A\rangle_T$ and $\langle\Delta_B\rangle_T$ have a fairly simple structure. In particular, we can combine the factors $\langle\Delta_B\rangle_T$ and $\langle\Delta_A\rangle_T$ that come from the expansion of the $S$-matrix with the rest of the expression. This can be done by inspection.
Let us define the following quantities:
\bea\label{psi}
\Psi(k,l,p;\alpha)&\equiv&     [\phi (k+l,p;\alpha)-\phi(k, p-l;\alpha)],\nn\\
 \Psi(k,l,p,\bar p;\alpha)&\equiv&  \Psi(k,l,p;\alpha) \, 
 \delta^{(2)}(\bar p-k-l+p),\nn\\
 \bar\Psi(k,l,p;\alpha)&\equiv&[\phi (k+l,p;\alpha)-\phi(k, p;\alpha)],\nn\\
 \bar\Psi(k,l,p,\bar p;\alpha)&\equiv&\bar\Psi(k,l,p;\alpha)
\,  \delta^{(2)}(\bar p-k-l+p).
\eea
We can write the $A$-type contribution to the cross section as
\bea\label{A}
&&
\hspace{-0.1cm}
A=-\frac{g^{12}N_c^5}{16}  \frac{1}{(2\pi)^4}\int_0^1 \,
{d\alpha\,d\beta\over (\beta+\bar\beta e^{\eta_1-\eta_2})  (\alpha+\bar\alpha e^{\eta_2-\eta_1})}\int
d^2k \, d^2 \bar k \, d^2 l \, d^2 \bar{l} \, d^2\bar p \, d^2\bar q \,  (2\pi)^8 \, 
%\frac{d^2k}{(2\pi)^2} \frac{d^2 \bar k}{(2\pi)^2} \frac{d^2 l}{(2\pi)^2} \frac{d^2\bar l}{(2\pi)^2}
%\frac{d^2\bar p}{(2\pi)^2} \frac{d^2\bar q}{(2\pi)^2}
\frac{\mu^2(k)\mu^2(\bar k)\lambda^2(l)\lambda^2(\bar l)}{l^4\bar l^4}
\nn\\
&&
%\hspace{-1cm}
\times
%\nn\\
%&&\times
{\rm tr }\Big\{\Big[\bar \Psi(k,l,p,\bar p;\alpha)\bar\Psi^*(k,l,q,\bar p;\alpha)+ \Psi(k,l,p,\bar p;\alpha)\Psi^*(k,l,q,\bar p;\alpha)\Big]\nn\\
&&
\hspace{1cm}
\times\Big[\bar \Psi (\bar k,\bar l,q,\bar q; \beta)\bar\Psi^*(\bar k,\bar l,p,\bar q;\beta)+ \Psi(\bar k,\bar l,q,\bar q;\beta)\Psi^*(\bar k,\bar l,p,\bar q;\beta)\Big]\Big\}\
 \\
%\eea
%to get
%\bea
%A
&=&
-\delta^{(2)}(0)\delta^{(2)}(p-q)\frac{g^{12}N_c^5}{16} \int_0^1 \,
{d\alpha\,d\beta\over (\beta+\bar\beta e^{\eta_1-\eta_2})  (\alpha+\bar\alpha e^{\eta_2-\eta_1})}
%\nn\\
%&&\times 
(2\pi)^4 \int d^2k \; d^2\bar k \; d^2l \; d^2\bar l \; \frac{\mu^2(k)\mu^2(\bar k)\lambda^2(l)\lambda^2(\bar l)}{l^4\bar l^4}\nn\\
&&\times{\rm tr}\Big\{ \Big[\bar \Psi(k,l,p;\alpha)\bar\Psi^*(k,l,p;\alpha)+ \Psi(k,l,p;\alpha)\Psi^*(k,l,p;\alpha)\Big]
%\nn\\
%&&
%\hspace{2.5cm}
%\times
\Big[\bar \Psi (\bar k,\bar l,p;\beta)\bar\Psi^*(\bar k,\bar l,p;\beta)+ \Psi(\bar k,\bar l,p;\beta)\Psi^*(\bar k,\bar l,p;\beta)\Big]\Big\}.\nn
\eea
Analogously, for the $B$-type we have
\bea\label{B}
B&=&-\frac{g^{12}N_c^5}{16} \frac{1}{(2\pi)^4}\int_0^1 \,
{d\alpha\,d\beta\over (\beta+\bar\beta e^{\eta_1-\eta_2})  (\alpha+\bar\alpha e^{\eta_2-\eta_1})}
%\nn\\
%&&\times
\int
d^2k \, d^2\bar k \, d^2l \, d^2\bar l \, d^2\bar p \, d^2\bar q  \; (2\pi)^8
%\frac{d^2k}{(2\pi)^2}\frac{d^2\bar k}{(2\pi)^2}\frac{d^2l}{(2\pi)^2} \frac{d^2\bar l}{(2\pi)^2}\frac{d^2\bar p}{(2\pi)^2}\frac{d^2\bar q}{(2\pi)^2} 
\frac{\mu^2(k)\mu^2(\bar k)\lambda^2(l)\lambda^2(\bar l)}{l^4\bar l^4}
\nn\\
&&
%\hspace{-1cm}
\times
%
%\\
%&&
%\times
 {\rm tr}\Big\{\Big[\bar \Psi(k,l,p,\bar p;\alpha)\bar\Psi^*(k,l,p,\bar q;\beta)+ \Psi(k,l,p,\bar p;\alpha)\Psi^*(k,l,p,\bar q;\beta)\Big]\nn\\
&&
\hspace{2cm}
\times\Big[\bar \Psi (\bar k,\bar l,q,\bar q;\beta)\bar\Psi^*(\bar k,\bar l,q,\bar p;\alpha)+ \Psi(\bar k,\bar l,q,\bar q;\beta)\Psi^*(\bar k,\bar l,q,\bar p;\alpha)\Big]\Big\}\,,
\nonumber\\
%\eea
%so
%\bea
%B
&=&
-\delta^{(2)}(0)\frac{g^{12}N_c^5}{16} \int_0^1 \,
{d\alpha\,d\beta\over (\beta+\bar\beta e^{\eta_1-\eta_2})  (\alpha+\bar\alpha e^{\eta_2-\eta_1})}
%\nn\\
%&&
%\hspace{-0.5cm}
%\times
(2\pi)^4
\int d^2k\; d^2\bar k\, d^2l\, d^2\bar l\,
\frac{\mu^2(k)\mu^2(\bar k)\lambda^2(l)\lambda^2(\bar l)}{l^4\bar l^4}\nn\\
&&\times
\delta^{(2)}(k+l-p-\bar k-\bar l+q)\ 
{\rm tr}\Big\{\Big[\bar \Psi(k,l,p;\alpha)\bar\Psi^*(k,l,p;\beta)+ \Psi(k,l,p;\alpha)\Psi^*(k,l,p;\beta)\Big]\nn\\
&&
\hspace{4.8cm}
\times\Big[\bar \Psi (\bar k,\bar l,q;\beta)\bar\Psi^*(\bar k,\bar l,q;\alpha)+ \Psi(\bar k,\bar l,q;\beta)\Psi^*(\bar k,\bar l,q;\alpha)\Big]\Big\}.
\eea
In both equations ${\rm tr}$ denotes now the spin trace. Besides, we have used Eqn. (\ref{alpha_to_rho}) in order to write the target color field in terms of the color charge density of the target and we have used
\beq
\langle \rho_T^a(k)\rho_T^b(p)\rangle_T=(2\pi)^2\lambda^2(k)\delta^{ab}\delta^{(2)}(k+p)
\eeq
which corresponds to the McLerran-Venugolapan model to contract the color charge densities of the target.

Note that the $A$-type contribution to pair production cross section  has the $\delta^{(2)}(p-q)$ structure, just  like the quark pair density in the wave function. This is somewhat surprising, since one may expect any sharp maximum in a distribution  in the projectile wave function  to be smeared by a momentum transfer from the target. However, in the present case one is dealing with a wave function and final sates with four particles - two quarks and two antiquarks. It is possible to produce the two quarks without changing their momenta by scattering the antiquarks out of the incoming wave function. We believe that this is the reason why the $\delta$-function is not smeared in the scattering process. Of course, as stressed above, if we take into account the finite size of the incoming projectile, this $\delta$-function will be smeared on the scale of the inverse proton radius. Note that this contribution is not due to the Hanbury-Brown--Twiss (HBT) effect, so the radius of the proton would be reflected in the final state radiation without the HBT effect.

\subsection{The estimates}

Like in the previous Section, we now estimate the correlated contribution to production for $\eta_1-\eta_2\gg 1$.
We will consider the situation when the saturation momentum of the target is smaller than that of the projectile, $Q_T<Q_s$. This is the regime where the correlations existing in the wave function of the projectile are not strongly distorted by the momentum transfer from the target. We thus expect these correlations to be reflected in quark pair production.

The calculations are performed in Appendix E.  There is one interesting element in these calculations which was not present in the calculations in the previous section. To understand it, consider the explicit expressions for the amplitudes which enter Eqns. (\ref{A},\ref{B}) at large rapidity separations:
\begin{eqnarray}
\bar\Psi(k,l,p;0)&=&-\frac{(k+l)\cdot p}{p^2(k+l)^2}+\frac{k\cdot p}{p^2k^2}+
2i\sigma^3\left\{\frac{(k+l)\times p}{p^2(k+l)^2}-\frac{k\times p}{p^2k^2}\right\},\\
\bar\Psi(k,l,p;1)&=&-\frac{(k+l)\cdot(k+l-p)}{(k+l)^2(k+l-p)^2} +\frac{k\cdot(k-p)}{k^2(k-p)^2} 
 %\nn \\
%&&
-2i\sigma^3\left\{\frac{(k+l)\times (k+l-p)}{(k+l)^2(k+l-p)^2}-\frac{k\times(k- p)}{k^2(k-p)^2}\right\},\nn\\
\Psi(k,l,p;0)&=&-\frac{(k+l)\cdot p}{p^2(k+l)^2}+\frac{k\cdot( p-l)}{(p-l)^2k^2}
+2i\sigma^3\left\{\frac{(k+l)\times p}{p^2(k+l)^2}-\frac{k\times (p-l)}{(p-l)^2k^2}\right\},\nn\\
\Psi(k,l,p;1)&=&-\frac{(k+l)\cdot(k+l-p)}{(k+l)^2(k+l-p)^2}+\frac{k\cdot (k+l-p)}{k^2(k+l-p)^2}
%\nn\\
%&&
-2i\sigma^3\left\{\frac{(k+l)\times(k+l- p)}{(k+l)^2(k+l-p)^2}-\frac{k\times(k+l-p)}{k^2(k+l-p)^2}\right\}.\nn
\end{eqnarray}
%\textcolor{red}{I think it is the one in red. Below it is the expression that you get directly from (2.8), above is now the same - the minus in red previously read plus.
%\begin{eqnarray}
%\bar\Psi(k,l,p;0)&=&-\frac{(k+l)\cdot p}{p^2(k+l)^2}+\frac{k\cdot p}{p^2k^2}+
%2i\sigma^3\left\{\frac{(k+l)\times p}{p^2(k+l)^2}-\frac{k\times p}{p^2k^2}\right\},\\
%\bar\Psi(k,l,p;1)&=&-\frac{(k+l)\cdot(k+l-p)}{(k+l)^2(k+l-p)^2} +\frac{k\cdot(k-p)}{k^2(k-p)^2}  \nn \\
%&+&2i\sigma^3\left\{\frac{(k+l)\times p}{(k+l)^2(k+l-p)^2}-\frac{k\times p}{k^2(k-p)^2}\right\},\nn\\
%\Psi(k,l,p;0)&=&-\frac{(k+l)\cdot p}{p^2(k+l)^2}+\frac{k\cdot( p-l)}{(p-l)^2k^2}
%+2i\sigma^3\left\{\frac{(k+l)\times p}{p^2(k+l)^2}-\frac{k\times (p-l)}{(p-l)^2k^2}\right\},\nn\\
%\Psi(k,l,p;1)&=&-\frac{(k+l)\cdot(k+l-p)}{(k+l)^2(k+l-p)^2}+\frac{k\cdot (k+l-p)}{k^2(k+l-p)^2}\nn\\
%&+&2i\sigma^3\left\{\frac{(k+l)\times p}{(k+l)^2(k+l-p)^2}-\frac{k\times(p-l)}{k^2(k+l-p)^2}\right\}.\nn
%\end{eqnarray}}
These expressions have several poles which give significant contributions upon momentum integrations. The poles at $k=0$ and $l=0$ are regulated by the vanishing of $\mu^2(0)$ and $\lambda^2(0)$ respectively. However, clearly the divergence at $k+l=0$  cannot be regulated by prescribing the behavior of $\mu^2$ or $\lambda^2$. The reason for the appearance of this divergence is quite clear. As explained above, requiring the vanishing of $\mu^2(k^2<Q_s^2)$  is equivalent to a condition of global color neutrality of the projectile on transverse distance scales larger than $Q^{-1}_S$. The same goes for the target. However, our eikonal scattering process is equivalent to double gluon exchange in the amplitude without restriction of color neutrality. Thus, after the scattering, the valence charge of the wave function  is not color neutral anymore. Such scattered colored projectile, when reconstituting its dressed wave function, emits gluons with the perturbative spectrum in the infrared (IR) which does not know about the color neutrality of the original projectile. This perturbative Weisz\"acker-Williams field of the colored outgoing projectile is the origin of the pole at $k+l=0$. It is clear, therefore, that the existence of finite $Q_s$ cannot regulate this divergence and it can only be regulated by genuine nonperturbative effects at the nonperturbative IR scale $\Lambda\sim \Lambda_{QCD}$.
Since the divergence is only logarithmic, the sensitivity to the IR is not too bad, and we will simply cut off this divergence at $\Lambda$ by hand.

The results of the explicit calculation in Appendix E are the following: for the $A$-type contribution,
\beq
A=-S(2\pi)^2\frac{50\pi^4g^{12}N_c^5}{16}\frac{\mu^4}{Q_s^4}\frac{\lambda^4}{Q_T^4}\frac{Q_s^2Q_T^2}{p^4}e^{\eta_2-\eta_1}(\eta_1-\eta_2)^2\ln\frac{Q_T^2}{\Lambda^2}\ln\frac{Q_s^4}{Q_T^2\Lambda^2}\delta^{(2)}(q-p).
\eeq
%{\color{red}If we define in the standard way $Q_s^2=g^4\mu^2$, $Q_T^2=g^4\lambda^2$, the result can be written as
%\beq
%A=-S\frac{50\pi^4N_c^5}{16 g^4}\frac{Q_s^2Q_T^2}{p^4}e^{\eta_2-\eta_1}(\eta_1-\eta_2)^2\ln\frac{Q_T^2}{\Lambda^2}\ln\frac{Q_s^4}{Q_T^2\Lambda^2}\delta^{(2)}(p-q).
%\eeq}
%This  has a whooping factor of $N_c$ in front. This has to be understood, especially in comparison to the gluon production.

The calculation for the $B$-type contribution is rather more lengthy. In Appendix E we present the calculation of all four terms keeping the leading logarithmic contributions and our final result for the $B$-type terms reads
%The other three terms are similar in nature, and we therefore expect them only to modify the overall numerical coefficient. Since in this paper we are interested only in the qualitative features of the effect, we do not pursue the complete calculation of this numerical factor.  We thus obtain
%\beq
%B=Kg^{12}N_c^5\frac{\mu^4\lambda^4}{p^2q^2(p-q)^4}   \frac{1}{Q_T^2}e^{\eta_2-\eta_1}(\eta_1-\eta_2)^2 \ln\frac{(p-q)^2}{Q_s^2}\ln\frac{Q_T^2}{\Lambda^2}
%\eeq
%where $K$ is a pure number.
\bea
B&=&-S(2\pi)^2g^{12}N_c^5\frac{9\pi^3}{p^4q^4}\Bigg\{ \frac{2(p^2+q^2)^2+p^2q^2}{(p-q)^4}\ln\left[\frac{(p-q)^2}{Q_s^2}\right]+
\frac{1}{2}\left[ \ln\left(\frac{q^2}{Q_s^2}\right)+\ln\left(\frac{p^2}{Q_s^2}\right) \right] \Bigg\}\nonumber\\
&&\times \frac{1}{Q_T^2}\ln\left(\frac{Q_T^2}{\Lambda^2}\right) \mu^4\lambda^4 e^{\eta_2-\eta_1}(\eta_1-\eta_2)^2.
\eea
Thus, our final result in the regime $p\sim q\sim p-q\gg Q_s \gg Q_T\gg \Lambda$ is
%\bea\label{nprod}
%&&\Big(\frac{dN}{d^2pd^2qd\eta_1d\eta_2}\Big)_{correlated}=\\
%&&-g^{12}N^5_cS\frac{\mu^4\lambda^4}{p^2q^2Q_T^2Q_s^2}e^{\eta_2-\eta_1}(\eta_1-\eta_2)^2\ln\frac{Q_T^2}{\Lambda^2}\Bigg[\frac{25}{16}\ln\frac{Q_s^4}{Q_T^2\Lambda^2}\delta^2(p-q)+K\frac{Q_s^2}{(p-q)^4} \ln\frac{(p-q)^2}{Q_s^2}\Bigg]\nonumber
%\eea
\bea
\label{nprod}
&&\left[\frac{d\sigma}{d^2pd^2qd\eta_1d\eta_2}\right]_{\rm correlated}=-S(2\pi)^2 \, g^{12}N_c^5\frac{\mu^4\lambda^4}{Q_s^2Q_T^2}e^{\eta_2-\eta_1}(\eta_1-\eta_2)^2
\ln\left(\frac{Q_T^2}{\Lambda^2}\right)\frac{\pi^3}{p^4}\\
&&\times
\Bigg\{ \frac{50\pi}{16}\ln\left(\frac{Q_s^4}{Q_T^2\Lambda^2}\right)\delta^{(2)}(q-p)+\frac{9Q_s^2}{q^4}\left[\frac{2(p^2+q^2)^2+p^2q^2}{(p-q)^4}\right]\ln\left[\frac{(p-q)^2}{Q_s^2}\right]
%\nonumber\\
%&&\hspace{0.7cm}
+
\frac{9Q_s^2}{2q^4}\left[\ln\left(\frac{q^2}{Q_s^2}\right)+\ln\left(\frac{p^2}{Q_s^2}\right)\right]\Bigg\}.\nn
\eea
If we define in the standard way $Q_s^2=g^4\mu^2$, $Q_T^2=g^4\lambda^2$, our final result can be written as
\bea
%\label{nprod}
&&\left[\frac{d\sigma}{d^2pd^2qd\eta_1d\eta_2}\right]_{\rm correlated}=-S(2\pi)^2 N_c^5  \frac{Q_s^2Q_T^2}{4g^4}
%\frac{\mu^4\lambda^4}{Q_s^2Q_T^2}
e^{\eta_2-\eta_1}(\eta_1-\eta_2)^2
\ln\left(\frac{Q_T^2}{\Lambda^2}\right)\frac{\pi^3}{p^4}
\\
&&\times
\Bigg\{ \frac{50\pi}{16}\ln\left(\frac{Q_s^4}{Q_T^2\Lambda^2}\right)\delta^{(2)}(q-p)+\frac{9Q_s^2}{q^4}\left[\frac{2(p^2+q^2)^2+p^2q^2}{(p-q)^4}\right]\ln\left[\frac{(p-q)^2}{Q_s^2}\right]
%\nonumber\\
%&&\hspace{0.7cm}
+
\frac{9Q_s^2}{2q^4}\left[\ln\left(\frac{q^2}{Q_s^2}\right)+\ln\left(\frac{p^2}{Q_s^2}\right)\right]\Bigg\}. \nn
\eea

The $\delta$ -function in the first term is an artefact of our use of the translationally invariant approximation for the projectile proton wave function. In a more careful treatment we expect it to be smeared over the scale of the inverse  proton size.

Our result for particle production Eqn. (\ref{nprod}) has a similar structure to the pair density in the projectile wave function Eqn. (\ref{npairs}). However, it has some significant differences. The first thing  to note is that, although Eqn. (\ref{npairs})  at large $\Delta\eta\equiv\eta_1-\eta_2$ has an enhancement factor $(\Delta\eta)^4$, the production cross section Eqn. (\ref{nprod}) has only a factor $(\Delta\eta)^2$. The second important difference is that the decrease at large transverse momentum is faster for the production  cross section. The second contribution in Eqn. (\ref{nprod}) has the overall power $p^{-8}$, as opposed to $p^{-6}$ in Eqn. (\ref{npairs}). The first, $\delta$-function term has the same power dependence $p^{-4}$, but the prefactor in Eqn. (\ref{nprod}) is proportional to $\mu^2\lambda^2$, as opposed to $\mu^4$ in Eqn. (\ref{npairs}). These general features are quite expected, since the number of correlated pairs produced in the final state has to be smaller than the number of pairs present in the incoming wave function. Recall that, similarly, the single inclusive particle production decreases at large momentum as $p^{-4}$, while the number of partons in the wave function decreases only as $p^{-2}$. In this sense, our results are consistent with expectations.

\section{Conclusions}

In this paper we calculate, for the first time, quark-quark correlated production in the CGC approach. We find that there is a depletion of pair production at like transverse momenta due to the Pauli blocking effect. A parallel quantum statistics effect  for gluons, the Bose enhancement, was discussed previously in connection to the ridge correlation.

 In contradistinction with the Bose enhancement for gluons,  Pauli blocking is short range in rapidity. The effect decays exponentially with the rapidity difference between the two produced quarks. This exponential decrease, however, is tempered somewhat by a factor quadratic in the rapidity difference, resulting in a dip at $\Delta \eta \sim 2$. Besides, the effect turns out to be parametrically ${\cal O}(\alpha_s^2 N_c)$ relative to gluon-gluon correlations, which for realistic values of $\alpha_s\sim 0.2$ and $N_c=3$ results in a mild suppression factor. Thus, it  is possible that the effect is big enough to be observable.

 It would be extremely interesting to device a measurement which could separate the part of the particle production which originates predominantly from the quarks in the wave function. One possibility that comes to mind would be to measure open charm-open charm correlations. The two charmed hadrons in the final state are more likely to originate from the charm component in the incoming hadron wave function rather than from hadronization of gluons. It is thus likely that the weight of the Pauli blocking effect in  such an observable is more significant than for unidentified charged particles. Whether it is possible to separate this short range in rapidity effect from the jet fragmentation contribution is another important question. Although the nature of the two effects is very distinct, it may be experimentally challenging to distinguish between the two. Similar considerations hold for the difference between the azimuthal correlations of equal and opposite sign charged particles.

Even though the measurement of the Pauli correlations may require a considerable effort, to our mind this effort is well worth making. Given that our knowledge of the hadronic wave function is rather rudimentary, this seems to be a very interesting opportunity to probe its structure well beyond the average observables that determine parton density functions, transverse momentum distributions and generalized parton densities.

\section*{Acknowledgments}
%\textcolor{red}{ADD GRANTS, CORRECT.}\\
NA thanks the Department of Physics of the University of Connecticut for warm hospitality during stays when part of this work was done.
The research was supported by the EU FP7 IRSES network "High-Energy QCD for Heavy Ions" under REA grant agreement \#318921;
the NSF grant 1614640 (AK);
the  ISRAELI SCIENCE FOUNDATION grants \# 1635/16 and \# 147/12 (ML) and the
 BSF grants \#2012124 and \#2014707 (AK,ML); the European Research Council grant HotLHC ERC-2011-StG-279579, Ministerio de Ciencia e Innovaci\'on of Spain under project FPA2014-58293-C2-1-P, and Xunta de Galicia (Conseller\'{\i}a de Educaci\'on) within the Strategic Unit AGRUP2015/11 (NA); and Funda\c{c}\~{a}o para a Ci\^{e}ncia e a Tecnologia (Portugal) under projects CERN/FIS-NUC/0049/2015 and  SFRH/BPD/112655/2015 (TA).

\appendix

\section{Light Cone Hamiltonian} \label{sec:A}
In this Appendix we present the Light Cone Hamiltonian calculation of the dressed perturbative state used in Section 2. In our notation, see \cite{Lublinsky:2016meo}, the light-cone components of four-vectors read $p^\mu\equiv (p^+, p^-,p)$, so $p$ represents the transverse momentum.

The free part of the Light Cone Hamiltonian (LCH, see \cite{LCH})  is
\bea
H_0&=&\int_{k^+>0}{dk^+\over 2\pi}{d^2k\over(2\pi)^2}{k^2\over 2k^+}\ a^{\dagger a}_i(k^+,k)\ a^{a}_i(k^+,k) \\ \nn
\nonumber
&+&\sum_{s}\,\int_{p^+>0}\frac{dp^+ d^2\,p}{(2\pi)^3}\,
\frac{p^2}{2\,p^+} \,\left[d_{\alpha\,s}^\dagger(p^+,p)\,d_{\alpha\,s}(p^+,p)\,+\,\bar d_{\alpha\,s}^\dagger(p^+,p)\,\bar d_{\alpha\,s}(p^+,p)
\right],
\eea
where $a,a^\dagger$ are gluon annihilation and creation operators, $a$ and $\alpha$ are color indices in the adjoint and fundamental representations, respectively, and $i$ and $s$ polarisation and helicity.
This defines the standard free dispersion relations:
\begin{equation}
E_g=k^-={k^2\over 2k^+}, \ \
E_q=p^-={p^2\over 2\,p^+}.
\end{equation}

To zeroth order the vacuum of the LCH is simply the zero energy Fock space vacuum of the operators  $a$, $d$ and $\bar d$:
$$a_q|0\rangle =0,\ \  d_p|0\rangle=0,\ \  \bar d_p|0\rangle=0,\ \
E_0=0.$$
The normalized one-particle states to zeroth order are
\bea
|k^+,k,a,i\rangle&=&{1\over (2\pi)^{3/2}}\, a^{a\,\dagger}_i(k^+,k)\,|0\rangle,\nn \\
\langle k_1^+,k_1,a,i|k_2^+,k_2,b,j\rangle\,&=&\, \delta^{ab}\, \delta_{ij}\, \delta^{(2)}(k_1 -k_2)\delta(k_1^+ -k_2^+), \nn \\
|p^+,p,\alpha,s\rangle&=&{1\over (2\,\pi)^{3/2}}\, d^{\dagger}_{\alpha,s}(p^+,p)\,|0\rangle,\nn\\
\langle p_1^+,p_1,\alpha,s_1|p_2^+,p_2,\beta,s_2\rangle\,&=&\, \delta_{\alpha\beta}\,\delta_{s_1 s_2} \, \delta^{(2)}(p_1 -p_2)\delta(p_1^+ -p_2^+).
\eea

The full Hamiltonian contains several types of perturbations,
$$\delta H\,=\,\delta H^\rho\,+\,\delta H^{g\,qq}\,+\,\cdots .$$
By $\cdots$ we denote terms that include the soft gluon sector, which is of no relevance for the present work. $\rho$ denotes the color density of the background field.

\subsubsection*{Interaction with the background field}
Recall that we are interested in approximate eigenstates of the Hamiltonian in the presence of the background color charge density  due to valence partons. The interaction with the background charge is comprised of three terms
\beq
\delta H^{\rho}\,=\,\delta H^{\rho\,g}\,+\, \delta H^{\rho\,qq}\,+\, \delta H^{\rho\,gg}\ .
\eeq
The last term is of no interest to us since it does not involve quarks. The remaining ones are
\bea
\delta H^{\rho\,g}&=&
\int_0^\infty {dk^+\over 2\pi}{d^2k\over (2\pi)^2}{g\,k_i\over \sqrt{2}\,|k^{+}|^{3/2}}\
\Big[a^{\dagger a}_i(k^+,\,k)\  \rho^a(-k) 
%\nonumber \\
%&&
+ a^a_i(k^+,\,k)\,\  \rho^a(k)\Big] , \\
\delta H^{\rho\,qq}&=&\sum_{s}
\int {dk^+d^2k\,dp^+d^2p\over (2\pi)^6}{g^2\over \,(k^{+})^2}\
\Big[d^{\dagger}_{\alpha\,s}(p^+,\,p)\ \tau^a_{\alpha\beta}\ \bar d^{\dagger}_{\beta\,s}(k^+\,-\,p^+,\,k-p)\
\rho^a(-k)
%\nonumber \\
%&&
+   h.c.\Big].
\eea
Here $\rho$ is a charge density operator, corresponding to the valence or hard degrees of freedom and depending only on transverse coordinates, and $\tau^a_{\alpha\beta}$ the color matrices in the fundamental representation. These charges satisfy the $SU(N)$ algebra:
\beq
[\rho^a(x),\, \rho^b(y)]\,=\,i\,f^{abc}\, \rho^c(x)\,\delta^{(2)}(x\,-\,y),\ \
[\rho^a(k),\, \rho^b(p)]\,=\,i\,\,f^{abc}\, \rho^c(k\,+\,p).
\eeq

\subsubsection*{Quark-gluon interaction}

The quark-gluon interaction responsible for quark production reads
\bea
\delta H^{g\,qq}&=& g\,\tau^a_{\alpha\beta}\,\sum_{s_1,s_2}\,\int {dp^+\,d^2p\, dk^+\,d^2k
\over 2^{3/2}\,(2\pi)^6\,(k^+)^{1/2}}\,\theta(k^+\,-\,p^+)\,\Gamma^i_{s_1\,s_2}(k^+,k,p^+,p)\nn\\
&\times& \Big[{a}_{i}^a(k^+,k)\,d^\dagger_{\alpha,\,s_1}(p^+,p)\,\bar
d^\dagger_{\beta,s_2}(k^+-p^+,k-p)
+
h.c.\Big], \eea
with the vertex $\Gamma^i$ defined as
\bea
\Gamma^i_{s_1s_2}(k^+,k,p^+,p)&=&\chi_{s_2}^\dagger\left[2\frac{k_i}{k^+}-
\frac{\sigma\cdot p}{p^+}\sigma^i-
\sigma^i\frac{\sigma\cdot (k-p)}{(k^+-p^+)}\right]\chi_{s_1} \\
&=&\chi_{s_2}^\dagger\left[2\frac{k_i}{k^+}-\left({p_i\over p^+}+{k_i-p_i\over k^+-p^+}\right)
+i\epsilon^{im}\sigma^3\left({p_m\over p^+}-{k_m-p_m\over k^+-p^+}\right)\right]\chi_{s_1} \nn \\
&=&\delta_{s_1s_2}\left[2\frac{k_i}{k^+}-\left({p_i\over p^+}+{k_i-p_i\over k^+-p^+}\right)
+2is_1\epsilon^{im}\left({p_m\over p^+}-{k_m-p_m\over k^+-p^+}\right)\right],\nn
\eea
and the spinors $\chi_{s=1/2}=(1,0)$ and $\chi_{s=-1/2}=(0,1)$ satisfying
\beq
\chi^\dagger_{s_1}{\bf 1}\chi_{s_2}=\delta_{s_1s_2},\ \
\chi^\dagger_{s_1}\sigma^3\chi_{s_2}=2s_1\delta_{s_1s_2}.
\eeq
%We will not be interested in this paper in observables which depend on spin. The only quantity which will enter
%is the matrix $M$ defined as
%\bea
%M^{ij}(k,p;\bar k, \bar p)
%&=& \sum_{s_1,s_2}\Gamma^i_{s_1s_2}(k,p)\Gamma^{j*}_{s_2s_1}(\bar k,\bar p)\nn \\
%&=&{8k_i\bar k_j\over k^+\bar k^+ }-
%4{k_i\over k^+}\left({\bar p_j\over \bar p^+}+{\bar k_j-\bar p_j\over \bar k^+-\bar p^+}\right)-
%4{\bar k_j\over \bar k^+}\left({p_i\over p^+}+{k_i-p_i\over k^+-p^+}\right)\nn \\
%&+&
%2\left({p_i\over p^+}+{k_i-p_i\over k^+-p^+}\right)
%\left({\bar p_j\over \bar p^+}+{\bar k_j-\bar p_j\over \bar k^+-\bar p^+}\right)\nn \\
%&+&2\epsilon^{im}\epsilon^{jn}
%\left({p_m\over p^+}-{k_m-p_m\over k^+-p^+}\right)\left({\bar p_n\over \bar p^+}-{\bar k_n-\bar
%p_n\over \bar k^+-\bar p^+}\right),
%\eea
%with
%\beq\label{epsilon}
%\epsilon^{ij}\epsilon^{mn}=\delta^{im}\delta^{jn}-\delta^{in}\delta^{mj}\ .
%\eeq

\subsection{Matrix elements}

In order to calculate the perturbative wave function one needs the following matrix elements:
\bea
\langle g|\delta H^{\rho\, g}|0\rangle&=&{\langle 0|a_i^a(k^+,k)\ \delta H^{\rho g}|0\rangle\over (2\pi)^{3/2}}
={gk_i\over 4\pi^{3/2}|k^{+}|^{3/2}}\rho^a(-k), \\
\langle q\bar q|\delta H^{\rho qq}|0\rangle&=&{\langle 0| d_{\alpha s_1}(q^+,q)\;\bar d_{\beta s_2}(p^+,p)
\delta H^{\rho\, qq}|0\rangle\over (2\pi)^3}
={g^2\tau^a_{\alpha\beta}\over (2\pi)^{3}(p^{+}+q^+)^{2}}\rho^a(-p-q)\delta_{s_1s_2}\ ,
\nn \\
\langle q\bar q|\delta H^{g\, qq}|g\rangle&=&
{\langle 0 |d_{\alpha s_1}(p^+,p)\bar d_{\beta s_2}(q^+,q)\delta H^{q}a^{a\dagger}_i(k^+,k)|0\rangle \over (2\pi)^{9/2}}\nn\\
&=&
g\tau^a_{\alpha\beta}{\Gamma^i_{s_1s_2}(k^+,k,p^+,p)\over 8\pi^{3/2}(k^+)^{1/2}}\delta^{(2)}(p+q-k)  \delta(p^++q^+-k^+) .\nn
\eea
With these matrix elements, using the standard perturbation theory we obtain the wave functions, Eqns.  (\ref{vd},\ref{v4})

%%%%%%%%%%%%%%%%%%%%%%%%%%%%%%%%%%%%%%%%%%%%%%%%%%%%%%%%%%%%%%%
%%%%%%%%%%%%%%%%%%%%%%%%%%%%%%%%%%%%%%%%%%%%%%%%%%%%%%%%%%%%%%%

\section{Derivation of the pair density}\label{sec:B}
In this Appendix we present the derivation of the pair density and show how we define $\Phi_2$ and $\Phi_4$ that is used in the calculations.

Let us start with the formal definition of  the pair density which is given in Eqn. (\ref{Npair}). Here, $p^+$ and $q^+$ are the longitudinal momenta, $p$ and $q$ are the transverse momenta of the quark pair in the wave function. First, we need to calculate the action of two quark annihilation operators on the dressed state which is given explicitly in Eqn. (\ref{v4}): 
\bea
\label{Npair_1}
d^{\kappa}_{s_2}(q^+,q)d^{\omega}_{s_1}(p^+,p) |v\rangle_4^D&=& \frac{1}{2}g^4\int
d\alpha \frac{dk^+d^2p'd^2\bar{p}'}{(2\pi)^3}
d\beta \frac{d\bar{k}^+d^2q'd^2\bar{q}'}{(2\pi)^3} 
\zeta^{\epsilon\iota}_{s'_1s'_2}(k^+,p',\bar{p}';\alpha)
\zeta^{\gamma\delta}_{r_1r_2}(\bar{k}^+,q',\bar{q}';\beta)\nonumber\\
&&
\times
d^{\kappa}_{s_2}(q^+,q)d^{\omega}_{s_1}(p^+,p) 
{d^{\dagger}}^{\epsilon}_{s'_1}(\bar{\alpha}k^+,p')
{d^{\dagger}}^{\gamma}_{r_1}(\bar{\beta}\bar{k}^+,q')
\bar{d}^{\dagger \iota}_{s'_2}(\alpha k^+,\bar{p}')
\bar{d}^{\dagger \delta}_{r_2}(\beta \bar{k}^+,\bar{q}') |v\rangle \, .
\eea
We use the anticommutation relations for the quark creation and annihilation operators
\bea
\{ d^{\omega}_{s_1}(k^+,k), d^{\dagger \zeta}_{s_2}(q^+,q)\}= (2\pi)^3\delta^{\omega \zeta}\delta_{s_1s_2} \delta(k^+-q^+) \delta^{(2)}(k-q)
\eea
in order to simplify Eqn. (\ref{Npair_1}) and we get
\bea
\label{Npair_2}
d^{\kappa}_{s_2}(q^+,q)d^{\omega}_{s_1}(p^+,p) |v\rangle_4^D&=& \frac{1}{2}g^4\int
d\alpha \frac{dk^+d^2p'd^2\bar{p}'}{(2\pi)^3}
d\beta \frac{d\bar{k}^+d^2q'd^2\bar{q}'}{(2\pi)^3} 
\zeta^{\epsilon\iota}_{s'_1s'_2}(k^+,p',\bar{p}';\alpha)
\zeta^{\gamma\delta}_{r_1r_2}(\bar{k}^+,q',\bar{q}';\beta)\nonumber\\
&&
\times
\Big\{ 
(2\pi)^3\delta^{\omega\epsilon}\delta_{s_1s'_1}\delta(p^+-\bar{\alpha}k^+)\delta^{(2)}(p-p')
(2\pi)^3\delta^{\kappa\gamma}\delta_{s_2r_1}\delta(q^+-\bar{\beta}\bar{k}^+)\delta^{(2)}(q-q')
\nonumber\\
&&
- 
(2\pi)^3\delta^{\kappa\epsilon}\delta_{s_2s'_1}\delta(q^+-\bar{\alpha}k^+)\delta^{(2)}(q-p')
(2\pi)^3\delta^{\omega\gamma}\delta_{s_1r_1}\delta(p^+-\bar{\beta}\bar{k}^+)\delta^{(2)}(p-q')\Big\}
\nonumber\\
&&
\times
\bar{d}^{\dagger \iota}_{s'_2}(\alpha k^+,\bar{p}')
\bar{d}^{\dagger \delta}_{r_2}(\beta \bar{k}^+,\bar{q}') |v\rangle \; .
\eea
It is now straight forward to calculate the quark pair density by simply calculating the overlap of Eqn. (\ref{Npair_2}) with its Hermitian conjugate which gives 
\bea
\label{Npair_squarep}
\frac{dN}{dp^+d^2pdq^+d^2q}&=&\frac{1}{(2\pi)^6}
%^{D}_{4}\langle v| d^{\dagger \omega}_{s_1}(p^+,p)d^{\dagger \kappa}_{s_2}(q^+,q)d^{\kappa}_{s_2}(q^+,q)d^{\omega}_{s_1}(p^+,p) |v\rangle_4^D =\nonumber\\
%&&
\frac{1}{4}g^8 \int
d\alpha \frac{dk^+d^2p'd^2\bar{p}'}{(2\pi)^3}
d\beta \frac{d\bar{k}^+d^2q'd^2\bar{q}'}{(2\pi)^3} 
d\alpha'' \frac{dl^+d^2p''d^2\bar{p}''}{(2\pi)^3}
d\beta'' \frac{d\bar{l}^+d^2q''d^2\bar{q}''}{(2\pi)^3}
\nonumber\\
&&
\times
\zeta^{\epsilon\iota}_{s'_1s'_2}(k^+,p',\bar{p}';\alpha)
\zeta^{\gamma\delta}_{r_1r_2}(\bar{k}^+,q',\bar{q}';\beta)
\zeta^{*\epsilon' \iota'}_{s''_1s''_2}(l^+,p'',\bar{p}'';\alpha'')
\zeta^{*\gamma' \delta' }_{r''_1 r''_2}(\bar{l}^+,q'',\bar{q}'';\beta'') (2\pi)^{12}
\nonumber\\
&&
\times
\Big\{ 
\delta^{\omega\epsilon}\delta_{s_1s'_1}\delta(p^+-\bar{\alpha}k^+)\delta^{(2)}(p-p')
\delta^{\kappa\gamma}\delta_{s_2r_1}\delta(q^+-\bar{\beta}\bar{k}^+)\delta^{(2)}(q-q')
\nonumber\\
&&
-
\delta^{\kappa\epsilon}\delta_{s_2s'_1}\delta(q^+-\bar{\alpha}k^+)\delta^{(2)}(q-p')
\delta^{\omega\gamma}\delta_{s_1r_1}\delta(p^+-\bar{\beta}\bar{k}^+)\delta^{(2)}(p-q')\Big\}
\nonumber\\
&&
\times
\Big\{
\delta^{\omega\epsilon'}\delta_{s_1s''_1}\delta(p^+-\bar{\alpha}''l^+)\delta^{(2)}(p-p'')
\delta^{\kappa\gamma'}\delta_{s_2r''_1}\delta(q^+-\bar{\beta}''\bar{l}^+)\delta^{(2)}(q-q'')
\nonumber\\
&&
-
\delta^{\kappa\epsilon'}\delta_{s_2s''_1}\delta(q^+-\bar{\alpha}''l^+)\delta^{(2)}(q-p'')
\delta^{\kappa\gamma'}\delta_{s_1r''_1}\delta(p^+-\bar{\beta}''\bar{l}^+)\delta^{(2)}(p-q'')\Big\}
\nonumber\\
&&
\times
\langle v|
\bar{d}^{\delta'}_{r''_2}(\beta''\bar{l}^+,\bar{q}'')
\bar{d}^{\iota'}_{s''_2}(\alpha''l^+,\bar{p}'')
\bar{d}^{\dagger \iota}_{s'_2}(\alpha k^+,\bar{p}')
\bar{d}^{\dagger \delta}_{r_2}(\beta \bar{k}^+,\bar{q}') |v\rangle.
\eea
Using the anticommutation relations for the antiquark creation and annihilation operators, we get another set of $\delta$-functions from the last line of Eqn. (\ref{Npair_squarep}). Hence, the quark pair density reads
\bea
\label{Npair_square}
\frac{dN}{dp^+d^2pdq^+d^2q}&=& \frac{1}{(2\pi)^6}
%^{D}_{4}\langle v| d^{\dagger \omega}_{s_1}(p^+,p)d^{\dagger \kappa}_{s_2}(q^+,q)d^{\kappa}_{s_2}(q^+,q)d^{\omega}_{s_1}(p^+,p) |v\rangle_4^D =\nonumber\\
%&&
\frac{1}{4}g^8 \int
d\alpha \frac{dk^+d^2p'd^2\bar{p}'}{(2\pi)^3}
d\beta \frac{d\bar{k}^+d^2q'd^2\bar{q}'}{(2\pi)^3} 
d\alpha'' \frac{dl^+d^2p''d^2\bar{p}''}{(2\pi)^3}
d\beta'' \frac{d\bar{l}^+d^2q''d^2\bar{q}''}{(2\pi)^3}
\nonumber\\
&&
\times
\zeta^{\epsilon\iota}_{s'_1s'_2}(k^+,p',\bar{p}';\alpha)
\zeta^{\gamma\delta}_{r_1r_2}(\bar{k}^+,q',\bar{q}';\beta)
\zeta^{*\epsilon' \iota'}_{s''_1s''_2}(l^+,p'',\bar{p}'';\alpha'')
\zeta^{*\gamma' \delta' }_{r''_1 r''_2}(\bar{l}^+,q'',\bar{q}'';\beta'') (2\pi)^{18}
\nonumber\\
&&
\times
\Big\{ 
\delta^{\omega\epsilon}\delta_{s_1s'_1}\delta(p^+-\bar{\alpha}k^+)\delta^{(2)}(p-p')
\delta^{\kappa\gamma}\delta_{s_2r_1}\delta(q^+-\bar{\beta}\bar{k}^+)\delta^{(2)}(q-q')
\nonumber\\
&&
-
\delta^{\kappa\epsilon}\delta_{s_2s'_1}\delta(q^+-\bar{\alpha}k^+)\delta^{(2)}(q-p')
\delta^{\omega\gamma}\delta_{s_1r_1}\delta(p^+-\bar{\beta}\bar{k}^+)\delta^{(2)}(p-q')\Big\}
\nonumber\\
&&
\times
\Big\{
\delta^{\omega\epsilon'}\delta_{s_1s''_1}\delta(p^+-\bar{\alpha}''l^+)\delta^{(2)}(p-p'')
\delta^{\kappa\gamma'}\delta_{s_2r''_1}\delta(q^+-\bar{\beta}''\bar{l}^+)\delta^{(2)}(q-q'')
\nonumber\\
&&
-
\delta^{\kappa\epsilon'}\delta_{s_2s''_1}\delta(q^+-\bar{\alpha}''l^+)\delta^{(2)}(q-p'')
\delta^{\kappa\gamma'}\delta_{s_1r''_1}\delta(p^+-\bar{\beta}''\bar{l}^+)\delta^{(2)}(p-q'')\Big\}
\nonumber\\
&&
\times
\Big\{
\delta^{\iota'\iota}\delta_{s''_2s'_2}\delta(\alpha''l^+-\alpha k^+)\delta^{(2)}(\bar{p}'-\bar{p}'')
\delta^{\delta'\delta}\delta_{r''_2r_2}\delta(\beta''\bar{l}^+-\beta\bar{k}^+)\delta^{(2)}(\bar{q}'-\bar{q}'')
\nonumber\\
&&
-
\delta^{\iota'\delta}\delta_{s''_2r_2}\delta(\alpha''l^+-\beta\bar{k}^+)\delta^{(2)}(\bar{q}'-\bar{p}'')
\delta^{\delta' \iota}\delta_{r''_2s'_2}\delta(\beta''\bar{l}^+-\alpha k^+)\delta^{(2)}(\bar{p}'-\bar{q}'')\Big\}.
\eea
We now substitute the definition of the $\zeta$-functions that is given in Eqn. (\ref{zeta_def}) and integrate over all the longitudinal momenta, the longitudinal momentum fractions $\alpha''$ and $\beta''$, and all the transverse momenta except $\bar{p}'$ and $\bar{q}'$. After all this, the quark pair density reads
\bea
\label{Npair_square_3}
&&
\frac{dN}{dp^+d^2pdq^+d^2q}=g^8\int d\alpha d\beta 
\frac{d^2k}{(2\pi)^2} \frac{d^2\bar{k}}{(2\pi)^2} \frac{d^2l}{(2\pi)^2} \frac{d^2\bar{l}}{(2\pi)^2}
d^2\bar{p}' \,  d^2\bar{q}'
\rho^a(k) \rho^b(\bar{k}) \rho^c(l) \rho^d(\bar{l})\nonumber\\
&&
\times
\Bigg\{\frac{1}{\bar{\alpha}\bar{\beta}}
\frac{1}{p^+\left(1+\frac{\alpha}{\bar{\alpha}} \right)}
\frac{1}{q^+\left(1+\frac{\beta}{\bar{\beta}} \right)}
{\rm tr}(\tau^a\tau^c){\rm tr}(\tau^b\tau^d)
\phi_{s_1s_2}(k,p,\bar{p}';\alpha) \phi_{r_1r_2}(\bar{k}, q,\bar{q}'; \beta)
\phi^*_{s_1s_2}(l,p,\bar{p}'; \alpha) \phi^*_{r_1r_2}(\bar{l},q,\bar{q}'; \beta)\nonumber\\
&&
-\frac{1}{\bar{\alpha}\bar{\beta}} \frac{1}{q^++\frac{\alpha}{\bar{\alpha}}p^+} \frac{1}{p^++\frac{\beta}{\bar{\beta}}q^+}
{\rm tr}(\tau^a\tau^c\tau^b\tau^d)
\phi_{s_1s_2}(k,p,\bar{p}';\alpha)\phi_{r_1r_2}(\bar{k},q,\bar{q}';\beta)
\nonumber\\
&&
\hspace{4cm}
\times
\phi^*_{r_1s_2}\left(l,q,\bar{p}; \frac{\alpha p^+}{\bar{\alpha}q^++\alpha p^+}\right)
\phi^*_{s_1r_2}\left(\bar{l},p,\bar{q}'; \frac{\beta q^+}{\bar{\beta}p^++\beta q^+}\right)\Bigg\}.
\eea
At this point one should note that the momentum fractions that appear in second term can be further simplified by realizing that 
\bea
\frac{q^+}{\bar{p}^+}=\frac{1- \frac{\alpha p^+}{\bar{\alpha}q^++\alpha p^+}}{\frac{\alpha p^+}{\bar{\alpha}q^++\alpha p^+}}=\frac{\bar{\alpha}q^+}{\alpha p^+}
\eea  
which simply shows that the momentum fraction  between the pairs is indeed $\alpha$. A similar argument is also true for the other momentum fraction that appears in the last line of Eqn. (\ref{Npair_square_3}). Then, we can write the quark pair density in the wave function as
\bea
\label{Npair_square_4}
&&
\frac{dN}{dp^+d^2pdq^+d^2q}=\frac{1}{(2\pi)^4}g^8\frac{1}{p^+q^+}\int
d^2k \, d^2\bar k \, d^2l \, d^2\bar l \, 
%\frac{d^2k}{(2\pi)^2} \frac{d^2\bar{k}}{(2\pi)^2} \frac{d^2l}{(2\pi)^2} \frac{d^2\bar{l}}{(2\pi)^2}
\rho^a(k) \rho^b(\bar{k}) \rho^c(l) \rho^d(\bar{l})\\
&&
\times
\Bigg\{ 
\Bigg[{\rm tr}(\tau^a\tau^c)\int d\alpha \int \frac{d^2\bar{p}'}{(2\pi)^2}\phi_{s_1s_2}(k,p,\bar{p}';\alpha)\phi^*_{s_1s_2}(l,p,\bar{p}';\alpha)\Bigg]
\Bigg[{\rm tr}(\tau^b\tau^d)\int d\beta \int \frac{d^2\bar{q}'}{(2\pi)^2} \phi_{r_1r_2}(\bar{k},q,\bar{q}';\beta)\phi^*_{r_1r_2}(\bar{l},q,\bar{q}';\beta)\Bigg]\nonumber\\
&&
-
{\rm tr}(\tau^a\tau^c\tau^b\tau^d) \int \frac{d\alpha d\beta}{\left(\alpha+\bar{\alpha}\frac{q^+}{p^+}\right) \left( \beta+\bar{\beta}\frac{p^+}{q^+}\right)} 
\int \frac{d^2\bar{p}'}{(2\pi)^2}\frac{d^2\bar{q}'}{(2\pi)^2}
\phi_{s_1s_2}(k,p,\bar{p}';\alpha)\phi_{r_1r_2}(\bar{k},q,\bar{q}';\beta)\phi^*_{r_1s_2}(l,q,\bar{p};\alpha)\phi^*_{s_1r_2}(\bar{l},p,\bar{q}; \beta)\Bigg\}.\nonumber
\eea
After defining $\eta_1=\ln (p^+_0/p^+)$ and $\eta_2=\ln (p^+_0/q^+)$, and using   Eqns. (\ref{Phi_2_def}, \ref{Phi_4_def}) for the definitions of $\Phi_2$ and $\Phi_4$ respectively, one gets Eqn. (\ref{Npair_square_final}).
%\bea
%\label{Npair_square_final}
%\frac{dN}{d\eta_1d^2pd\eta_2d^2q}&=&\frac{1}{(2\pi)^6}g^8\int 
%\frac{d^2k}{(2\pi)^2} \frac{d^2\bar{k}}{(2\pi)^2} \frac{d^2l}{(2\pi)^2} \frac{d^2\bar{l}}{(2\pi)^2}
%\rho^a(k) \rho^b(\bar{k}) \rho^c(l) \rho^d(\bar{l}) \nonumber\\
%&&
%\times
%\Bigg\{ {\rm tr}(\tau^a\tau^c) {\rm tr}(\tau^b\tau^d) \Phi_2(k,l; p; \alpha) \Phi_2(\bar{k},\bar{l}; q; \beta)%\nonumber\\
%%&&
%-{\rm tr}(\tau^a\tau^c\tau^b\tau^d) \Phi_4(k,\bar{l},\bar{k},l;p,q)\Bigg\}
%\eea

%%%%%%%%%%%%%%%%%%%%%%%%%%%%%%%%%%%%%%%%%%%%%%%%%%%%%%%%%%%%%%%%%%%%%%
\section{Estimate of the pair density in the wave function}\label{sec:C}

In this Appendix we present the details of the calculation of the quark pair density in the CGC wave function discussed in Section \ref{sec:Pauli}.

Consider first $\Phi^A_4$:
%\bea
%\int_{k,\bar k ,l ,\bar l}\Phi_4^A&\simeq &\,\delta^{(2)}(p-q)\delta^{(2)}(0) e^{\eta_2-\eta_1}\int_0^1 \,
%\frac{d\alpha\,d\beta}{ \alpha\bar \beta} \int_{k,\bar k}\frac{\mu^2(k)\mu^2(\bar k)}{k^4\bar k^4}\\
%&\times&\frac{2\{(k\cdot p)^2+4[k^2p^2-(k\cdot p)^2]\}\{[\bar k\cdot(\bar k-p)]^2+4[\bar k^2p^2-(\bar k\cdot p)^2]\}}{p^4(\bar k-p)^4}\,.\nn
%\eea
%
\bea
\hspace{-0.5cm}
\Phi_4^A(p,q)&\simeq &\,\delta^{(2)}(p-q)\delta^{(2)}(0) e^{\eta_2-\eta_1}\int_0^1 \,
\frac{d\alpha\,d\beta}{ \alpha\bar \beta} \int \frac{d^2k}{(2\pi)^2} \frac{d^2\bar k}{(2\pi)^2} \frac{(2\pi)^2\mu^2(k) (2\pi)^2\mu^2(\bar k)}{k^4\bar k^4}\\
&\times&\frac{2\{(k\cdot p)^2+4[k^2p^2-(k\cdot p)^2]\}\{[\bar k\cdot(\bar k-p)]^2+4[\bar k^2p^2-(\bar k\cdot p)^2]\}}{p^4(\bar k-p)^4}\,.\nn
\eea
The  integral naively is quite badly divergent. Let us  understand what regulates the divergencies:

\noindent $\bullet$
 The $\alpha$ integral. This  logarithmically divergent integral is clearly regulated at $\alpha\sim e^{\eta_2-\eta_1}$. Thus it yields a factor $\eta_1-\eta_2$.

\noindent $\bullet$ The $\beta$ integral is clearly regulated at $\bar\beta\sim e^{\eta_2-\eta_1}$, and results in an identical factor
$\eta_1-\eta_2$.

\noindent $\bullet$  The $\bar k$ integral
$$I_{\bar k}=\int d^2\bar k \; \frac{\mu^2(\bar k)}{\bar k^4(\bar k-p)^4}\{[\bar k\cdot(\bar k-p)]^2+4[\bar k^2p^2-(\bar k\cdot p)^2]\}.$$
This diverges logarithmically at $\bar k=p$ and $\bar k=0$. As it is clear from Eqn. (\ref{phi4A}), the divergence at $\bar k=p$ is regulated at $(\bar k-p)^2\sim \bar \beta p^2\sim e^{\eta_2-\eta_1} p^2$. It is cut off in the ultraviolet (UV) by the values $(\bar k-p)^2\sim p^2$. Thus the "pole" at $\bar k-p=0$ in actual fact gives the contribution to the integral of order
$$I_{\bar k}^1\simeq \frac{5\pi}{2}\frac{\mu^2}{p^2}\ln \frac{p^2}{e^{\eta_2-\eta_1}p^2}=\frac{5\pi}{2}\frac{\mu^2}{p^2}(\eta_1-\eta_2),$$
where the numerical factor follows from the angular integration in the terms involving $\bar k\cdot p$.
The pole at $\bar k=0$ is regulated by vanishing of  $\mu^2(0)$ and is cut off  at $\bar k^2=Q_s^2$. In the UV the integral  is cut off at $\bar k^2\sim p^2$. Then this pole contributes
$$I_{\bar k}^2\simeq \frac{5\pi}{2} \frac{\mu^2}{p^2}\ln \frac{p^2}{Q_s^2}\, ,$$
so that the total result is
$$I_{\bar k}\simeq \frac{5\pi }{2}\frac{\mu^2}{p^2}\left(\eta_1-\eta_2+\ln\frac{p^2}{Q_s^2}\right).$$

\noindent $\bullet$ The $k$ integral
$$I_k=\int d^2 k\; \frac{\mu^2(k)}{p^4k^4}\{(k\cdot p)^2+4[k^2p^2-(k\cdot p)^2]\}.$$
This diverges at $k\rightarrow 0$ and $k\rightarrow\infty$. The IR divergence is again regulated by $Q_s$, while the UV divergence, as is clear from Eqn.(\ref{phi4A}) is regulated at $k^2\sim \frac{1}{\alpha}p^2\sim e^{\eta_1-\eta_2}p^2$. With the same angular integral as before, we find
$$I_{ k}\simeq \frac{5\pi}{2}\frac{\mu^2}{p^2}\left(\eta_1-\eta_2+\ln\frac{p^2}{Q_s^2}\right).$$
Overall, we find that, to leading logarithmic accuracy at large $\eta_1-\eta_2$,
\begin{equation}\label{eta4}
\Phi_4^A(p,q)\simeq \frac{25\pi^2}{2}\delta^{(2)}(p-q)\delta^{(2)}(0) e^{\eta_2-\eta_1}\frac{\mu^4}{p^4}(\eta_1-\eta_2)^2\left[\eta_1-\eta_2+\ln\frac{p^2}{Q_s^2}\right]^2.
\end{equation}
Interestingly, although the correlation decreases with rapidity, the exponential  decrease is dampened by the fourth power of the rapidity difference. It therefore could be numerically quite significant up to relatively large rapidity differences.

Now let us consider the $\Phi^B_4$ term. In the same kinematic regime, we have
\bea
\Phi^B_4(p,q)&\simeq&2\,\delta^{(2)}(0)\int \frac{d^2k}{(2\pi)^2} \frac{d^2\bar k}{(2\pi)^2}\delta^{(2)}(\bar k -k-q+p)e^{\eta_2-\eta_1}
%&\times&
\int\frac{d\alpha d\beta}{\alpha\bar\beta}\frac{(2\pi)^2\mu^2(k) (2\pi)^2\mu^2(k+q-p)}{k^4(k+q-p)^4(k-p)^4p^2q^2}\nn\\
&&\times
\left[k^2(k\cdot p+4p^2)-5(k\cdot p)^2\right]
\Big\{(k+q-p)^2\left[(k+q-p)\cdot q+4q^2\right]-5[(k+q-p)\cdot q]^2\Big\} \, .
%&-&4(2k\cdot p-k^2)[(k+q-p)\cdot(k-q-p)]\nn \\
%&\times& \Big\{[k\cdot(k-p)][p\cdot q]-[k\cdot q][p\cdot(k-p)]\Big\}\Bigg\}.\nn
\eea
The difference now is that there is only one integral over $k$. This integral gets contributions from three poles: $k=0,\ p-q,\ p$. The first two are regulated by the appropriate $\mu^2$, while the last one, as before, is regulated by the denominator at $(k-p)^2\sim e^{\eta_2-\eta_1}\max(p^2,q^2)$. In the UV all the integrals are regulated by a scale of order $p-q$. The contributions of the first two poles  give
\beq
\pi e^{\eta_2-\eta_1}(\eta_1-\eta_2)^2\frac{\mu^4}{p^4q^4}\frac{3(p^2+q^2)}{(p-q)^4}
\left\{5[p^2q^2-(p\cdot q)^2]-(p-q)^2p\cdot q\right\}\ln\frac{(p-q)^2}{Q_s^2}\, .\nn
\eeq
The third pole gives
\beq
\pi e^{\eta_2-\eta_1}(\eta_1-\eta_2)^3\frac{\mu^4}{p^4q^4}p\cdot q\, .\nn
\eeq
Thus, finally,
\bea
\hspace{-0.4cm}
\label{phib1}
\Phi^B_4(p,q)\simeq \delta^{(2)}(0)e^{\eta_2-\eta_1}(\eta_1-\eta_2)^2\frac{\pi\mu^4}{p^4q^4}
\!\Bigg[\frac{3(p^2+q^2)}{(p-q)^4}
\!\!
\left\{5[p^2q^2-(p\cdot q)^2]\!-\!(p-q)^2p\cdot q\right\}\!\!\!\
%\nn\\
%&\times& 
\ln\frac{(p-q)^2}{Q_s^2} +(\eta_1-\eta_2)\, p\cdot q\Bigg].
\eea
 Putting together Eqn. (\ref{eta4}) and Eqn. (\ref{phib1}) gives Eqn. (\ref{npairs}).

%%%%%%%%%%%%%%%%%%%%%%%%%%%%%%%%%%%%%%%%%%%%%%%%%%%%%%%%%%%%%%%%%
%%%%%%%%%%%%%%%%%%%%%%%%%%%%%%%%%%%%%%%%%%%%%%%%%%%%%%%%%%%%%%%%%

\section{The diagonalizing operator $\Omega$}
To calculate particle production in the CGC approach one requires the knowledge of the operator $\Omega$, which diagonalizes the LCH to a given order in perturbation theory \cite{Kovner:2007zu}. The operator $\Omega$ in our case can be represented as
\beq
{\Omega}\,=\,\Omega_g \,\Omega_{qq}\,\Omega_{gqq},
\eeq
where  $\Omega_g$ and $\Omega_{qq}$ come from the diagonalization of the perturbations   $\delta H^{\rho\, g}$ and $\delta H^{\rho\, qq}$ respectively:
\bea
\Omega_g&=&{\rm exp}\left\{\,-\,
%i\int d^2x\,b_i^a(x)\,\int_{\Lambda e^{Y_0}}^{\Lambda e^Y}{dk^+\over \sqrt{2}\, \pi| k^+|^{1/2}}
i\int d^2x\,b_i^a(x)\,\int{dk^+\over \sqrt{2}\, \pi| k^+|^{1/2}}
\left[a^{ a}_i(k^+, x)\,+\,a^{\dagger a}_i(k^+, x)\right]\right\}
%\Omega_g&=&{\rm Exp}\left\{\,-\,
%i\int {d^2k\over (2\,\pi)^2}\,b_i^a(k)\,\int_{e^{Y_0}\,\Lambda}^{e^Y\,\Lambda}{dk^+\over \sqrt{2}\,\pi| k^+|^{1/2}}
%\left[a^{\dagger\, a}_i(k^+, k)\,-\,a^{ a}_i(k^+, k)\right]\right\}\,
 \label{C}
\eea
and
\bea
%&&|\theta\rangle_2=\Omega_{qq}\ |0\rangle. \nn \\
&&\Omega_{qq}={\rm exp}\,\Bigg\{g^2\,\tau^a_{\alpha\beta} \,
\int {dk^+\over (2\,\pi)^2 }\,\int_0^1\,d\alpha\,
\int_{z,\bar z,x}\,\rho^a(x)\, \phi^{(1)}_{s_1,s_2}(x,z,\bar z;\alpha)
%\nn \\
%&& \hspace{1cm} \times 
\left[d^\dagger_{\alpha,s_1}(\alpha\,k^+,z)\,\bar d^\dagger_{\beta,s_2}(\bar\alpha k^+,\bar z)\,-\,h.c.\right ] \Bigg\}.
\eea
 In these expressions, the integration over the +-momenta has to be done in a region $[k_0 e^{Y_0},k_0 e^Y]$ \cite{Kovner:2007zu} with $k_0$ some cutoff that separates soft from fast modes, and  the ``classical'' field $b_i$ is the Weizs\"aker-Williams field of the color charge density $\rho^a$:
\begin{equation}
b^a_i(k)\,=\,g\ {-i\,k_i\over k_\perp^2}\,\rho^a(-k)\,,\ \
b^a_i(x)\,=\,{g\over 2\pi}\ \int d^2y{(x-y)_i\over (x-y)^2}\,\rho^a(y)\,.
\end{equation}
Since the perturbations $H^{\rho\, g}$ and $H^{\rho\, qq}$ involve different degrees of freedom, and to leading order these degrees of freedom do not interact, at this level the diagonalizing operator is simply the product of the two.

Finally, the operator $\Omega_{gqq}$ diagonalizes the gluon-quark interaction. This is performed perturbatively with the result
\bea
%&&|\theta\rangle_2=\Omega_{qq}\ |0\rangle. \nn \\
\Omega_{gqq}&=&{\rm exp}\,\Bigg\{
g\,\tau^a_{\alpha\beta}
%\sum_{s_1,s_2}
\,\int {dp^+\,d^2p\, dk^+\,d^2k
\over 2^{3/2}\,(2\pi)^6\,(k^+)^{1/2}}\,\theta(k^+ - p^+)\,{\Gamma^i_{s_1\,s_2}\over E_{p}+E_{k-p}}\,
\nn \\
&
\times&
 \left[{a}_{i}^a(k^+,k)\,d^\dagger_{\alpha,\,s_1}(p^+,p)\,\bar
d^\dagger_{\beta,\,s_2}(k^+-p^+,k-p)\,+\,
\,h.c.\right]
\Bigg\}.
\eea
As explained in the text, we do not take into account in the production cross section the contributions from two gluons splitting into two quark-antiquark pairs after scattering from the target. For that reason, we do not need to include the perturbations $H^{\rho\, gg}$
% and $H^{ggg}$
and the entire gluon sector in the diagonalization process.
%\beq
%Q_A^{ab}(z_1,\bar z)\equiv tr[\tau^a S_F(z_1)\tau^b S^\dagger_F(\bar z)]
%\eeq

%%%%%%%%%%%%%%%%%%%%%%%%%%%%%%%%%%%%%%%%%%%%%%%%%%%%%%%%%%%%%%%%%%%%%
%%%%%%%%%%%%%%%%%%%%%%%%%%%%%%%%%%%%%%%%%%%%%%%%%%%%%%%%%%%%%%%%%%%%%

\section{Estimate for pair production cross section}
In this appendix we present the calculation of the pair production cross section discussed in Section 3. As indicated before, our estimates are valid in the kinematics $\eta_1\gg\eta_2$, $|q|\sim |p|\sim |q-p|\gg Q_s\gg Q_T\gg\Lambda$, with $\Lambda$ some nonperturbative scale.

\subsection{The  {\it A}-term}
It is simplest to look at the $A$-term, Eqn. (\ref{A}). There are four integrals involved, and each one factorizes into the product of $(k,l)$ and $(\bar k,\bar l)$ integrals. Let us consider them separately.\footnote{In the rest of the appendix we introduce a shorthand notation: $\int_k\equiv\int d^2k$.}

First, we consider
\bea
I_1&=&\int_{k,l}\frac{\mu^2(k)\lambda^2(l)}{l^4}|\bar\Psi(k,l,p,0)|^2\\
&=& \int_{k,l}\frac{\mu^2(k)\lambda^2(l)}{l^4}\, \frac{2}{p^4}\,\Bigg\{\frac{[(k+l)\cdot p]^2}{(k+l)^4}+\frac{(k\cdot p)^2}{k^4}-2\frac{[(k+l)\cdot p]\,(k\cdot p)}{k^2(k+l)^2}
\nn\\
&&+
4\Bigg[\frac{(k+l)^2p^2-[(k+l)\cdot p]^2}{(k+l)^4}+\frac{k^2p^2-(k\cdot p)^2}{k^4}
%\nn\\
%&&\hspace{5mm}
-2\frac{[(k+l)\cdot k]\,p^2-[(k+l)\cdot p]\, (k\cdot p)}{k^2(k+l)^2}\Bigg]\Bigg\}.\nn
\eea

The integral is dominated by the "poles" at $k=0$, $l=0$ and $k+l=0$. The first two divergences are regulated, as before by the vanishing of $\mu^2$ and $\lambda^2$ below their respective saturation momenta.
 The third pole is quite interesting and it has some physics in it. Its origin is explained in the text. This divergence is regulated by the genuine nonperturbative scale $\Lambda$.

Let us first integrate over the part of the phase space $l^2<Q_s^2$. In this regime we can expand the integrand in $l/k$.
We have
\bea
|\bar \Psi(k,l,p,0)|^2&\simeq &2\Bigg\{\frac{(l\cdot p)^2}{p^4k^4}+4\frac{(k\cdot p)^2(k\cdot l)^2}{p^4k^8}-4\frac{(k\cdot p)( k\cdot l)( p\cdot l)}{p^4k^6}
\nn \\
&&
+\,4\frac{(l\times p)^2}{p^4k^4}+16\frac{(k\times p)^2(k\cdot l)^2}{p^4k^8}-16\frac{(l\times p)(k\times p)(k\cdot l)}{p^4k^6}\Bigg\}\, ,
\eea
\beq \int_{Q_T^2<l^2<Q_s^2}\frac{\lambda^2(l)}{l^4}|\bar \Psi(k,l,p,0)|^2\simeq\frac{5 \pi\lambda^2}{p^2k^4}\ln\frac{Q_s^2}{Q_T^2}
\eeq
and
\beq\int_k \int_{Q_T^2<l^2<Q_s^2}\frac{\mu^2(k)\lambda^2(l)}{l^4}|\bar \Psi(k,l,p,0)|^2\simeq\frac{5 \pi^2\mu^2\lambda^2}{Q_s^2}\frac{1}{p^2}\ln\frac{Q_s^2}{Q_T^2}\,.
\eeq

In the rest of the phase space we perform the $k$ integral first. It is saturated by the two poles, $k=0$ an $k=-l$. Each  one of the terms also is formally UV divergent, but this divergence cancels between all the terms. We approximate the integrals by
\bea
\int_k\mu^2(k)\frac{[(k+l)\cdot p]^2}{p^4(k+l)^4}&\approx& \frac{\pi\mu^2}{2p^2}\ln\frac{e^{\eta_1-\eta_2}p^2}{\Lambda^2}\,,\\
\int_k\mu^2(k)\frac{(k\cdot p)^2}{p^4k^4}&\approx& \frac{\pi\mu^2}{2p^2}\ln\frac{e^{\eta_1-\eta_2}p^2}{Q_s^2}\,,\nn\\
\int_k\mu^2(k)\frac{[(k+l)\cdot p]\,[k\cdot p]}{p^4(k+l)^2k^2}&\approx& \frac{\pi\mu^2}{2p^2}\ln\frac{e^{\eta_1-\eta_2}p^2}{l^2}\,.\nn
\eea
Thus we find
\beq
I^{l>Q_s}_1\simeq \frac{5\pi\mu^2}{p^2}\int_{l>Q_s}\frac{\lambda^2(l)}{l^4}\ln \frac{l^4}{\Lambda^2Q_s^2}\simeq\frac{5\pi^2\mu^2\lambda^2}{p^2Q_s^2}\ln\frac{Q_s^2}{\Lambda^2}\,.
\eeq
All in all,
\beq
I_1\simeq\frac{5 \pi^2\mu^2\lambda^2}{p^2Q_s^2}\ln\frac{Q_s^4}{Q_T^2\Lambda^2}\,.
\eeq

Now let consider the second integral
\bea
I_2&=&\int_{k,l}\frac{\mu^2(k)\lambda^2(l)}{l^4}|\Psi(k,l,p,0)|^2\\
&=& \int_{k,l}\frac{\mu^2(k)\lambda^2(l)}{l^4}\,2\,\Bigg\{\frac{\left[(k+l)\cdot p\right]^2}{p^4(k+l)^4}+\frac{\left[k\cdot(p- l)\right]^2}{(p-l)^4k^4}-2\frac{\left[(k+l)\cdot p\right]\left[k\cdot( p-l)\right]}{p^2(p-l)^2k^2(k+l)^2}\nn\\
&+&4\Bigg[
\frac{(k+l)^2p^2-\left[(k+l)\cdot p\right]^2}{p^4(k+l)^4}+\frac{k^2(p-l)^2-[k\cdot( p-l)]^2}{(p-l)^4k^4}\nn\\
&-&2\frac{[(k+l)\cdot k]\,[p\cdot (p-l)]-[(k+l)\cdot (p-l)]\,[ k\cdot p]}{p^2(p-l)^2k^2(k+l)^2}\Bigg]\Bigg\}\,.\nn
\eea
Again, first we consider $l^2<Q_s^2$. The algebra is longer, but the final result is the same:
\beq\int_k \int_{Q_T^2<l^2<Q_s^2}\frac{\mu^2(k)\lambda^2(l)}{l^4}| \Psi(k,l,p,0)|^2\simeq\frac{5\pi^2\mu^2\lambda^2}{Q_s^2}\frac{1}{p^2}\ln\frac{Q_s^2}{Q_T^2}\,.
\eeq

In the rest of the integral, integrating over $k$ we obtain
\bea
I^{l>Q_s}_2&\simeq&5\pi\mu^2\int_{l>Q_s}\frac{\lambda^2(l)}{l^4}\Bigg[\frac{1}{p^2}\ln\frac{e^{\eta_1-\eta_2}p^2}{\Lambda^2}+\frac{1}{(p-l)^2}\ln\frac{e^{\eta_1-\eta_2}p^2}{Q_s^2}\nn\\
&-&2\frac{p\cdot(p-l)}{p^2(p-l)^2}\ln\frac{e^{\eta_1-\eta_2}p^2}{l^2}\Bigg].
\eea
Since the integral is dominated by $l\sim Q_s\ll p$, the difference between $I_2$ and $I_1$  is negligible, and we obtain
\beq
I^{l>Q_s}_2\simeq I^{l>Q_s}_1\simeq \frac{5 \pi^2 \mu^2\lambda^2}{p^2Q_s^2}\ln\frac{Q_s^2}{\Lambda^2}
\eeq
and, thus,
\beq
I_2\simeq I_1\simeq \frac{5\pi^2 \mu^2\lambda^2}{p^2Q_s^2}\ln\frac{Q_s^4}{Q_T^2\Lambda^2}\,.
\eeq

Now it is the turn of
\bea\label{i3}
I_3&=&\int_{k,l}\frac{\mu^2(k)\lambda^2(l)}{l^4}|\bar\Psi(k,l,p,1)|^2\\
&=& \int_{k,l}\frac{\mu^2(k)\lambda^2(l)}{l^4}\,2 \Bigg\{\frac{[(k+l)\cdot(k+l- p)]^2}{(k+l-p)^4(k+l)^4}+\frac{[k\cdot(k-p)]^2}{(k-p)^4k^4}
%\nn\\
%&&
-2\frac{[(k+l)\cdot(k+l- p)]\,[k\cdot(k- p)]}{(k+l-p)^2(k+l)^2k^2(k-p)^2}\nn\\
&+&4\Bigg[\frac{(k+l)^2(k+l-p)^2-[(k+l)\cdot(k+l- p)]^2}{(k+l-p)^4(k+l)^4}+\frac{k^2(k-p)^2-[k\cdot(k- p)]^2}{(k-p)^4k^4}\nn\\
&&-2\frac{[(k+l)\cdot k]\ [(k+l-p)\cdot (k-p)]-[(k+l)\cdot (k-p)]\,[(k+l-p)\cdot k]}{(k+l-p)^2(k-p)^2k^2(k+l)^2}\Bigg]\Bigg\}.\nn
\eea
We have seen that $I_1$ did not have a term proportional to $1/Q_T^2$, which means that the integral over $l$ did not receive a large contribution from the region $l\sim Q_T$ despite the factor $1/l^4$ in the integrand. The reason was that the rest of the integrand vanished at $l=0$. The integral $I_3$ superficially has the same property. However one has to be more careful. Expanding the integrand of $I_1$ in powers of $l$ was justified for $l<Q_s$, since it was equivalent to expansion in $l/k$ and by definition $k>Q_s$. However in $I_3$ this is not the case, since $k-p$ is not bounded from below by $Q_s$, but instead by $\Lambda$. Thus even if $l\sim Q_T$ and $Q_T<Q_s$, we cannot formally expand the integrand of $I_3$ in powers of $l$.
We have to examine the range $l\sim Q_T$ separately.

Let us consider the second and third  lines in Eqn. (\ref{i3}). The first and second terms are equal to each other, since one can change variables $k\rightarrow k+l$ , and this does not affect $\mu^2$ for values of $k$ close to $p$ that dominate the integral. These two integrals in $k$ are logarithmic in the whole range $|k-p|>\Lambda$. On the other hand, the last integral in line three is only logarithmic for $|k-p|>Q_T$, assuming that $l\sim Q_T$. Thus $Q_T$ provides a UV cutoff on the logarithmic integral in the first two terms. Therefore the region  $l\sim Q_T$ does give the leading contribution in this integral. The same is true for the last two lines in Eqn. (\ref{i3}), since the integrals are very similar.
We thus obtain
\beq
I_3\simeq\int_{l}\frac{5\pi\mu^2\lambda^2(l)}{p^2l^4} 2\ln\frac{Q_T^2}{\Lambda^2}=\frac{10 \pi^2\mu^2\lambda^2}{p^2Q_T^2}\ln\frac{Q_T^2}{\Lambda^2}\,.
\eeq

%The first two terms in the second line are similar. They are both logarithmic integrals in the IR, cutoff in the UV at $k\sim p$. %The IR cutoffs are different on different poles. The $k=0$ pole is cutoff by $Q_s^2$, while $k=-l$ and $k=p$ poles are %cutoff at $\Lambda$. The third term is a little different. It is logarithmic in the range $p>k>l$ and also $p>k-p>l$. If we %assume that $l$ is always small $l<Q_s$, then the lower cutoff on the first logarithmic range is in fact at $Q_s$, while the %second logarithmic range is cutoff by the smallest value sf $l$ which is $Q_T$. In total we obtain
%\beq
%I_3=\frac{ 5\mu^2}{2p^2}\int_l\frac{\lambda^2(l)}{l^4}\Big[3\ln\frac{p^2}{\Lambda^2}-\ln\frac{p^2}{Q_s^2}-2\ln %%%\frac{p^2}{Q_T^2}\big] =\frac {5\mu^2\lambda^2}{2p^2Q_T^2}\ln\frac{Q_T^4Q_s^2}{\Lambda^6}\eeq

Finally, the last integral:
\bea\label{i4}
I_4&=&\int_{k,l}\frac{\mu^2(k)\lambda^2(l)}{l^4}|\Psi(k,l,p,1)|^2\\
&=& \int_{k,l}\frac{\mu^2(k)\lambda^2(l)}{l^4}\,2\,\Bigg\{\frac{[(k+l)\cdot(k+l- p)]^2}{(k+l-p)^4(k+l)^4}+\frac{[(k\cdot(k+l-p)]^2}{(k+l-p)^4k^4}
%\nn\\
%&&
-2\frac{[(k+l)\cdot(k+l- p)]\,[k\cdot(k+l- p)]}{(k+l-p)^4(k+l)^2k^2}\nn\\
&+&4\Bigg[\frac{(k+l)^2(k+l-p)^2-[(k+l)\cdot(k+l- p)]^2}{(k+l-p)^4(k+l)^4}+\frac{k^2(k+l-p)^2-[k\cdot(k+l- p)]^2}{(k+l-p)^4k^4}\nn\\
&&-2\frac{[(k+l)\cdot k]\,(k+l-p)^2-[(k+l)\cdot (k+l-p)]\,[ (k+l-p)\cdot k]}{(k+l-p)^4k^2(k+l)^2}\Bigg]\Bigg\}\,.\nn
\eea
In this expression, clearly the pole at $k+l-p=0$   does not give a contribution when $l\sim Q_T$, since in this case $k+l\approx k$, and the three terms in the second and third  lines of Eqn.(\ref{i4}) cancel each other. The contribution will be proportional to $l^2$, which means the result will not have a factor $1/Q_T^2$. It is thus parametrically smaller than $I_3$, and can be neglected,

%With our preset accuracy, this integral is equal to the previous one
\beq
I_4\ll I_3.
\eeq

Thus, for the $A$-contribution we get
\beq
A=-\frac{S}{(2\pi)^2}\frac{50\pi^4g^{12}N_c^5}{16}\frac{\mu^4}{Q_s^4}\frac{\lambda^4}{Q_T^4}\frac{Q_s^2Q_T^2}{p^4}e^{\eta_2-\eta_1}(\eta_1-\eta_2)^2\ln\frac{Q_T^2}{\Lambda^2}\ln\frac{Q_s^4}{Q_T^2\Lambda^2}\delta^{(2)}(q-p).
\eeq
%Incidentally, if we define in the standard way $Q_s^2=g^4\mu^2$, $Q_T^2=g^4\lambda^2$, the result can be written as
%\beq
%A=\frac{25N_c^5}{16 g^4}\frac{Q_s^2Q_T^2}{p^4}e^{\eta_2-\eta_1}(\eta_1-\eta_2)^2\ln\frac{Q_T^2}{\Lambda^2}\ln\frac{Q_s^4}{Q_T^2\Lambda^2}\delta^2qk-p)
%\eeq
%This  has a whooping factor of $N_c$ in front. This has to be understood, especially in comparison to the gluon production.

\subsection{The $B$-term}

Now let us analyse the $B$-term, Eqn. (\ref{B}). This calculation is more cumbersome. We need to analyze all four terms in Eqn. (\ref{B}).
%However all four of them seem to be of the same nature and appear to give the results which differ by a constant. Since the present calculation is only intended to explore the qualitative features of the result, we will limit ourselves to estimating only
\subsubsection{$B_1$}
The first term to be estimated reads
\bea\label{j1}
J_1&=&\int_{k,\bar k,l,\bar l}\frac{\mu^2(k)\mu^2(\bar k)\lambda^2(l)\lambda^2(\bar l)}{l^4\bar l^4}\delta^{(2)}(k+l-p-\bar k-\bar l+q)
%\nn\\
%&\times&
{\rm tr}\{\bar \Psi(k,l,p;0)\bar\Psi^*(k,l,p;1)\bar\Psi(\bar k,\bar l,q;1)\bar \Psi^*(\bar k,\bar l,q;0)\}\nn\\
&=&2\int_{k,\bar k,l,\bar l}\frac{\mu^2(k)\mu^2(\bar k)\lambda^2(l)\lambda^2(\bar l)}{l^4\bar l^4}\delta^{(2)}(k+l-p-\bar k-\bar l+q)\\
&\times&\Bigg\{\Bigg(
\left[\frac{(k+l)\cdot p}{(k+l)^2p^2}-\frac{k\cdot p}{k^2p^2}\right]
\left[\frac{(k+l)\cdot(k+l-p)}{(k+l)^2(k+l-p)^2} -\frac{k\cdot(k-p)}{k^2(k-p)^2} \right]\nn\\
&&-4
\left[\frac{(k+l)\times p}{(k+l)^2p^2}-\frac{k\times p}{k^2p^2}\right]
\left[\frac{(k+l)\times (k+l-p)}{(k+l)^2(k+l-p)^2}-\frac{k\times(k- p)}{k^2(k-p)^2}\right]\Bigg)\nn\\
&&\times\Bigg(
\left[\frac{(\bar k+\bar l)\cdot q}{(\bar k+\bar l)^2q^2}-\frac{\bar k\cdot q}{\bar k^2q^2}\right]
\left[\frac{(\bar k+\bar l)\cdot(\bar k+\bar l-q)}{(\bar k+\bar l)^2(\bar k+\bar l-q)^2} -\frac{\bar k\cdot(\bar k-q)}{\bar k^2(\bar k-q)^2} \right]\nn\\
&&-4
\left[\frac{(\bar k+\bar l)\times q}{(\bar k+\bar l)^2q^2}-\frac{\bar k\times q}{\bar k^2q^2}\right]
\left[\frac{(\bar k+\bar l)\times (\bar k+\bar l-q)}{(\bar k+\bar l)^2(\bar k+\bar l-q)^2}-\frac{\bar k\times(\bar k-q)} {\bar k^2(\bar k-q)^2}\right]\Bigg)\nn\\
&+&4\Bigg(
\left[\frac{(k+l)\cdot p}{(k+l)^2p^2}-\frac{k\cdot p}{k^2p^2}\right]
\left[\frac{(k+l)\times (k+l-p)}{(k+l)^2(k+l-p)^2}-\frac{k\times(k- p)}{k^2(k-p)^2}\right]\nn\\
&&+
\left[\frac{(k+l)\cdot(k+l-p)}{(k+l)^2(k+l-p)^2} -\frac{k\cdot(k-p)}{k^2(k-p)^2} \right]
\left[\frac{(k+l)\times p}{(k+l)^2p^2}-\frac{k\times p}{k^2p^2}\right]\Bigg)\nn\\
&&\times \Bigg(
\left[\frac{(\bar k+\bar l)\cdot q}{(\bar k+\bar l)^2q^2}-\frac{\bar k\cdot q}{\bar k^2q^2}\right]
\left[\frac{(\bar k+\bar l)\times (\bar k+\bar l-q)}{(\bar k+\bar l)^2(\bar k+\bar l-q)^2}-\frac{\bar k\times(\bar k-q)} {\bar k^2(\bar k-q)^2}\right]\nn\\
&&+
\left[\frac{(\bar k+\bar l)\cdot(\bar k+\bar l-q)}{(\bar k+\bar l)^2(\bar k+\bar l-q)^2} -\frac{\bar k\cdot(\bar k-q)}{\bar k^2(\bar k-q)^2} \right]
\left[\frac{(\bar k+\bar l)\times q}{q^2(\bar k+\bar l)^2}-\frac{\bar k\times q}{q^2\bar k^2}\right]\Bigg)\Bigg\}.\nn
\eea

First, one can see that this contains no leading contribution from $l,\bar l\sim Q_T$.
Consider for example the first factor:
\beq
\left[\frac{(k+l)\cdot p}{(k+l)^2p^2}-\frac{k\cdot p}{k^2p^2}\right]
\left[\frac{(k+l)\cdot(k+l-p)}{(k+l)^2(k+l-p)^2} -\frac{k\cdot(k-p)}{k^2(k-p)^2} \right].
\eeq
For $l\sim Q_T$, we can expand in $l/k$. The first factor then is immediately proportional to $l$. To this order in $l/k$ we can also take $k+l=k$ in the first factor of the first term in the brackets. In the remainder of the terms, as long as $k$ is far from the pole at $p$, we can set $k=p$, since the only contribution can come from the pole at $k=p$. The factor then becomes
\bea
&&
\left[\frac{(k+l)\cdot p}{(k+l)^2p^2}-\frac{k\cdot p}{k^2p^2}\right]
 \left[\frac{(k+l)\cdot(k+l-p)}{(k+l)^2(k+l-p)^2} -\frac{k\cdot(k-p)}{k^2(k-p)^2} \right]
% \\
%&&
\approx
-2\,\frac{l\cdot p}{p^4}\left[\frac{p\cdot(k+l-p)}{p^2(k+l-p)^2}-\frac{p\cdot(k-p)}{p^2(k-p)^2}\right].
\eea
The same can be done with the $\bar l, \bar k$ dependent factor
\bea
&&\left[\frac{(\bar k+\bar l)\cdot q}{(\bar k+\bar l)^2q^2}-\frac{\bar k\cdot q}{\bar k^2q^2}\right]
\left[\frac{(\bar k+\bar l)\cdot(\bar k+\bar l-q)}{(\bar k+\bar l)^2(\bar k+\bar l-q)^2} -\frac{\bar k\cdot(\bar k-q)}{\bar k^2(\bar k-q)^2} \right]
%\\
%&&
\approx
-2\frac{\bar l\cdot q}{q^4}\left[\frac{q\cdot (k+l-p)}{(k+l-p)^2q^2}-\frac{q\cdot(k+l-\bar l-p)}{q^2(k+l-\bar l-p)^2}\right]
\eea
where we have used the constraint imposed by the $\delta$-function.
We can shift the integration variable $k\rightarrow k-l$, and the $k$-integral then becomes
\bea
&&
\hspace{-1cm}
4\int_k
\frac{l\cdot p}{p^4}\frac{\bar l\cdot q}{q^4}\left[\frac{p\cdot(k-p)}{p^2(k-p)^2}-\frac{p\cdot(k-l-p)}{p^2(k-l-p)^2}\right]\left[\frac{q\cdot (k-p)}{q^2(k-p)^2}-\frac{q\cdot(k-\bar l-p)}{q^2(k-\bar l-p)^2}\right]
%\\
%&&
\approx 2\pi\,\frac{l\cdot p}{p^4}\frac{\bar l\cdot q}{q^4}\frac{q\cdot p}{p^2q^2}\ln\left(\frac{\min\{l^2,\bar l^2\}}{\Lambda^2}\right)\,.
\eea
In this symmetric form, it is clear that the logarithmic behavior of the integrand at $k\approx p$ is cutoff in the UV by the smallest of $l$ and $\bar l$. However, the subsequent integral over $l$ and $\bar l$ vanishes because, apart from the explicit factor $(l\cdot p)(\bar l\cdot q)$, the rest of the integrand is invariant under independent rotations of $l$ and $\bar l$. This, of course, does not mean that no contribution at all comes from the region $l^2,\bar l^2<Q_s^2$. To obtain such a contribution one needs to expand one order further in $l/k$ and $\bar l/\bar k$, and it therefore can result, at most, in a logarithmic dependence on $Q_T$. Nevertheless, there is still a possibility that $l>Q_s$, but $\bar l \sim Q_T$, which would contribute to order $1/Q_T^2$. In fact, these are exactly the terms that are interesting to us, since they give a contribution comparable to those from the $A$-term.

Now we integrate over $\bar k$ first, and $k$ second. The first integral is trivial - it just realizes the $\delta$-function. After that we are left with integrals that, as before, have poles. The poles for the $k$ integration are:\\
$\bullet$  $P_1$:  $k=0$,\\
$\bullet$ $P_2$:  $k+l=0$,\\
$\bullet$ $P_3$:  $k+l-p=0$,\\
$\bullet$ $P_4$:  $\bar k+\bar l=k+l-p+q=0$,\\
$\bullet$ $P_5$: $\bar k=k+l-\bar l-p+q=0$.\\
Let us be very schematic.

\noindent $\bullet$ {\bf The $k=0$ pole}

\noindent Computing the coefficient of the $k=0$ pole (as usual assuming $k_ik_j\rightarrow \frac{k^2}{2}\delta_{ij}$), we get
\bea\label{p1}
P_1&=&\int_{k,l,\bar l}2\frac{3}{2}\frac{1}{l^4 \bar l ^4}\frac{1}{p^2}\left(\frac{1}{k^2}\right)_{l}
%\nn\\
%&\times&
\Bigg\{\left[\frac{(l-p+q)\cdot q}{(l-p+q)^2q^2}-
\frac{(l-\bar l-p+q)\cdot q}{(l-\bar l-p+q)^2q^2}\right]
\\
&&\times
\left[\frac{(l-p+q)\cdot (l-p)}{(l-p+q)^2(l-p)^2}-
\frac{(l-\bar l-p+q)\cdot(l-\bar l-p)}{(l-\bar l-p+q)^2(l-\bar l-p)^2}\right]\nn\\
&&-4
\left[\frac{(l-p+q)\times q}{(l-p+q)^2q^2}-
\frac{(l-\bar l-p+q)\times q}{(l-\bar l-p+q)^2q^2}\right]
%\nn\\
%&&\times
\left[\frac{(l-p+q)\times (l-p)}{(l-p+q)^2(l-p)^2}-
\frac{(l-\bar l-p+q)\times(l-\bar l-p)}{(l-\bar l-p+q)^2(l-\bar l-p)^2}\right]\Bigg\}\nn\\
&\rightarrow& \int_{k,l,\bar l}2\frac{9}{4}\frac{1}{p^2q^2}\frac{1}{l^4 \bar l ^4}\left(\frac{1}{k^2}\right)_{l}
\left\{\left[\frac{1}{(l-p+q)^2}\right]_{\bar l}+\left[\frac{1}{(l-\bar l-p+q)^2}\right]_{\bar l}\right\}\nn.
\eea
Here, the subscript denotes the scale of the integrand at which the logarithmic integral is cutoff in the UV.

The $k$ integral yields
\beq
\label{k_int}
\left(\frac{1}{k^2}\right)_{l}\rightarrow \pi \ln\left(\frac{l^2}{Q_s^2}\right)\,.
\eeq
The integral over $l$ now picks the two poles  in the parenthesis in Eqn. (\ref{p1}). The result reads
\bea &&\int_l\frac{1}{l^4}\ln\left(\frac{l^2}{Q_s^2}\right)\left\{\left[\frac{1}{(l-p+q)^2}\right]_{\bar l}+\left[\frac{1}{(l-\bar l-p+q)^2}\right]_{\bar l}\right\}\\
&\approx& \frac{2\pi}{(p-q)^4}\ln\left[\frac{(p-q)^2}{Q_s^2}\right]\ln\left(\frac{\bar l^2}{\Lambda^2}\right)\,.\nonumber
\eea
The last integral over $\bar l$ yields
\beq
\label{lbar_int}
\int_{\bar l}\frac{1}{\bar l^4}\ln\frac{\bar l^2}{\Lambda^2}\approx\frac{\pi}{Q_T^2}\ln\left(\frac{Q_T^2}{\Lambda^2}\right)\,.
\eeq
There is an additional contribution to the $l$ integral, coming from $l\sim Q_s$. However, this contribution is of order $\bar l^2$ as is obvious from the first line in Eqn. (\ref{p1}), and  therefore is not going to yield any $1/Q_T^2$ term. We will ignore similar contributions in the following.

Finally,
\beq
P_1\approx 2\frac{9\pi^3}{2}\frac{1}{p^2q^2}\frac{1}{(p-q)^4} \ln\left[\frac{(p-q)^2}{Q_s^2}\right]   \frac{1}{Q_T^2}  \ln\left(\frac{Q_T^2}{\Lambda^2}\right)\,.
\eeq
Note that we get no contribution of order $1/(Q_s^2Q_T^2)$, but only $1/Q_T^2$. On the other hand, for $p=q$ our calculation yields a strong peak. We have assumed here that $|p-q|\sim |p|\sim |q|$, and thus the exact form of the contribution at $|q-p|\sim Q_s$ is beyond the present accuracy.

\noindent $\bullet$ {\bf The $k+l=0$ pole}

\noindent The corresponding coefficient reads
\bea\label{p2}
P_2&=&\int_{k,l,\bar l} 2\frac{3}{2}\frac{1}{p^2}\frac{1}{l^4 \bar l ^4}\left(\frac{1}{(k+l)^2}\right)_{l}\nn\\
&\times&\Bigg\{\Big[\frac{(q-p)\cdot q}{(q-p)^2q^2}-
\frac{(q-p-\bar l)\cdot q}{(q-p-\bar l)^2q^2}\Big]
\Big[\frac{(q-p)\cdot (-p)}{(q-p)^2p^2}-
\frac{(q-p-\bar l)\cdot(-p-\bar l)}{(q-p-\bar l)^2(p+\bar l)^2}\Big]\nn\\
&-&4
\Big[\frac{(q-p)\times q}{(q-p)^2q^2}-
\frac{(q-p-\bar l)\times q}{(q-p-\bar l)^2q^2}\Big]
\Big[\frac{(q-p)\times (-p)}{(q-p)^2p^2}-
\frac{(q-p-\bar l)\times(-p-\bar l)}{(q-p-\bar l)^2(p+\bar l)^2}\Big]\Bigg\}
\nn\\
&\rightarrow&\int_{k,l,\bar l} 2\frac{9}{4}\frac{1}{p^2q^2}\frac{1}{l^4 \bar l ^4}\left(\frac{1}{(k+l)^2}\right)_{l}\left(\frac{1}{(\bar l+p-q)^2}\right)_{Q_s, p-q}\,,
\eea
where the lower limit in the second integral is $Q_s$, since the pole is in $\bar k$, which is limited by $\mu^2(\bar k)$.
Here the $\bar l$ integral is pinned to the pole and not to $\bar l=0$, however the $l$ integral is free to wander all the way down to $Q_T$. Thus we get for $P_2$ the result up to a factor of $1/2$ identical to $P_1$,
\beq
P_2=\frac{1}{2}P_1\,.
\eeq

\noindent $\bullet$ {\bf The $k+l-p=0$ pole}

\noindent This pole is a little different, since the contribution comes from different terms. Recall that this also corresponds to $\bar k+\bar l-q=0$. It reads
\bea
P_3&=&2 \int_{k,l,\bar l} \frac{1}{l^4 \bar l ^4}\left[\left(\frac{1}{p^2}-\frac{(p-l)\cdot p}{p^2(p-l)^2}\right)\frac{p\cdot (k+l-p)}{p^2(k+l-p)^2}-4\frac{l\times p}{p^2(p-l)^2}\frac{p\times(k+l-p)}{p^2(k+l-p)^2}\right]\nn\\
&\times&\left[\left(\frac{1}{q^2}-\frac{(q-\bar l)\cdot q}{q^2(q-\bar l)^2}\right)\frac{q\cdot(k+l-p)}{q^2(k+l-p)^2}-4\frac{\bar l\times q}{q^2(\bar l-q)^2}\frac{q\times(k+l-p)}{q^2(k+l-p)^2}\right]\nn\\
&+&4\left[\left(\frac{1}{p^2}-\frac{(p-l)\cdot p}{p^2(p-l)^2}\right)\frac{p\times(k+l-p)}{p^2(k+l-p)^2}+\frac{l\times p}{p^2(p-l)^2}\frac{p\cdot (k+l-p)}{p^2(k+l-p)^2}\right]\nn\\
&\times&\left[\left(\frac{1}{q^2}-\frac{(q-\bar l)\cdot q}{q^2(q-\bar l)^2}\right)\frac{q\times(k+l-p)}{q^2(k+l-p)^2}+
\frac{\bar l\times q}{q^2(\bar l-q)^2}\frac{q\cdot(k+l-p)}{q^2(k+l-p)^2}\right]\\
&\approx&2 \int_{k,l,\bar l} \frac{1}{l^4 \bar l ^4} \left(\frac{1}{(k+l-p)^2}\right)_{p}
%\nn\\
%&\times&
\left[\frac{5}{2}\frac{p\cdot q}{p^4q^4}\left\{\Big[1-\frac{(p-l)\cdot p}{(p-l)^2}\Big]\Big[1-\frac{(q-\bar l)\cdot q}{(q-\bar l)^2}\Big]+4\frac{l\times p}{(p-l)^2}\frac{\bar l\times q}{(q-\bar l)^2}\right\}\right].\nn
\eea
This expression has the following redeeming feature:
%features: \textcolor{red}{It has only one logarithmic factor, as opposed to the product of two.} But, more importantly, it
It is clear that it does not bring any factors of the form $1/Q_T^2$ or even $1/Q_s^2$, since, for any $l<p,\ \bar l<q$, the integrand is proportional to $l^2\bar l^2$. Thus, this contribution can be neglected relative to $P_1$ and $P_2$,
\beq
P_3\ll P_1,P_2\,.
\eeq

\noindent $\bullet$ {\bf The $\bar k=0$ and $\bar k+\bar l=0$ poles}

\noindent The contribution of these poles is, by symmetry, identical to $P_1$ and $P_2$ respectively.

Thus, our result for $J_1$ is
%\beq
%J_1=3\mu^4\lambda^4P_1= 27\pi^3 \frac{\mu^4\lambda^4}{p^2q^2(p-q)^4}   \frac{1}{Q_T^2} \ln\frac{(p-q)^2}{Q_s^2}\ln\frac{Q_T^2}{\Lambda^2}\,.
%\eeq
\bea
J_1=\pi^3\frac{9}{2}\frac{6}{p^2q^2}\frac{1}{(p-q)^4}\ln\left[\frac{(p-q)^2}{Q_s^2}\right]\frac{1}{Q_T^2}\ln\left(\frac{Q_T^2}{\Lambda^2}\right)\mu^4\lambda^4.
\eea

\subsubsection{$B_2$}

The second term in the $B$-type contribution reads
\bea
J_2&=&\int_{ k,\bar{k} ,l,\bar{l}}\frac{\mu^2(k)\mu^2(\bar{k})\lambda^2(l)\lambda^2(\bar{l})}{l^4\bar{l}^4}\delta^{(2)}(k+l-p-\bar{k}-\bar{l}+q)\nonumber\\
&&\times
 {\rm tr}\left\{ \bar{\Psi}(k,l,p;0) \bar{\Psi}^*(k,l,p;1) \Psi(\bar k, \bar l, q; 1)\Psi^*(\bar k, \bar l, q; 0)
\right\} \nonumber\\
&=& 2 \int_{ k,\bar{k}, l,\bar{l}}\frac{\mu^2(k)\mu^2(\bar{k})\lambda^2(l)\lambda^2(\bar{l})}{l^4\bar{l}^4}\delta^{(2)}(k+l-p-\bar{k}-\bar{l}+q)\\
&\times&\Bigg\{\Bigg(
\left[ \frac{(k+l)\cdot p}{(k+l)^2p^2}-\frac{k\cdot p}{k^2p^2} \right]
\left[ \frac{(k+l)\cdot (k+l-p)}{(k+l)^2(k+l-p)^2} - \frac{k\cdot(k-p)}{k^2(k-p)^2} \right]\nonumber\\
&&\hspace{1cm}-4
\left[ \frac{(k+l) \times p}{(k+l)^2p^2}-\frac{k \times p}{k^2p^2} \right]
\left[ \frac{(k+l) \times (k+l-p)}{(k+l)^2(k+l-p)^2} - \frac{k \times (k-p)}{k^2(k-p)^2} \right] \Bigg)\nonumber\\
&&\times
\Bigg(
\left[ \frac{(\bar k+\bar l)\cdot q}{(\bar k+\bar l)^2q^2}-\frac{\bar k\cdot (q-\bar l)}{\bar k^2(q-\bar l)^2} \right]
\left[ \frac{(\bar k+\bar l)\cdot (\bar k+\bar l-q)}{(\bar k+\bar l)^2(\bar k+\bar l-q)^2} - \frac{\bar k\cdot(\bar k+\bar l-q)}{\bar k^2(\bar k+\bar l-q)^2} \right]\nonumber\\
&&\hspace{1cm}-4
\left[ \frac{(\bar k+\bar l)\times q}{(\bar k+\bar l)^2q^2}-\frac{\bar k\times (q-\bar l)}{k^2(q-\bar l)^2} \right]
\left[ \frac{(\bar k+\bar l)\times (\bar k+\bar l-q)}{(\bar k+\bar l)^2(\bar k+\bar l-q)^2} - \frac{\bar k\times(\bar k+\bar l-q)}{\bar k^2(\bar k+\bar l-q)^2} \right] \Bigg)\nonumber\\
&&+ 4 \Bigg(
\left[ \frac{(k+l)\cdot p}{(k+l)^2p^2}-\frac{k\cdot p}{k^2p^2} \right]
\left[ \frac{(k+l)\times (k+l-p)}{(k+l)^2(k+l-p)^2} - \frac{k\times (k-p)}{k^2{k-p}^2} \right]\nonumber\\
&&\hspace{1cm}+
\left[ \frac{(k+l)\cdot (k+l-p)}{(k+l)^2(k+l-p)^2} - \frac{k\cdot(k-p)}{k^2{k-p}^2} \right]
\left[ \frac{(k+l) \times p}{(k+l)^2p^2}-\frac{k \times p}{k^2p^2} \right] \Bigg) \nonumber\\
&&\times
\Bigg(
\left[ \frac{(\bar k+\bar l)\cdot q}{(\bar k+\bar l)^2q^2}-\frac{\bar k\cdot (q-\bar l)}{\bar k^2(q-\bar l)^2} \right]
\left[ \frac{(\bar k+\bar l)\times (\bar k+\bar l-q)}{(\bar k+\bar l)^2(\bar k+\bar l-q)^2} - \frac{\bar k\times(\bar k+\bar l-q)}{\bar k^2(\bar k+\bar l-q)^2} \right]\nonumber\\
&&\hspace{1cm}+
\left[ \frac{(\bar k+\bar l)\cdot (\bar k+\bar l-q)}{(\bar k+\bar l)^2(\bar k+\bar l-q)^2} - \frac{\bar k\cdot(\bar k+\bar l-q)}{\bar k^2(\bar k+\bar l-q)^2} \right]
\left[ \frac{(\bar k+\bar l)\times q}{(\bar k+\bar l)^2q^2}-\frac{\bar k\times (q-\bar l)}{\bar k^2(q-\bar l)^2} \right]
\Bigg)\Bigg\}.\nonumber
\eea

We first integrate over $\bar k$ and then over $k$. The first integral is trivial to perform by using the $\delta$-function. In the second integral, the leading contribution comes from four different poles: $P_1: k=0$, $P_2: k+l=0$, $P_3: \bar k+\bar l=0$ and $P_4: \bar k=0$.

The contribution arising from the first pole reads
\bea
P_1&=&2\int_{k,l,\bar l}\frac{3}{2}\frac{1}{l^4{\bar l}^4} \frac{1}{p^2}  \left( \frac{1}{k^2}\right)_l
\Bigg\{ \left[ \frac{(l-p+q)\cdot q}{(l-p+q)^2q^2}-\frac{(l-\bar l-p+q)\cdot(q-\bar l)}{(l-\bar l-p+q)^2(q-\bar l)^2}\right] \\
&&
\hspace{3.8cm}
\times
\left[ \frac{(l-p+q)\cdot(l-p)}{(l-p+q)^2(l-p)^2}-\frac{(l-\bar l-p+q)\cdot(l-p)}{(l-\bar l-p+q)^2(l-p)^2}\right]\nonumber\\
&&
\hspace{3.4cm}
-4\left[ \frac{(l-p+q)\times q}{(l-p+q)^2q^2}-\frac{(l-\bar l-p+q)\times(q-\bar l)}{(l-\bar l-p+q)^2(q-\bar l)^2}\right]\nonumber\\
&&
\hspace{3.8cm}
\times
\left[ \frac{(l-p+q)\times(l-p)}{(l-p+q)^2(l-p)^2}-\frac{(l-\bar l-p+q)\times(l-p)}{(l-\bar l-p+q)^2(l-p)^2}\right]\Bigg\}.\nonumber
\eea
The integration over $k$ is given by Eqn. (\ref{k_int}). On the other hand, the integration over $l$ picks up two poles:
\beq
P_1= 2\pi\int_{l,\bar l}\frac{9}{4}\frac{1}{p^2}\frac{1}{l^4{\bar l}^4}\ln\left( \frac{l^2}{Q_s^2}\right)
\Bigg\{ \frac{1}{q^2}\left[\frac{1}{(l-p+q)^2}\right]_{\bar l}+\frac{1}{(q-\bar l)^2}\left[\frac{1}{(l-\bar l-p+q)^2}\right]_{\bar l}\Bigg\}.
\eeq
Finally, the integration over $l$ gives
\bea
P_1&\approx& 2{\pi}^2\int_{\bar l}\frac{9}{4}\frac{1}{p^2q^2}\frac{1}{(p-q)^4}\ln \left[ \frac{(p-q)^2}{Q_s^2}\right]\frac{1}{{\bar l}^4}\ln\left( \frac{{\bar l}^2}{\Lambda^2}\right)
%\nonumber\\
%&&
+\,2{\pi}^2\int_{\bar l}\frac{9}{4}\frac{1}{p^2}\frac{1}{(p-q)^4}\ln \left[ \frac{(p-q)^2}{Q_s^2}\right]\frac{1}{{\bar l}^4} \frac{1}{(\bar l-q)^2}\ln\left( \frac{{\bar l}^2}{\Lambda^2}\right).
\eea
The integration over $\bar l$ for the first term is exactly the same as Eqn. (\ref{lbar_int}). However, in the second term, the integration over $\bar l$ picks up the pole at $\bar l=q$ and gets an extra factor  $q^2$ instead of $Q_T^2$ in the denominator. Thus, it is suppressed with respect to the first term and can be neglected at the accuracy that we perform the calculation. Then, the $P_1$ contribution to the $B_2$-type terms reads
\beq
P_1\approx 2{\pi}^3\frac{9}{4}\frac{1}{p^2q^2}\frac{1}{(p-q)^4}\ln \left[ \frac{(p-q)^2}{Q_s^2}\right] \frac{1}{Q_T^2} \ln\left( \frac{Q_T^2}{\Lambda^2}\right).
\eeq

The contribution from the pole at $k+l=0$ to $B_2$-type terms is very similar to the contribution of the same pole to the $B_1$-type terms and it reads
\bea
P_2&=&2\int_{k,l,\bar l}\frac{3}{2}\frac{1}{l^4{\bar l}^4}\frac{1}{p^2}\left[\frac{1}{(k+l)^2}\right]_l 
%\\
%&&\times
\Bigg\{ \left[ \frac{(q-p)\cdot q}{(q-p)^2 q^2}-\frac{(q-p-\bar l)\cdot (q-\bar l)}{(q-p-\bar l)^2 (q-\bar l)^2}\right]
%\nonumber\\
%&&\times
\left[ \frac{(q-p)\cdot (-p)}{(q-p)^2p^2}- \frac{(q-p-\bar l)\cdot (-p)}{(q-p-\bar l)^2 p^2}\right]\nonumber\\
&&-4
\left[ \frac{(q-p)\times q}{(q-p)^2 q^2}-\frac{(q-p-\bar l)\times (q-\bar l)}{(q-p-\bar l)^2 (q-\bar l)^2}\right]
%\nonumber\\
%&&\times
\left[ \frac{(q-p)\times (-p)}{(q-p)^2p^2}- \frac{(q-p-\bar l)\times (-p)}{(q-p-\bar l)^2 p^2}\right]\Bigg\}.
\eea
Integration over $\bar l$ picks up a pole at $(\bar l+p-q)=0$ and one gets
\bea
P_2&=&2\int_{k,l,\bar l}\frac{9}{4}\frac{1}{p^4}\frac{1}{l^4{\bar l}^4}\left[ \frac{1}{(k+l)^2} \right]_l \left[ \frac{1}{(\bar l+p-q)^2}\right]_{Q_s, p-q}\nonumber\\
&=& 2{\pi}^3\frac{9}{4}\frac{1}{p^4}\frac{1}{(p-q)^4}\ln\left[ \frac{(p-q)^2}{Q_s^2}\right] \frac{1}{Q_T^2}\ln \left( \frac{Q_T^2}{\Lambda^2}\right).
\eea
For the $B_2$-type terms the pole $P_2: k+l=0$ and $P_3: \bar k+\bar l=0$ are symmetric under the exchange $p\leftrightarrow q$. Thus, we can immediately write the $P_3$ contribution to  these terms as
\beq
P_3=2{\pi}^3\frac{9}{4}\frac{1}{q^4}\frac{1}{(p-q)^4}\ln\left[ \frac{(p-q)^2}{Q_s^2}\right] \frac{1}{Q_T^2}\ln \left( \frac{Q_T^2}{\Lambda^2}\right).
\eeq
The last pole that contributes to the $B_2$-type terms is $P_4: \bar k=0$, and it reads
\bea
P_4&=&2\int_{\bar k, l,\bar l} \frac{3}{2}\frac{1}{l^4{\bar l}^4}\frac{1}{(q-\bar l)^2}\left(\frac{1}{{\bar k}^2} \right)_{\bar l}
\Bigg\{ \left[ \frac{(\bar l-q+p)\cdot p}{(\bar l-q+p)^2p^2} - \frac{(\bar l-q-l+p)\cdot p}{(\bar l -q-l+p)^2p^2}\right]\nonumber\\
&&
\times
\left[ \frac{(\bar l-q+p)\cdot (\bar l-q)}{(\bar l-q+p)^2(\bar l-q)^2} - \frac{(\bar l-q-l+p)\cdot (\bar l-q-l)}{(\bar l-q-l+p)^2(\bar l-q-l)^2}\right]\nonumber\\
&&
-4
\left[ \frac{(\bar l-q+p)\times p}{(\bar l-q+p)^2p^2} - \frac{(\bar l-q-l+p)\times p}{(\bar l -q-l+p)^2p^2}\right]\nonumber\\
&&
\times
\left[ \frac{(\bar l-q+p)\times (\bar l-q)}{(\bar l-q+p)^2(\bar l-q)^2} - \frac{(\bar l-q-l+p)\times(\bar l-q-l)}{(\bar l-q-l+p)^2(\bar l-q-l)^2}\right]\Bigg\}.
\eea
After integrating over $\bar k$ and renaming $l\leftrightarrow \bar l$, one realizes that the integration over $l$ picks up three poles:
\bea
\label{P_4_lbar}
P_4&=&2\pi\int_{l,\bar l} \frac{3}{2}\frac{1}{l^4{\bar l}^4}\ln\left( \frac{l^2}{Q_s^2}\right)
\Bigg\{ \frac{3}{2}\frac{1}{p^4}\left[\frac{1}{(l+p-q)^2}\right]_{\bar l}+\frac{3}{2}\frac{1}{p^2}\frac{1}{(\bar l-p)^2}\left[ \frac{1}{(l-\bar l-q+p)^2}\right]_{\bar l}
\\
&+&
\left[\frac{1}{(l-q)^2}\right]_{\bar l}
\left[ \left( \frac{1}{p^2}-\frac{(p-\bar l)\cdot p}{(p-\bar l)^2p^2}\right) \left( \frac{1}{p^2}+\frac{(p-\bar l)\cdot \bar l}{(p-\bar l)^2{\bar l}^2}\right)
-4\frac{(p-\bar l)\times p}{(p-\bar l)^2p^2}\frac{(p-\bar l)\times (-\bar l)}{(p-\bar l)^2{\bar l}^2}\right]\Bigg\}.
\nonumber
\eea
Note that after integrating over $l$ and $\bar l$, the leading contribution will come from the terms where there are no extra poles in the integration over $\bar l$. Thus, the leading contribution of $P_4$ comes from the pole at $l+p-q=0$ and the pole at $l-q=0$,
\bea
P_4\approx2{\pi}^2\int_{\bar l}\frac{9}{4}\frac{1}{p^4}
\left\{ \frac{1}{(p-q)^4}\ln\left[ \frac{(p-q)^2}{Q_s^2}\right]
+\frac{1}{q^4}\ln \left( \frac{q^2}{Q_s^2}\right)\right\}
\frac{1}{{\bar l}^4}\ln\left( \frac{{\bar l}^2}{\Lambda^2}\right).
\eea
Finally, the integration over $\bar l$ is straightforward to perform and the $P_4$ contribution to the $B$-type terms reads
\beq
P_4\approx2{\pi}^3\frac{9}{4}\frac{1}{p^4}\left\{ \frac{1}{(p-q)^4}\ln\left[\frac{(p-q)^2}{Q_s^2}\right]+\frac{1}{q^4}\ln\left(\frac{q^2}{Q_s^2}\right)\right\}\frac{1}{Q_T^2}\ln\left( \frac{Q_T^2}{\Lambda^2}\right).
\eeq
Adding all contributions, we get
%\bea
%J_2&\approx&\pi^3\frac{9}{2}\left[\frac{1}{q^2p^2}+\frac{2}{p^4}+\frac{1}{q^4}\right]\frac{1}{(p-q)^4}\ln\left[\frac{(p-q)^2}{Q_s^2}\right]\frac{1}{Q_T^2}\ln\left(\frac{Q_T^2}{\Lambda^2}\right)\mu^4\lambda^4\nonumber\\
%&&+
%\pi^33\frac{1}{p^4q^4}\ln\left(\frac{q^2}{Q_s^2}\right)\frac{1}{Q_T^2}\ln\left(\frac{Q_T^2}{\Lambda^2}\right)\mu^4\lambda^4
%\eea
%
\bea
J_2&\approx&\pi^3\frac{9}{2}
\Bigg\{
\left[\frac{1}{q^2p^2}+\frac{2}{p^4}+\frac{1}{q^4}\right]\frac{1}{(p-q)^4}\ln\left[\frac{(p-q)^2}{Q_s^2}\right]
+\frac{1}{p^4q^4}\ln\left(\frac{q^2}{Q_s^2}\right)
\Bigg\}
%\nonumber \\
%&&\times \, 
\frac{1}{Q_T^2}\ln\left(\frac{Q_T^2}{\Lambda^2}\right)\mu^4\lambda^4.
\eea

\subsubsection{$B_3$}
The third term in the $B$-type contribution reads
\bea
J_3&=&\int_{ k,\bar{k} ,l,\bar{l}}\frac{\mu^2(k)\mu^2(\bar{k})\lambda^2(l)\lambda^2(\bar{l})}{l^4\bar{l}^4}\delta^{(2)}(k+l-p-\bar{k}-\bar{l}+q)\nonumber\\
&&\times
 {\rm tr}\left\{ \Psi(k,l,p;0) \Psi^*(k,l,p;1) \bar \Psi(\bar k, \bar l, q; 1) \bar \Psi^*(\bar k, \bar l, q; 0)
\right\} \nonumber\\
&=& 2 \int_{ k,\bar{k}, l,\bar{l}}\frac{\mu^2(k)\mu^2(\bar{k})\lambda^2(l)\lambda^2(\bar{l})}{l^4\bar{l}^4}\delta^{(2)}(k+l-p-\bar{k}-\bar{l}+q)\\
&\times&\Bigg\{\Bigg(
\left[ \frac{(k+l)\cdot p}{(k+l)^2p^2}-\frac{k\cdot (p-l)}{k^2(p-l)^2} \right]
\left[ \frac{(k+l)\cdot (k+l-p)}{(k+l)^2(k+l-p)^2} - \frac{k\cdot(k+l-p)}{k^2(k+l-p)^2} \right]\nonumber\\
&&\hspace{1cm}-4
\left[ \frac{(k+l) \times p}{(k+l)^2p^2}-\frac{k \times (p-l)}{k^2(p-l)^2} \right]
\left[ \frac{(k+l) \times (k+l-p)}{(k+l)^2(k+l-p)^2} - \frac{k \times (k+l-p)}{k^2(k+l-p)^2} \right] \Bigg)\nonumber\\
&&\times
\Bigg(
\left[ \frac{(\bar k+\bar l)\cdot q}{(\bar k+\bar l)^2q^2}-\frac{\bar k\cdot q}{\bar k^2q^2} \right]
\left[ \frac{(\bar k+\bar l)\cdot (\bar k+\bar l-q)}{(\bar k+\bar l)^2(\bar k+\bar l-q)^2} - \frac{\bar k\cdot(\bar k-q)}{\bar k^2(\bar k -q)^2} \right]\nonumber\\
&&\hspace{1cm}-4
\left[ \frac{(\bar k+\bar l)\times q}{(\bar k+\bar l)^2q^2}-\frac{\bar k\times q}{\bar k^2q^2} \right]
\left[ \frac{(\bar k+\bar l)\times (\bar k+\bar l-q)}{(\bar k+\bar l)^2(\bar k+\bar l-q)^2} - \frac{\bar k\times(\bar k-q)}{\bar k^2(\bar k-q)^2} \right] \Bigg)\nonumber\\
&&+ 4 \Bigg(
\left[ \frac{(k+l)\cdot p}{(k+l)^2p^2}-\frac{k\cdot (p-l)}{k^2(p-l)^2} \right]
\left[ \frac{(k+l)\times (k+l-p)}{(k+l)^2(k+l-p)^2} - \frac{k\times (k+l-p)}{k^2(k+l-p)^2} \right]\nonumber\\
&&\hspace{1cm}+
\left[ \frac{(k+l)\cdot (k+l-p)}{(k+l)^2(k+l-p)^2} - \frac{k\cdot(k+l-p)}{k^2(k+l-p)^2} \right]
\left[ \frac{(k+l) \times p}{(k+l)^2p^2}-\frac{k \times (p-l)}{k^2(p-l)^2} \right] \Bigg) \nonumber\\
&&\times
\Bigg(
\left[ \frac{(\bar k+\bar l)\cdot q}{(\bar k+\bar l)^2q^2}-\frac{\bar k\cdot q}{k^2q^2} \right]
\left[ \frac{(\bar k+\bar l)\times (\bar k+\bar l-q)}{(\bar k+\bar l)^2(\bar k+\bar l-q)^2} - \frac{\bar k\times(\bar k-q)}{\bar k^2(\bar k-q)^2} \right]\nonumber\\
&&\hspace{1cm}+
\left[ \frac{(\bar k+\bar l)\cdot (\bar k+\bar l-q)}{(\bar k+\bar l)^2(\bar k+\bar l-q)^2} - \frac{\bar k\cdot(\bar k-q)}{\bar k^2(\bar k-q)^2} \right]
\left[ \frac{(\bar k+\bar l)\times q}{(\bar k+\bar l)^2q^2}-\frac{\bar k\times q}{k^2q^2} \right]
\Bigg)\Bigg\}.\nonumber
\eea
As in the case of $B_1$-type and $B_2$-type terms, we also integrate over $\bar k$ by using the $\delta$-function to calculate the $B_3$-type terms. Then, the integration over $k$ picks up four poles: $P_1: k=0$, $P_2: k+l=0$, $P_3: \bar k +\bar l=0$ and $P_4: \bar k=0$.

The contribution from $P_1$ reads
\bea
P_1&=& 2\int_{k,l,\bar l}\frac{3}{2}\frac{1}{l^4{\bar l}^4}  \frac{1}{(p-l)^2}   \left(\frac{1}{k^2}\right)_l
\Bigg\{ \left[\frac{(l-p+q)\cdot q}{(l-p+q)^2q^2}-\frac{(l-p-\bar l +q)\cdot q}{(l-p-\bar l+q)^2q^2}\right]\\
&&\times
\left[ \frac{(l-p+q)\cdot(l-p)}{(l-p+q)^2(l-p^2)}-\frac{(l-p-\bar l+q)\cdot (l-p-\bar l)}{(l-p-\bar l+q)^2(l-p-\bar l)^2}\right]\nonumber\\
&&-4
\left[\frac{(l-p+q)\times q}{(l-p+q)^2q^2}-\frac{(l-p-\bar l +q)\times q}{(l-p-\bar l+q)^2q^2}\right]
%\nonumber\\
%&&\times
\left[ \frac{(l-p+q)\times(l-p)}{(l-p+q)^2(l-p^2)}-\frac{(l-p-\bar l+q)\times (l-p-\bar l)}{(l-p-\bar l+q)^2(l-p-\bar l)^2}\right]\Bigg\}.
\nonumber
\eea
The integration over $k$ is straight forward to perform. On the other hand, integration over $l$ picks up three poles:
\bea
P_1&=&2\pi\int_{l,\bar l} \frac{3}{2} \frac{1}{l^4{\bar l}^4} \ln \left( \frac{l^2}{Q_s^2}\right)
\Bigg\{ \frac{3}{2} \frac{1}{q^4} \left[ \frac{1}{(l-p+q)^2}\right]_{\bar l} + \frac{3}{2}\frac{1}{q^2}\frac{1}{(\bar l - q)^2}\left[ \frac{1}{(l-p-\bar l+q)^2}\right]_{\bar l}\\
&&+\left[\frac{1}{(p-l)^2}\right]_{\bar l}
\left[ \left( \frac{1}{q^2}-\frac{(q-\bar l)\cdot q}{(q-\bar l)^2 q^2}\right)\left( \frac{1}{q^2}+\frac{(q-\bar l)\cdot \bar l}{(q-\bar l)^2{\bar l}^2}\right)
-4\frac{(q-\bar l)\times q}{(q-\bar l)^2 q^2}\frac{{q-\bar l}\times(-\bar l)}{(q-\bar l)^2{\bar l}^2}\right]\Bigg\}.\nonumber
\eea
Note that, as in the case of $B_2$-type terms, the leading contribution will come from the terms with no extra poles for the $\bar l$ integration. Thus, after performing the $l$ integration, the $P_1$ contribution reads
\bea
P_1&\approx& 2{\pi}^2\int_{\bar l} \frac{9}{4} \frac{1}{q^4} \left\{ \frac{1}{(p-q)^4}\ln\left[\frac{(p-q)^2}{Q_s^2}\right] +\frac{1}{p^4}\ln\left( \frac{p^2}{Q_s^2}\right)\right\} \frac{1}{{\bar l}^4}\ln\left( \frac{{\bar l}^2}{\Lambda^2}\right).
\eea
Finally, the integration over $\bar l$ is straightforward to perform and the result gives
\bea
P_1&\approx& 2{\pi}^3 \frac{9}{4} \frac{1}{q^4} \left\{ \frac{1}{(p-q)^4}\ln\left[\frac{(p-q)^2}{Q_s^2}\right] +\frac{1}{p^4}\ln\left( \frac{p^2}{Q_s^2}\right)\right\} \frac{1}{Q_T^2}\ln\left( \frac{{Q_T}^2}{\Lambda^2}\right).
\eea
Now, let us calculate the contribution from the pole at $P_2: k+l=0$ to the $B_3$-type terms,
\bea
P_2&=&2\int_{k,l,\bar l}\frac{3}{2}\frac{1}{l^4\bar l^4}\frac{1}{p^2} \left[ \frac{1}{(k+l)^2} \right]_l\\
&&\times
\Bigg\{ \left[ \frac{(q-p)\cdot q}{(q-p)^2 q^2}-\frac{(q-p-\bar l)\cdot q}{(q-p-\bar l)^2q^2}\right]
%\\
%&&\times
\left[ \frac{(q-p)\cdot (-p)}{(q-p)^2p^2}-\frac{(q-p-\bar l)\cdot (-p-\bar l)}{(q-p-\bar l)^2(p+\bar l)^2}\right] \nonumber\\
&&-4
\left[ \frac{(q-p)\times q}{(q-p)^2 q^2}-\frac{(q-p-\bar l)\times q}{(q-p-\bar l)^2q^2}\right]
\left[ \frac{(q-p)\times (-p)}{(q-p)^2p^2}-\frac{(q-p-\bar l)\times (-p-\bar l)}{(q-p-\bar l)^2(p+\bar l)^2}\right] \Bigg\}.\nonumber
\eea
The integration over $\bar l$ picks up one pole:
\bea
P_2&\approx&2\int_{k,l,\bar l}\frac{9}{4}\frac{1}{l^4\bar l^4}\frac{1}{p^2q^2} \left[ \frac{1}{(k+l)^2} \right]_l \left[\frac{1}{(\bar l+p-q)^2}\right]_{Q_s, p-q}.
\eea
After performing all the integrals we get
\bea
\label{B3P2}
P_2\approx2{\pi}^3\frac{9}{4}\frac{1}{p^2q^2}\frac{1}{(p-q)^4}\ln\left[ \frac{(p-q)^2}{Q_s^2}\right]\frac{1}{Q_T^2}\ln \left( \frac{Q_T^2}{\Lambda^2} \right).
\eea

The contribution from the pole $P_3: \bar k+\bar l=0$, can be obtained directly from the result of $P_2$ with the exchange of $p\leftrightarrow q$ due to  symmetry.  However, Eqn. (\ref{B3P2}) is symmetric under the exchange of $p$ and $q$. Thus, the contribution from the pole $P_3$ is equal to the contribution from the pole $P_2$.

The last contribution to the $B_3$-type terms is coming from the pole $P_4: \bar k=0$ and it reads
\bea
P_4&=&2\int_{\bar k ,l ,\bar l} \frac{3}{2} \frac{1}{l^4\bar l^4}\frac{1}{q^2}\left(\frac{1}{\bar k^2}\right)_{\bar l}
\Bigg\{ \left[ \frac{(\bar l-q+p)\cdot p}{(\bar l-q+p)^2p^2}-\frac{(\bar l-l-q+p)\cdot(p-l)}{(\bar l-l-q+p)^2(p-l)^2} \right]\\
&&\times \left[ \frac{(\bar l-q+p)\cdot (\bar l-q)}{(\bar l-q+p)^2(\bar l- q)^2}-\frac{(\bar l-l-q+p)\cdot (\bar l-q)}{(\bar l-l -q+p)^2(\bar l-q)^2}\right]\nonumber\\
&&-4 \left[ \frac{(\bar l-q+p)\times p}{(\bar l-q+p)^2p^2}-\frac{(\bar l-l-q+p)\times(p-l)}{(\bar l-l-q+p)^2(p-l)^2} \right]
%\nonumber \\
%&&\times 
\left[ \frac{(\bar l-q+p)\times (\bar l-q)}{(\bar l-q+p)^2(\bar l- q)^2}-\frac{(\bar l-l-q+p)\times (\bar l-q)}{(\bar l-l -q+p)^2(\bar l-q)^2}\right]\Bigg\}.\nonumber
\eea
After performing the integration over $\bar k$ and renaming $\bar l\leftrightarrow l$, one can easily see that the integration over $l$ picks up two poles:
\bea
P_4&=&2\pi\int_{l,\bar l}\frac{3}{2}\frac{1}{q^2}\frac{1}{l^4\bar l^4}\ln\left( \frac{l^2}{Q_s^2}\right)
%\\
%&&\times
 \,\left\{ \frac{3}{2}\frac{1}{p^2}\left[\frac{1}{(l-q+p)^2}\right]_{\bar l}+\frac{3}{2}\frac{1}{(\bar l-p)^2}\left[ \frac{1}{(l-\bar l-q+p)^2}\right]_{\bar l}\right\}.
 %\nonumber
\eea
The second term in the brackets picks up two poles when integrating over $\bar l$ and it gives a suppressed contribution with respect to the first term, thus can be neglected. Then, the leading contribution comes from the first term in the brackets and after integrating over $l$ and $\bar l$, we get
\bea
P_4\approx 2{\pi}^3\frac{9}{4}\frac{1}{p^2q^2}\frac{1}{(p-q)^4}\ln\left[\frac{(p-q)^2}{Q_s^2}\right]\frac{1}{Q_T^2}\ln\left( \frac{Q_T^2}{\Lambda^2}\right).
\eea
Adding all contributions together, we get
\bea
J_3&\approx&\pi^3\frac{9}{2}
\Bigg\{
\left[\frac{3}{q^2p^2}+\frac{1}{q^4}\right]\frac{1}{(p-q)^4}\ln\left[\frac{(p-q)^2}{Q_s^2}\right]
+\frac{1}{p^4q^4}\ln\left(\frac{p^2}{Q_s^2}\right)
\Bigg\}
%\nonumber\\&&\times \,
 \frac{1}{Q_T^2}\ln\left(\frac{Q_T^2}{\Lambda^2}\right)\mu^4\lambda^4.
\eea

\subsubsection{$B_4$}

The fourth term in the $B$-type contribution reads
\bea
J_4&=&\int_{ k,\bar{k}, l,\bar{l}}\frac{\mu^2(k)\mu^2(\bar{k})\lambda^2(l)\lambda^2(\bar{l})}{l^4\bar{l}^4}\delta^{(2)}(k+l-p-\bar{k}-\bar{l}+q)\nonumber\\
&&\times
 {\rm tr}\left\{ \Psi(k,l,p;0) \Psi^*(k,l,p;1) \Psi(\bar k, \bar l, q; 1)  \Psi^*(\bar k, \bar l, q; 0)
\right\} \nonumber\\
&=& 2 \int_{ k,\bar{k} ,l,\bar{l}}\frac{\mu^2(k)\mu^2(\bar{k})\lambda^2(l)\lambda^2(\bar{l})}{l^4\bar{l}^4}\delta^{(2)}(k+l-p-\bar{k}-\bar{l}+q)\\
&\times&\Bigg\{\Bigg(
\left[ \frac{(k+l)\cdot p}{(k+l)^2p^2}-\frac{k\cdot (p-l)}{k^2(p-l)^2} \right]
\left[ \frac{(k+l)\cdot (k+l-p)}{(k+l)^2(k+l-p)^2} - \frac{k\cdot(k+l-p)}{k^2(k+l-p)^2} \right]\nonumber\\
&&\hspace{1cm}-4
\left[ \frac{(k+l) \times p}{(k+l)^2p^2}-\frac{k \times (p-l)}{k^2(p-l)^2} \right]
\left[ \frac{(k+l) \times (k+l-p)}{(k+l)^2(k+l-p)^2} - \frac{k \times (k+l-p)}{k^2(k+l-p)^2} \right] \Bigg)\nonumber\\
&&\times
\Bigg(
\left[ \frac{(\bar k+\bar l)\cdot q}{(\bar k+\bar l)^2q^2}-\frac{\bar k\cdot (q-\bar l)}{\bar k^2(q-\bar l)^2} \right]
\left[ \frac{(\bar k+\bar l)\cdot (\bar k+\bar l-q)}{(\bar k+\bar l)^2(\bar k+\bar l-q)^2} - \frac{\bar k\cdot(\bar k+\bar l-q)}{\bar k^2(\bar k+\bar l-q)^2} \right]\nonumber\\
&&\hspace{1cm}-4
\left[ \frac{(\bar k+\bar l)\times q}{(\bar k+\bar l)^2q^2}-\frac{\bar k\times (q-\bar l)}{\bar k^2(q-\bar l)^2} \right]
\left[ \frac{(\bar k+\bar l)\times (\bar k+\bar l-q)}{(\bar k+\bar l)^2(\bar k+\bar l-q)^2} - \frac{\bar k\times(\bar k+\bar l-q)}{\bar k^2(\bar k+\bar l-q)^2} \right] \Bigg)\nonumber\\
&&+ 4 \Bigg(
\left[ \frac{(k+l)\cdot p}{(k+l)^2p^2}-\frac{k\cdot (p-l)}{k^2(p-l)^2} \right]
\left[ \frac{(k+l)\times (k+l-p)}{(k+l)^2(k+l-p)^2} - \frac{k\times (k+l-p)}{k^2(k+l-p)^2} \right]\nonumber\\
&&\hspace{1cm}+
\left[ \frac{(k+l)\cdot (k+l-p)}{(k+l)^2(k+l-p)^2} - \frac{k\cdot(k+l-p)}{k^2(k+l-p)^2} \right]
\left[ \frac{(k+l) \times p}{(k+l)^2p^2}-\frac{k \times (p-l)}{k^2(p-l)^2} \right] \Bigg) \nonumber\\
&&\times
\Bigg(
\left[ \frac{(\bar k+\bar l)\cdot q}{(\bar k+\bar l)^2q^2}-\frac{\bar k\cdot (q-\bar l)}{k^2(q-\bar l)^2} \right]
\left[ \frac{(\bar k+\bar l)\times (\bar k+\bar l-q)}{(\bar k+\bar l)^2(\bar k+\bar l-q)^2} - \frac{\bar k\times(\bar k+\bar l-q)}{\bar k^2(\bar k+\bar l-q)^2} \right]\nonumber\\
&&\hspace{1cm}+
\left[ \frac{(\bar k+\bar l)\cdot (\bar k+\bar l-q)}{(\bar k+\bar l)^2(\bar k+\bar l-q)^2} - \frac{\bar k\cdot(\bar k+\bar l-q)}{\bar k^2(\bar k+\bar l-q)^2} \right]
\left[ \frac{(\bar k+\bar l)\times q}{(\bar k+\bar l)^2q^2}-\frac{\bar k\times (q-\bar l)}{k^2(q-\bar l)^2} \right]
\Bigg)\Bigg\}.\nonumber
\eea
There are again four pole contributions: $P_1: k=0$, $P_2: k+l=0$, $P_3: \bar k+\bar l=0$ and $P_4:\bar k=0$. However, $B_4$-type terms are symmetric under the exchange $(k,l,p)\leftrightarrow (\bar k, \bar l, q)$. Thus, for these terms we only need to calculate the $P_1$ and $P_2$ contributions. So, let us start with the $P_1$ contribution:
\bea
P_1&=&2\int_{k,l,\bar l}\frac{3}{2}\frac{1}{l^4\bar l^4}\frac{1}{(p-l)^2}\left(\frac{1}{k^2}\right)_l
\Bigg\{ \left[  \frac{(l-p+q)\cdot q}{(l-p+q)^2q^2}-\frac{(l-\bar l-p+q)\cdot (q-\bar l)}{(l-\bar l-p+q)^2(q-\bar l)^2}\right]\\
&&\times
\left[ \frac{(l-p+q)\cdot (l-p)}{(l-p+q)^2(l-p)^2}-\frac{(l-\bar l-p+q)\cdot(l-p)}{(l-\bar l-p+q)^2(l-p)^2}\right]\nonumber\\
&&-4
\left[  \frac{(l-p+q)\times q}{(l-p+q)^2q^2}-\frac{(l-\bar l-p+q)\times (q-\bar l)}{(l-\bar l-p+q)^2(q-\bar l)^2}\right]\nonumber\\
&&\times
\left[ \frac{(l-p+q)\times (l-p)}{(l-p+q)^2(l-p)^2}-\frac{(l-\bar l-p+q)\times(l-p)}{(l-\bar l-p+q)^2(l-p)^2}\right]\Bigg\}.\nonumber
\eea
The integration over $l$ picks up two poles:
\bea
P_1&=& 2\pi\int_{l,\bar l}\frac{3}{2}\frac{1}{l^4\bar l^4}\ln\left(\frac{l^2}{Q_s^2}\right)
\left\{ \frac{3}{2}\frac{1}{q^4}\left[\frac{1}{(l-p+q)^2}\right]_{\bar l}+\frac{3}{2}\frac{1}{(q-\bar l)^4}\left[\frac{1}{(l-\bar l-p+q)^2}\right]_{\bar l}\right\}\nonumber\\
&\approx& 2{\pi}^2\int_{\bar l}\frac{9}{4}\frac{1}{(p-q)^4}\ln\left[\frac{(p-q)^2}{Q_s^2}\right]\left[\frac{1}{q^4}+\frac{1}{(q-\bar l)^4}\right] \frac{1}{\bar l^4}\ln\left(\frac{\bar l^2}{\Lambda^2}\right).
\eea
Note that the second term in the brackets has a double pole when integrating over $\bar l$ and it does not pick up a factor of $Q_T^2$ in the denominator after the integration over $\bar l$. Thus, it is suppressed with respect to the first term in the bracket and can be neglected. Hence, the leading contribution comes from the first term and after performing the $\bar l$ integral, we get
\bea
P_1\approx 2{\pi}^3\frac{9}{4}\frac{1}{q^4}\frac{1}{(p-q)^4}\ln\left[\frac{(p-q)^2}{Q_s^2}\right]\frac{1}{Q_T^2}\ln\left(\frac{Q_T^2}{\Lambda^2}\right).
\eea
As we have argued before, the $P_4$ contribution is identical to $P_1$  when $p$ and $q$ are exchanged. Thus, we can  write the result of $P_4$ as
\bea
P_4\approx 2{\pi}^3\frac{9}{4}\frac{1}{p^4}\frac{1}{(p-q)^4}\ln\left[\frac{(p-q)^2}{Q_s^2}\right]\frac{1}{Q_T^2}\ln\left(\frac{Q_T^2}{\Lambda^2}\right).
\eea

The contribution from the pole at $k+l=0$ can be written as
\bea
P_2&=&2\int_{k,l,\bar l}\frac{3}{2}\frac{1}{l^4\bar l^4}\frac{1}{p^2}\left[\frac{1}{(k+l)^2}\right]_l \\
&&\times
\Bigg\{ \left[ \frac{(q-p)\cdot q}{(q-p)^2q^2}-\frac{(q-\bar l-p)\cdot(q-\bar l)}{(q-\bar l-p)^2(q-\bar l)^2}\right]
\left[\frac{(q-p)\cdot (-p)}{(q-p)^2p^2}-\frac{(q-\bar l-p)\cdot (-p)}{(q-\bar l-p)^2p^2}\right]\nonumber\\
&&-4
\left[ \frac{(q-p)\times q}{(q-p)^2q^2}-\frac{(q-\bar l-p)\times(q-\bar l)}{(q-\bar l-p)^2(q-\bar l)^2}\right]
\left[\frac{(q-p)\times (-p)}{(q-p)^2p^2}-\frac{(q-\bar l-p)\times (-p)}{(q-\bar l-p)^2p^2}\right]\Bigg\}.\nonumber
\eea
The integration over $\bar l$ picks up one pole:
\bea
P_2\approx2\int_{k,l,\bar l}\frac{9}{4}\frac{1}{l^4\bar l^4}\frac{1}{p^4}\left[\frac{1}{(k+l)^2}\right]_l\left[\frac{1}{(\bar l-q+p)^2}\right]_{Q_s, p-q}.
\eea
Integrating over all the variables we get
\bea
P_2\approx 2\pi^3\frac{9}{4}\frac{1}{p^4}\frac{1}{(p-q)^4}\ln\left[\frac{(p-q)^2}{Q_s^2}\right]\frac{1}{Q_T^2}\ln\left(\frac{Q_T^2}{\Lambda^2}\right).
\eea
The contribution from $P_3$ is identical to $P_2$  when $p$ and $q$ are exchanged. Thus, $P_3$  reads
\bea
P_3\approx 2\pi^3\frac{9}{4}\frac{1}{q^4}\frac{1}{(p-q)^4}\ln\left[\frac{(p-q)^2}{Q_s^2}\right]\frac{1}{Q_T^2}\ln\left(\frac{Q_T^2}{\Lambda^2}\right).
\eea
Adding all contributions together, we get
\bea
J_4&\approx&\pi^3\frac{9}{2}
\left[\frac{2}{q^4}+\frac{2}{p^4}\right]\frac{1}{(p-q)^4}\ln\left[\frac{(p-q)^2}{Q_s^2}\right] 
%\nonumber\\ &&\times\,
\frac{1}{Q_T^2}\ln\left(\frac{Q_T^2}{\Lambda^2}\right)\mu^4\lambda^4.
\eea

  \end{document}